\pdfoutput=1
\documentclass[11pt]{article}
\usepackage{cite}
\usepackage{graphicx}
\usepackage{latexsym}   
\usepackage{mathrsfs}
\usepackage[scr=rsfs,cal=boondox]{mathalfa}
\usepackage[overload]{textcase}

\font\grande=cmr9.5 scaled \magstep4
\font\medio=cmr9.5 scaled \magstep2
\outer\def\beginsection#1\par{\medbreak\bigskip
      \message{#1}\leftline{\bf#1}\nobreak\medskip
\vskip-\parskip
      \noindent}

\begin{document}
\bibliographystyle{unsrt}

\titlepage
\vspace{1cm}
\begin{center}
{\grande The hypermagnetic power spectra}\\
\vspace{0.5cm}
{\grande and the phases of Sakharov oscillations}\\
\vspace{1.5 cm}
Massimo Giovannini \footnote{e-mail address: massimo.giovannini@cern.ch}\\
\vspace{0.5cm}
{{\sl Department of Physics, CERN, 1211 Geneva 23, Switzerland }}\\
\vspace{0.5cm}
{{\sl INFN, Section of Milan-Bicocca, 20126 Milan, Italy}}\\
\vspace*{1cm}
\end{center}
\vskip 0.3cm
\centerline{\medio  Abstract}
\vskip 0.5cm
If the gauge fields are amplified from the inflationary vacuum, the quantum mechanical initial data correspond to travelling waves that turn asymptotically into standing waves whose phases only depend on the evolution of the gauge coupling. We point out that these gauge analogs of the Sakharov oscillations are exchanged by the duality symmetry and ultimately constrain both the relative scaling of the hypermagnetic power spectra and their final asymptotic values. Unlike the case of the density contrasts in a relativistic plasma, the standing oscillations never develop since they are eventually overdamped by the finite value of the conductivity as soon as the corresponding modes are comparable with the expansion rates after inflation. We show that the late-time value of the magnetic field is not determined at radiation dominance (and in spite of the value of the wavenumber) but  it depends on the moment when the  wavelengths (comparable with the Mpc) get of the order of the Hubble radius before equality. This means that the magnetogenesis requirements are only relaxed if the post-inflationary expansion rate is slower than radiation but the opposite is true when the plasma expands faster than radiation and the corresponding power spectra are further suppressed. After combining the present findings with the evolution of the gauge coupling we show that these results are consistent with a magnetogenesis scenario where the gauge coupling is always perturbative during the inflationary stage while, in the dual case, the same requirements cannot be satisfied. 
\noindent
\vspace{5mm}
\vfill
\newpage

\renewcommand{\theequation}{1.\arabic{equation}}
\setcounter{equation}{0}
\section{Introduction}
\label{sec1}
The early evolution of cosmological gauge fields is essential for large-scale magnetism \cite{one,two,three,four,five} and it has been closely scrutinized since the pioneering attempts of Hoyle and Zeldovich. If the space-time is homogeneous (but anisotropic) the gauge fields lead to Bianchi-type backgrounds \cite{six}. Conversely when the plasma is isotropic (but not completely homogeneous)  the gauge fields must not distort the peaks of the temperature and polarization observables of microwave background \cite{seven}. Already the first few WMAP data releases \cite{eight,nine} (eventually confirmed by the latest observations \cite{ten}) corroborated the adiabatic and Gaussian nature of large-scale inhomogeneities and strongly constrained the possibility of a large-scale anisotropy induced by the gauge fields. If a constant electric (or magnetic) field  gravitates there exist explicit anisotropic solutions where the symmetry of the metric mirrors the orientations of the gauge fields\footnote{For instance a constant magnetic field is polarized 
along a specific (Cartesian) direction the four-dimensional metric is generally anisotropic and the magnetic energy density scales according to the covariant conservation of the total energy-momentum tensor of the system.}.  This strategy breaks explicitly the isotropy of the background but it is still viable if we ought to infer the cosmological limits on slightly skew stresses, as originally discussed in \cite{eleven}. While not all the anisotropic 
background geometries are compatible with the presence of a uniform magnetic or electric field, the logic of the  type-I models can be generalized to different Bianchi classes \cite{six} (see, for instance,\cite{twelve,thirteen}). The current formulation of the concordance paradigm stipulates that the underlying geometry is conformally flat \cite{eight,nine,ten} and this occurrence is broadly consistent with the presence of an early inflationary stage of expansion that washes out the intrinsic curvature and the spatial gradients \cite{elevena,elevenb}. If we assume the strict validity of general relativity also during inflation, the evolution of  gauge fields is generally Weyl invariant \cite{five} and duality invariant \cite{fivea,fiveb}. As in the anisotropic case, the scaling of the gauge fields is therefore obtained from the evolution of the total energy-momentum tensor of the system as established long ago in the covariant treatment of charged fluids \cite{five}. 

While below the scale of the formation of the light nuclei the variation of the gauge couplings is severely constrained \cite{elevend,elevene,elevenf,eleveng}, there are no reasons why the same should also happen in the early Universe. On the other hand, if the gauge coupling evolves the Weyl invariance is broken and the gauge fields can be generated from their quantum fluctuations. Because of the late-time constraints on the variation of the gauge couplings it is natural to assume that the breaking of Weyl invariance takes place during inflation and immediately after it. In concrete terms the inflationary breaking of Weyl invariance may happen when the inflaton or some other spectator field inherits either a scalar or a pseudo-scalar coupling to the gauge fields. The axion-like couplings \cite{fourteen,fourteena,fifteena,fifteenb} have been first investigated with the purpose of breaking  Weyl invariance \cite{sixteen,seventeen,eighteen} (see also \cite{nineteen}) but more specific analyses demonstrated that the pseudoscalar vertex efficiently amplifies only the wavelengths that are localized around the comoving Hubble radius during inflation \cite{seventeen,eighteen}. This coupling is however relevant for the production of the magnetic helicity and of the magnetic gyrotropy (i.e. field configurations whose flux lines are linked or twisted). The  produced Chern-Simons condensate leads to a viable mechanism for baryogenesis via hypermagnetic knots \cite{nineteen} and plays a relevant role in anomalous magnetohydrodynamics where the evolution of the hypermagnetic fields at finite conductivity is analyzed in the presence of anomalous charges \cite{twentytwo}. In the collisions of heavy ions this phenomenon is often dubbed chiral magnetic effect \cite{twentythree,twentythreea}. Because of the relative inefficiency of the pseudo-scalar vertex, the attention has been moved to a second class of scenarios where the inflaton and the spectator fields couple predominantly (but not exclusively) to the kinetic term of the gauge fields. In this case large-scale magnetic fields could be produced both during inflation \cite{twentyfour} and in the context of specific bouncing scenarios \cite{twentyfive} with a number of features that have been investigated through the years (see, for instance, Refs. \cite{CC0,CC1,CC2,CC3,CC4,CC5,CC6,CC7,CC8,CC9,CC10,CC11,CC12} for an incomplete list of contributions). There are finally scenarios where both scalar and pseudoscalar 
couplings are included on equal footing and these models are relevant for baryogenesis \cite{CC13,CC14}, as originally suggested in \cite{nineteen}.

The purpose here is not to single out a specific scenario but rather to discuss the general 
scaling properties of the produced (hypercharge) fields and their impact on the final values of the (electromagnetic) fields. 
If the modes of the field are quantum mechanically generated during inflation their are asymptotically represented by travelling waves. However later on, thanks to the evolution of the gauge coupling, the travelling waves turn into standing waves. This 
occurrence is not surprising since, from a quantum mechanical viewpoint, the parametric amplification can be viewed as the creation of pairs of particles with equal energies and oppositely directed momenta. This aspect has been first pointed out by A. D. Sakharov so that the standing waves that normally arise in these processes are often dubbed Sakharov oscillations \cite{fortytwo,fortythree}. In the hot big-bang scenario the standing waves that characterize the density contrast and the matter power spectrum have been originally identified by Peebles and Yu \cite{SAK2} in their pioneering paper on the foundations of the so-called adiabatic paradigm. These gauge analogs of the Sakharov oscillations are exchanged by the duality symmetry \cite{fivea,fiveb} (see also \cite{fortyseven}) and ultimately constrain both the relative scaling of the hypermagnetic power spectra and their final asymptotic values in a way that has not been fully clarified so far.

One of the relevant aspects of this analysis is that the phases of the standing waves bear the mark of the dynamics of the gauge coupling. In particular we shall demonstrate that when the gauge coupling increases during inflation and then flattens out the hypermagnetic power spectra predominantly oscillate like sines in the asymptotic region after the end of inflation:
\begin{equation}
P_{B}(k,\tau)\, \propto \, \sin^2{k\,\tau}, \qquad \qquad P_{E}(k,\tau)\, \propto \, \cos^2{k\,\tau}.
\label{FOURb}
\end{equation}
In the dual dynamical evolution (i.e. when the gauge coupling first decreases and then flattens out) the corresponding phases are exchanged because 
of the duality symmetry of the problem. If $\overline{P}_{B}(k,\tau)$ and $\overline{P}_{E}(k,\tau)$ are the hypermagnetic and hyperelectric fields obtained when the gauge coupling decreases and then relaxes after inflation we have that the corresponding phases are exchanged in comparison with the result of Eq. (\ref{FOURb}):
\begin{equation}
\overline{P}_{B}(k,\tau)\, \propto \, \cos^2{k\,\tau}, \qquad \qquad \overline{P}_{E}(k,\tau)\, \propto \, \sin^2{k\,\tau}.
\label{FOURc}
\end{equation}
After symmetry breaking the electromagnetic power spectra at late times are computed from the physical power spectra. To evaluate the late-time power spectra in the current literature there are two complementary requirements that are often used interchangeably. In the first case the comoving power spectra are evaluated 
when radiation starts dominating the evolution of the plasma (i.e. $\tau_{r}$ in what 
follows);  another possibility is that the comoving 
power spectra must be evaluated at $\tau_{k}$ which is the moment 
at which the given scale associated with magnetogenesis 
crosses the (comoving) Hubble radius. These two time-scales are in general 
very different since they have to do with totally different physical considerations.
A related issue is therefore given by the final value of the magnetic power 
spectra when the Universe is not immediately dominated by radiation.
Since $\tau_{r}$ and $\tau_{k}$ are different the final results may also be very different
and this is related to the the Sakharov phases as they emerge from the inflationary 
stage of expansion. The 
analysis reported in this paper ultimately supports the idea that a stage of increasing 
gauge coupling satisfies the magnetogenesis requirements while the early 
evolution of the hypercharge fields is always perturbative. This actually true 
provided the late-time power spectra are correctly evaluated; as we shall see this 
is not always the case.

The layout of this paper is, in short, the following. The general perspective of the effective treatment of the magnetogenesis scenarios is swiftly recalled in section \ref{sec2} and, in particular, the 
interplay between the effective theories of inflation and the magnetogenesis scenarios is considered
in subsection \ref{subs21}. In subsection \ref{subs22} we introduce the quantum mechanical aspects of the problem and the definition of the gauge power spectra for the general class of scenarios where the hyperelectric and the hypermagnetic susceptibilities do not necessarily coincide. Finally, in the last part of the section we derive the general scaling of the gauge power 
spectra during inflation (subsection \ref{subs23}) and after inflation (subsection \ref{subs24}).
The results of section \ref{sec2} are eventually  applied in section \ref{sec3} where 
the gauge analog of the Sakharov oscillations are discussed in the case 
of increasing (see subsection \ref{subs31}) and decreasing gauge couplings. 
In subsection \ref{subs32} we  consider the role of the physical power 
spectra as opposed to the comoving ones. Section \ref{sec4} is devoted to the crossing 
time $\tau_{k}$ that is one of the essential scales of the problem but it is often overlooked.
While before $\tau_{k}$ the Sakharov oscillations can be expanded and do not 
modify the power spectra (see subsection \ref{subs42}) when 
$\tau\geq \tau_{k}$ the conductivity dominates, duality gets broken 
and this ultimately is the moment where the mode functions change their evolution (see subsection \ref{subs43}).
The implications of the results of sections \ref{sec3} and \ref{sec4} 
are finally discussed in section \ref{sec5} where  the final asymptotic values of the gauge power spectra are explicitly discussed with the secondary aim of addressing some recent claims.
In subsection \ref{subs51} we consider the simplest possibility of a post-inflationary 
radiation stage while in subsection \ref{subs52} we address the case of the intermediate 
phases different from radiation. The final part of the section deals with two 
complementary aspects. The first one (see subsection \ref{subs53}) 
considers the interplay between the scales of the problem and the duality symmetry; the 
question is if and how the crossing time $\tau_{k}$ could be mistaken for the 
time of radiation dominance. Finally subsection \ref{subs54} it will be shown that 
the Sakharov phases also arise when the pseudo-scalar couplings are 
present in the original action. 
Section \ref{sec6} contains the concluding remarks.
Two technical aspects of the problem have been relegated to the appendices and, 
in particular, the limits of the comoving power spectra (see appendix \ref{APPA}) 
and the general formulas arising in the discussion of multiple post-inflationary phases 
(see appendix \ref{APPB}).

\renewcommand{\theequation}{2.\arabic{equation}}
\setcounter{equation}{0}
\section{Evolution of the gauge fields and duality}
\label{sec2}
\subsection{Effective theories, inflation and magnetogenesis}
\label{subs21}
Even if similar discussions could be conducted in the case of scalar fields not directly connected with the inflaton, 
for the sake of concreteness we now focus on the dynamics of single field inflationary scenarios described in terms of a scalar-tensor theory for the field $\varphi$ with potential $V(\varphi)$; the effective action of $\varphi$ at tree level is given by\footnote{In what follows the Greek indices are four-dimensional while the lowercase Latin indices run over the 
three spatial dimensions. The signature of the metric is 
$(+,\, -,\, -,\,-)$. The reduced Planck mass $\overline{M}_{P} = M_{P}/\sqrt{8 \pi}$ is defined in terms of $M_{P}= 1/\sqrt{G}$.}
\begin{equation}
S= \int d^{4} x\,\,\sqrt{- G} \biggl[ - \frac{ \overline{M}_{P}^2\,\, R}{ 2} + 
\frac{1}{2} G^{\alpha\beta} \partial_{\alpha} \varphi \partial_{\beta} \varphi - V(\varphi)\biggr],
\label{2one}
\end{equation}
which is in fact the first term of a theory whose higher derivatives are suppressed by the (negative) powers of a (large) mass scale $M$. While the methods of effective field theory are customarily employed to classify single-field inflationary models \cite{fortyeight}, 
it has been pointed out that the same logic can be applied to the magnetogenesis scenarios \cite{fortynine}.  As there exist both generic and non-generic models of inflation based on the action (\ref{2one}), the same is true in the case of the magnetogenesis scenarios \cite{fortyeight,fortynine}.  If we introduce $\phi = \varphi/M$ (i.e. the inflaton field in units of the effective mass scale) the leading correction to Eq. (\ref{2one}) consists then of all possible terms (overall $12$ in four space-time dimensions) containing {\em four derivatives}:
\begin{itemize}
\item{} there are $3$ terms only involving the higher derivatives 
of the inflaton field [e.g. $(G^{\alpha\beta}\partial_{\alpha}\phi \partial_{\beta} \phi)^2$];
\item{} there are $3$ further corrections where the inflaton multiplies 
either a single power of the Ricci scalar (i.e. $R \, G^{\alpha\beta} \partial_{\alpha}\phi\partial_{\beta} \phi$) or the Ricci tensor (i.e.  
$R_{\alpha\beta} \partial^{\alpha}\phi \, \partial^{\beta} \phi$);
\item{} the remaining $6$ terms are quadratic in the curvature 
and involve not only $R^2$  or $R_{\mu\nu}\, R^{\mu\nu}$ but 
also $R_{\mu\alpha\nu\beta} \, \widetilde{\,R\,}^{\mu\alpha\nu\beta}$
and $C_{\mu\alpha\nu\beta} \, \widetilde{\,C\,}^{\mu\alpha\nu\beta}$
where $R_{\mu\alpha\nu\beta}$ and $C_{\mu\alpha\nu\beta}$ denote the Riemann and  Weyl tensors while 
$ \widetilde{\,R\,}^{\mu\alpha\nu\beta}$ and $\widetilde{\,C\,}^{\mu\alpha\nu\beta}$ are the corresponding duals.
\end{itemize}
Each of these $12$ corrections (listed and discussed in Ref. \cite{fortyeight}) is weighted by inflaton-dependent couplings that do not contain any derivative and the same 
logic can be extended to the magnetogenesis 
scenarios so that the effective action of Eq. (\ref{2one}) is now complemented by 
the contribution of the hypercharge fields where $\lambda= \lambda(\phi)$ and $\overline{\lambda}= \overline{\lambda}(\phi)$ 
 parametrize, respectively, the scalar and the pseudo-scalar interactions
 \begin{equation}
S= - \frac{1}{16 \pi} \int d^4 x\,\, \sqrt{-G} \,\, \biggl[ \lambda \,\, Y_{\mu\nu} \, Y^{\mu\nu} +
\overline{\lambda}  \,\, Y_{\mu\nu} \, \widetilde{\,Y\,}^{\mu\nu} \biggr],
\label{TWO}
\end{equation}
and $Y_{\mu\nu}$ denotes the gauge field strength while $\widetilde{\,Y\,}^{\mu\nu} = E^{\mu\nu\alpha\beta} \, Y_{\alpha\beta}/2$ is the dual field\footnote{Recall, in this respect, that $E^{\mu\nu\alpha\beta} = \epsilon^{\mu\nu\alpha\beta}/\sqrt{-G}$ where 
 $\epsilon^{\mu\nu\alpha\beta}$ is the Levi-Civita symbol in four dimensions.}.
In the notations of Eq. (\ref{TWO}) the gauge coupling is defined as $g^2 = 4\,\pi/\lambda$.  The corrections to Eq. (\ref{TWO}) consist of $14$ terms that also contain four derivatives and that are weighted by field-dependent couplings; these terms have been already studies in Ref. \cite{fortynine} 
and can be schematically written as:
\begin{eqnarray}
&& \Delta {\mathcal L}_{gauge} = \frac{\sqrt{-G}}{16 \, \pi\, M^2} \biggl[ \lambda_{1}(\phi) \, R\, Y_{\alpha\beta}\, Y^{\alpha\beta} + \,.\,.\,.\,.\,+ \lambda_{7}(\phi) \nabla_{\mu}\nabla^{\nu} \phi\,  Y_{\nu\alpha} Y^{\mu\alpha}
\nonumber\\
&& + \overline{\lambda}_{1}(\phi) \, R\, Y_{\alpha\beta}\, \widetilde{\, Y\,}^{\alpha\beta}  +  \,.\,.\,.\,.\,+ \overline{\lambda}_{7}(\phi) \nabla_{\mu}\nabla^{\nu} \phi\,  Y_{\nu\alpha} \, \widetilde{\,Y\,}^{\mu\alpha}
 \biggr].
\label{2two}
\end{eqnarray}
Concerning Eqs. (\ref{TWO})--(\ref{2two}) the following remarks are in order:
\begin{itemize}
\item{} of the $14$ different terms appearing in Eq. (\ref{2two}) the first $7$  are parity-even while the remaining  $7$are parity-odd; the contributions that do not break parity are associated with $\lambda_{i}(\phi)$  while the ones that break parity are multiplied by  $\overline{\,\lambda\,}_{i}(\phi)$ where, in both cases, $i =1,\,\, ...\,,\,\,7$;
\item{} in the absence 
of $\phi$-dependence the first term of Eq. (\ref{2two}) (together with the remaining two contributions i.e. $\lambda_{2}(\phi) \, R_{\mu}^{\,\,\,\,\,\nu} \, Y^{\mu\alpha}\, Y_{\alpha\nu} $ and $\lambda_{3}(\phi) \, R_{\mu\alpha\nu\beta} \, Y^{\mu\alpha} \, Y^{\nu\beta} $) have been analyzed in  Ref. \cite{fifty,fiftyone} 
and have been also applied to the analysis of large-scale magnetism long ago mostly with negative conclusions \cite{fiftytwo};
\item{} the covariant gradients of the inflaton  (e.g. 
 $\lambda_{5}(\phi) \Box \phi Y_{\alpha\beta} Y^{\alpha\beta}$ and $\lambda_{6}(\phi) \partial_{\mu}\phi \partial^{\nu}\phi 
Y^{\mu\alpha} \, Y_{\nu\alpha}$) together with their corresponding duals (i.e. $\overline{\lambda}_{5}(\phi) \Box \phi Y_{\alpha\beta} \widetilde{\,Y\,}^{\alpha\beta}$ and $\overline{\lambda}_{6}(\phi) \partial_{\mu}\phi \partial^{\nu}\phi 
\widetilde{\,Y\,}^{\mu\alpha} \, Y_{\nu\alpha}$) arise in the relativistic theory of Van der Waals (or Casimir-Polder) interactions in flat \cite{fiftythree,fiftyfour} and curved \cite{fiftyfive} backgrounds.
\end{itemize}
What matters, for the present ends, is that the full Lagrangian density (i.e. ${\mathcal L}_{gauge} + \Delta {\mathcal L}_{gauge}$) in a conformally 
flat background geometry  (i.e. $\overline{G}_{\mu\nu} = a^2(\tau) \, \, \eta_{\mu\nu}$) becomes 
\begin{eqnarray}
S_{gauge} &=& \int d^{3} x \int d\tau \biggl({\mathcal L}_{gauge} + \Delta {\mathcal L}_{gauge}\biggr) 
\nonumber\\
&=& \frac{1}{2}\int d^{3} x \int d\tau \biggl( E^2 - B^2 +  \frac{\overline{\chi}^2}{\chi_{B} \, \chi_{E}}\, \vec{E}\cdot\vec{B} \biggr),
\label{2three}
\end{eqnarray}
where $\vec{E}(\vec{x},\tau)$ and $\vec{B}(\vec{x},\tau)$ denote the comoving fields that are related to their physical counterparts $\vec{e}(\vec{x},\tau)$ and $\vec{{\mathcal b}}(\vec{x},\tau)$ as:
\begin{equation}
\vec{B}(\vec{x},\tau) = a^2(\tau)\,\, \chi_{B}(\tau) \,\, \vec{{\mathcal b}}(\vec{x},\tau), \qquad\qquad \vec{E}(\vec{x},\tau) = a^2(\tau) \,\,\chi_{E}(\tau) \,\, \vec{{\mathcal e}}(\vec{x},\tau).
\label{2four}
\end{equation}
To derive Eq. (\ref{2three}) from Eq. (\ref{2two}) we must recall that, in terms of the physical fields, the gauge field strengths are $Y^{i\,0} = {\mathcal e}^{i}/a^2$ and $Y^{i\,j} = - \epsilon^{i\,j\,k} {\mathcal b}_{k}/a^2$ and the equations for the comoving fields follow from Eq. (\ref{2three}):
\begin{eqnarray}
&& \vec{\nabla}\times \bigl( \chi_{B} \, \vec{B} \bigr) - \partial_{\tau} \bigl(\chi_{E} \, \vec{E}\bigr) +
\biggl(\frac{\partial_{\tau}\overline{\chi}^2}{\chi_{B}} \biggr)\vec{B}  =0,\qquad \vec{\nabla} \cdot \bigl( \chi_{E} \, \vec{E} \bigr) +\vec{\nabla} \cdot \biggl[\biggl(\frac{\overline{\chi}^2}{\chi_{B}}\biggr) \vec{B}\biggr] =0,
\label{2seven}\\
&& \partial_{\tau} \biggl(\frac{\vec{B}}{\chi_{B}}\biggr) + \vec{\nabla} \times \biggl(\frac{\vec{E}}{\chi_{E}}\biggr) =0,\qquad \vec{\nabla}\cdot\biggl(\frac{\vec{B}}{\chi_{B}}\biggr) =0.
\label{2eight}
\end{eqnarray}
In Eqs. (\ref{2three})--(\ref{2four}) $\chi_{E}$ and $\chi_{B}$ denote, respectively, the electric and the magnetic 
susceptibilities whose explicit expressions depends on all the effective couplings 
listed in Eq. (\ref{2two}). We note that when $\overline{\chi} \to  0$ (i.e. in the absence of parity-breaking terms) Eqs. (\ref{2seven})--(\ref{2eight}) are invariant for the standard duality\footnote{In the case $\overline{\chi} \neq 0$ the duality symmetry can be generalized but we shall not analyze the explicit transformation rules since they are not strictly essential in the forthcoming discussion.} symmetry \cite{fivea,fiveb}:  when the susceptibilities are exchanged (i.e. $\chi_{B}\to 1/\chi_{E}$) the underlying equations are invariant provided  $\vec{B} \to \vec{E}$ and $\vec{E} \to - \, \vec{B}$.  The explicit forms of the susceptibilities are all different however the corresponding expressions share similar features;  for instance, the expression of $\chi_{B}^2$ is:
\begin{equation}
\chi_{B}^2 = \frac{\lambda}{4 \pi} \biggl( 1 + \frac{H^2 }{M^2} d_{B}^{(1)} - \epsilon \frac{H^2}{M^2} d_{B}^{(2)} - \sqrt{\epsilon}\, \frac{H^2 M_{P}}{M^3} d_{B}^{(3)} +\sqrt{\epsilon} \,\, \eta\, \frac{H^2 M_{P}}{M^3} \,\, d_{B}^{(4)}\biggr),
\label{AA2}
\end{equation}
where $d_{B}^{(i)}$ with $i= 1,\, 2,\, 3,\, 4$ are $\phi$-dependent 
coefficients and similar explicit expressions hold for the other two susceptibilities $\chi_{E}^2$ and $\overline{\chi}^2$ (see \cite{fortynine} for the explicit expressions). However since $\epsilon = - \dot{H}/H^2$ and $\eta = - \ddot{\phi}/(H \, \dot{\phi})$ are the conventional slow-roll parameters, from Eq. (\ref{AA2}) (and from its 
analogs in the hyeperelectric and scalar cases) we can deduce the following three related observations:
\begin{itemize}
\item{} if  $\epsilon \leq 1$ 
the scale $M$ may coincide, for practical purposes, with the Planck scale (i.e. $M \simeq \sqrt{2 \epsilon} \,\overline{M}_{P}$): this is the case of generic theories of inflation (when $\phi$ is not constrained by symmetry principles) and $M= {\mathcal O}(M_{P})$  otherwise  $\dot{\phi}/H$ would diverge; this happens since, within the 
slow-roll approximation it follows that $ \dot{\phi}/H = \sqrt{2 \epsilon} \,\,\overline{M}_{P}/M$;
\item{} if $M= \sqrt{2 \epsilon} \,\,\overline{M}_{P}$ then $H/M$ is slightly larger than $H/\overline{M}_{P}$;
\item{} if $\epsilon\ll 1$ we should then require  that 
$M \gg \sqrt{2\epsilon} \, \overline{M}_{P}$
\end{itemize}
In the case of conventional inflationary scenarios the physical wavenumber $k/a$ and the Hubble rate coincide at horizon exit and, more precisely, we will have that 
\begin{equation}
\overline{M}_{P}^2 H^2/M_{P}^4 = \epsilon {\mathcal A}_{{\mathcal R}}/8 = {\mathcal O}(10^{-10}) \epsilon,
\label{DEFes}
\end{equation}
where ${\mathcal A}_{{\mathcal R}} = {\mathcal O}(2.41\times 10^{-9})$ is the amplitude of the curvature inhomogeneities assigned at the pivot scale $k_{p} = 0.002\, \mathrm{Mpc}^{-1}$. Keeping track of the various factors and introducing  
${\mathcal A}_{0} = 8 \pi^3 {\mathcal A}_{{\mathcal R}}$ the leading contributions to $\chi_{E}^2$, $\chi_{B}^2$ and $\overline{\chi}^2$ are all ${\mathcal O}({\mathcal A_{0}}/\epsilon)$ while the subleading terms are 
 ${\mathcal O}(\eta\, {\mathcal A}_{0}/\epsilon)$ and ${\mathcal O}({\mathcal A}_{0})$.
The observational determinations of the tensor to scalar ratio $r_{T}$ range between $r_{T} < 0.07$ \cite{RT0,RT2} and $r_{T}< 0.01$  
\cite{ten}. Since the consistency relations 
stipulate that $\epsilon \simeq r_{T}/16$, we have to acknowledge that  $\epsilon< 10^{-3} $ so that we are in the last case discussed above. We should then require that $M \gg \sqrt{2\epsilon} \, \overline{M}_{P}$,  $ M\simeq \overline{M}_{P}$ and $\epsilon \ll 1$ so that the leading contributions appearing in Eq. (\ref{AA2}) is associated with $d_{B}^{(1)}$ which is overall ${\mathcal O}(8 \pi {\mathcal A}_{0} \epsilon)$ and exactly the same result follows from the analysis of $\chi_{E}^2$ and $\overline{\chi}^2$.  All in all, for the present ends  we have two opposite cases:
\begin{itemize}
\item{} in the first situation the naturalness of the couplings would imply that the $\lambda_{i}(\phi)$ are all of the order of $\lambda(\phi)$ and similarly for the $\overline{\lambda}_{i}(\phi)$ which should all be ${\mathcal O}(\overline{\lambda})$; the same conclusion follows if $\lambda_{i}(\phi) \ll \lambda(\phi)$ and $\overline{\lambda}_{i}(\phi) \ll \overline{\lambda}(\phi)$; 
\item{} in the opposite situation $\lambda_{i}(\phi) \gg \lambda(\phi)$ and $\overline{\lambda}_{i}(\phi) \gg \overline{\lambda}(\phi)$ the hyperelectric and the hypermagnetic susceptibilities may evolve at different rates.  
\end{itemize}
Even if similar considerations  can be generalized to the case of non-generic models \cite{fiftyfive}, in what follows we shall prefrentially focus on the first case that is the one of {\em generic models} of inflation where the two susceptibilities $\chi_{B}$ and $\chi_{E}$ coincide.

\subsection{Field operators and gauge power spectra}
\label{subs22}
From Eq. (\ref{2three}) the action for the vector potential can be written
 in the Coulomb gauge:
\begin{equation}
S_{gauge} = \frac{1}{2}\int d^{3}x \, \int d\tau\, \biggl[ \chi_{E}^2 \bigl(\partial_{\tau} \, Y_{i}\bigr)^2 - \chi_{B}^2 (\partial_{k} Y_{i})^2 - 
\partial_{\tau} \overline{\chi}^2 Y_{k} \bigl(\partial_{i} \, Y_{j}) \epsilon^{i \, j\, k} \biggr].
\end{equation}
The previous action takes a simpler form if it is rephrased in terms of a new time parametrization and by rescaling 
the susceptibilities as follows:
\begin{equation}
\tau\to \eta = \eta(\tau), \qquad  d\tau= n(\eta)\, d\eta,\qquad \chi_{E} = \sqrt{n} \, \chi, \qquad \chi_{B} = \frac{\chi}{\sqrt{n}}.
\label{etatime}
\end{equation}
In the $\eta$-parametrization the previous action can then be expressed as
$\vec{{\mathcal A}}= \chi\, \vec{Y}$: 
\begin{equation}
S_{gauge}= \frac{1}{2} \int d^3 x\, \int \, d\eta \, \biggl[ \dot{{\mathcal A}}_{a}^2 + \biggl(\frac{\dot{\chi}}{\chi}\biggr)^2 {\mathcal A}_{a}^2  - 2  \biggl(\frac{\dot{\chi}}{\chi}\biggr) 
{\mathcal A}_{a} \, \dot{\mathcal A}_{a} - \partial_{i} {\mathcal A}_{a} \partial^{i} {\mathcal A}_{a} - 
\frac{\dot{\overline{\chi}}^2}{\chi^2}
{\mathcal A}_{a} \partial_{b} {\mathcal A}_{m} \, \epsilon^{a \, b\, m} \biggr],
\end{equation}
where the overdots now denote a derivation with respect to the new time coordinate $\eta$. The derivatives with respect to $\eta$ and the derivations with respect to the cosmic time coordinate $t$ never appear in this subsection\footnote{Note furthermore that the variable $\eta$ should not 
be confused with one of the slow-roll parameters introduced earlier on in Eq. (\ref{AA2}).}.
The classical fields and the conjugate momenta can then be promoted to the status of quantum operators so that the mode expansion of the hyperelectric and hypermagnetic fields in the circular basis turns out to be:
\begin{eqnarray}
&&\widehat{E}_{i}(\vec{x},\eta) = -  \int\frac{d^{3} k}{(2\pi)^{3/2} \, \sqrt{n(\eta)}} \sum_{\alpha=R,\,L} \, 
\biggl[ g_{k,\,\alpha}(\eta) \, \widehat{a}_{k,\alpha} \,\,  \hat{\varepsilon}^{(\alpha)}_{i}(\hat{k})\,\,e^{- i \vec{k} \cdot\vec{x}} + \mathrm{H.c.}\biggr],
\label{EIGHTEEN}\\
&&\widehat{B}_{k}(\vec{x}, \eta) =  - i   \int\, \frac{\,\epsilon_{i\,j\,k}\, k_{i} \,d^{3} k}{(2\pi)^{3/2}\, \sqrt{n(\eta)}} \sum_{\alpha=R,\,L}\,
\biggl[ f_{k,\, \alpha}(\eta) \, \widehat{a}_{k,\,\alpha}\, \,\,  \hat{\varepsilon}^{(\alpha)}_{j}(\hat{k})\, e^{- i \vec{k} \cdot\vec{x}} -\mathrm{H.c.}\biggr],
\label{NINETEEN}
\end{eqnarray}
where $\hat{\varepsilon}^{(R)}(\hat{k})$ and  $\hat{\varepsilon}^{(L)}(\hat{k})$ denote the two complex polarization obeying 
$\hat{k} \times \hat{\varepsilon}^{(R)} = - i\, \hat{\varepsilon}^{(R)}$ and $\hat{k} \times \hat{\varepsilon}^{(L)} =  i\, \hat{\varepsilon}^{(L)}$. In Eq. (\ref{NINETEEN}) we introduced the right-handed polarizations and their left-handed counterparts 
both defined in terms of the linear polarizations $\hat{e}_{\oplus}$ and $\hat{e}_{\otimes}$: 
\begin{equation}
 \hat{\varepsilon}^{(R)}(\hat{k})= (\hat{e}_{\oplus} + i\, \hat{e}_{\otimes})/\sqrt{2}, \qquad\qquad \hat{\varepsilon}^{(L)\ast}(\hat{k}) = \hat{\varepsilon}^{(R)}(\hat{k}), \qquad \qquad \hat{e}_{\oplus} \times \hat{e}_{\otimes} = \hat{k}.
\end{equation} 
 The creation and annihilation operators are directly defined in the circular basis and they obey the standard commutation relation $[\widehat{a}_{\vec{k}, \, \alpha}, \, \widehat{a}_{\vec{p}, \, \beta}] = \delta^{(3)}(\vec{k}- \vec{p})\, \delta_{\alpha\beta}$. The $R$-mode functions obey:
\begin{equation}
\dot{f}_{k,\, R} = g_{k,\,R} + (\dot{\chi}/\chi) f_{k,\, R},\qquad
\dot{g}_{k,\,R} = - k^2 \, f_{k,\, R} - (\dot{\chi}/\chi) \,g_{k,\,R}  +  \, (\dot{\overline{\chi}}^2/\chi^2) \, k \, f_{k,\, R},
\label{RMODE1}
\end{equation}
where, as already mentioned that overdot indicates a derivation with respect to the $\eta$-time. If the two susceptibilities coincide (as we shall be mostly discuss hereunder) we have that $n(\eta) \to 1$ in Eq. (\ref{etatime}) and 
the conformal time coordinate coincides with $\eta$; in this case the overdot and the prime (denoting a derivation with respect to $\tau$) can be used interchangeably.  The two equations appearing in (\ref{RMODE1}) 
can be combined in a single expression:
\begin{equation}
\ddot{f}_{k,\,R}+ [k^2 - \ddot{\chi}/\chi] f_{k,\,R} - (\dot{\overline{\chi}}^2/\chi^2) k f_{k,\, R} =0,
\label{RMODE2}
\end{equation}
Similarly, for the $L$-mode the equations are 
\begin{equation}
\dot{f}_{k,\, L} = g_{k,\,L} + (\dot{\chi}/\chi) f_{k,\, L},\qquad
\dot{g}_{k,\,L} = - k^2 \, f_{k,\, L} - (\dot{\chi}/\chi) \,g_{k,\,L}  -  \, (\dot{\overline{\chi}}^2/\chi^2) \, k \, f_{k,\, L},
\label{LMODE1}
\end{equation}
and their decoupled form can be written as:
\begin{equation}
 \ddot{f}_{k,\,L}+ [k^2 - \ddot{\chi}/\chi] f_{k,\,L} + (\dot{\overline{\chi}}^2/\chi^2) k f_{k,\, L} =0.
\label{LMODE2}
\end{equation}
From Eqs. (\ref{EIGHTEEN})--(\ref{NINETEEN}) the 
two-point functions in Fourier space become:
\begin{eqnarray}
&& \langle \widehat{E}_{i}(\vec{k},\,\eta)\, \widehat{E}_{j}(\vec{p},\,\eta) \rangle = \frac{ 2 \pi^2 }{k^3} \biggl[\, P_{E}(k,\eta) \, p_{ij}(\hat{k}) 
+ P_{E}^{(G)}(k,\eta)\, \, i\, \epsilon_{i\, j\, \ell} \, \hat{k}^{\ell}\biggr] \, \delta^{(3)}(\vec{p} + \vec{k}),
\label{TWENTY1}\\
&& \langle \widehat{B}_{i}(\vec{k},\,\eta)\, \widehat{B}_{j}(\vec{p},\,\eta) \rangle = \frac{ 2 \pi^2 }{k^3} \biggl[\, P_{B}(k,\eta) \, p_{ij}(\hat{k}) 
+ P_{B}^{(G)}(k,\eta)\, \,i\, \epsilon_{i\, j\, \ell} \, \hat{k}^{\ell}\biggr] \, \delta^{(3)}(\vec{p} + \vec{k}),
\label{TWENTY2}
\end{eqnarray}
where $p_{i j} = \delta_{i j} - \hat{k}_{i} \, \hat{k}_{j}$ is the usual divergenceless projector.
In Eqs. (\ref{TWENTY1})--(\ref{TWENTY2}) $P_{E}(k,\,\eta)$ and $P_{B}(k,\,\eta)$ denote the hyperelectric and the hypermagnetic power spectra
while $P_{E}^{(G)}(k,\,\eta)$ and $P_{B}^{(G)}(k,\,\eta)$ account for corresponding gyrotropic contributions:
\begin{eqnarray}
P_{E}(k,\,\eta) &=& \frac{k^{3}}{4 \pi^2\, n(\eta)} \biggl[ \bigl| g_{k,\,L}\bigr|^2 + \bigl| g_{k,\,R}\bigr|^2 \biggr], \qquad
P_{B}(k,\,\eta) = \frac{k^{5}}{4 \pi^2\, n} \biggl[ \bigl| f_{k,\,L}\bigr|^2 + \bigl| f_{k,\,R}\bigr|^2 \biggr],
\label{TWENTY3}\\
P_{E}^{(G)}(k,\,\eta) &=& \frac{k^{3}}{4 \pi^2\, n(\eta)} \biggl[  \bigl| g_{k,\,L}\bigr|^2 - \bigl| g_{k,\,R}\bigr|^2 \biggr],\qquad
P_{B}^{(G)}(k,\,\eta) =  \frac{k^{5}}{4 \pi^2\, n(\eta)} \biggl[ \bigl| f_{k,\,L}\bigr|^2 - \bigl| f_{k,\,R}\bigr|^2\biggr].
\label{TWENTY4}
\end{eqnarray}
We note that the presence of either $P_{E}^{(G)}(k,\,\eta)$ or  $P_{B}^{(G)}(k,\,\eta)$ implies that 
the expectation values of either the hyperelectric (i.e. $\langle \vec{E} \cdot \vec{\nabla} \times \vec{E} \rangle$) 
or of the hypermagnetic (i.e. $\langle \vec{B} \cdot \vec{\nabla} \times \vec{B} \rangle$) gyrotropies 
do not vanish.

\subsection{General scaling during inflation}
\label{subs23}
As established in the subsection \ref{subs21} for generic models of inflation we can safely assume 
that $M= {\mathcal O}(\overline{M}_{P})$ and $\epsilon \ll 1$ so that the possible anisotropies 
in the coupling constants can be ignored (i.e. $n(\tau) = n(\eta) \to 1$).  This means that during inflation the scaling of the gauge power spectra and of the spectral energy density depends on the rate of variation of the susceptibility $\chi$ (or of the gauge coupling). If $\overline{\lambda} \to 0$ 
and in the case of an {\em increasing gauge} coupling during a stage of quasi-de Sitter expansion 
\begin{equation}
g= \frac{1}{ \chi}= \sqrt{ \frac{4 \pi}{\lambda}}, \qquad \qquad \frac{g^{\prime}}{g} = \gamma {\mathcal H}, \qquad\qquad \gamma >0,
\label{ggg1}
\end{equation}
where $\gamma$ measures the rate of increase of the gauge-coupling; note, incidentally, that we shall be denoting one of the mode functions by $g_{k}$ but, because of the subscript, this notation cannot be confused with the gauge coupling 
$g(\tau)$. 

In an effective theory sense, for a generic model of inflation the power spectra of the hyperelctric and hypermagnetic fields ultimately depend on $\gamma$ and they follow from the solutions of Eqs. (\ref{RMODE1})--(\ref{LMODE1}) and from Eq. (\ref{ggg1}). If we now impose quantum mechanical initial conditions and take into account Eq. (\ref{ggg1}),  the comoving power spectra of  Eqs. (\ref{TWENTY3})  become:
\begin{eqnarray}
P_{E}(k, \tau) &=& {\mathcal H}^4 \, {\mathcal C}(|\gamma -1/2|) \, \biggl( \frac{k}{{\mathcal H}} \biggr)^{m_{E}}, \qquad \qquad m_{E} = 5 - | 2 \gamma -1|,
\label{PE1}\\
P_{B}(k, \tau) &=& {\mathcal H}^4\, {\mathcal C}(\gamma +1/2) \, \biggl( \frac{k}{{\mathcal H}} \biggr)^{m_{B}}, \qquad \qquad m_{B} = 4 - 2\gamma,
\label{PB1}
\end{eqnarray}
where ${\mathcal C}(x) = \Gamma^2(x) 2^{ 2 x - 3}/\pi^3$ and ${\mathcal H} = a^{\prime}/a$ denotes the expansion rate during inflation. In the case of slow-roll dynamics for a generic inflationary scenario where $ \epsilon \ll 1$  we have that $ -(1 -\epsilon) \tau \simeq {\mathcal H}^{-1}$ implying that ${\mathcal H} \simeq -1/\tau$ up to slow-roll corrections\footnote{The gyrotropic contributions of Eq. (\ref{TWENTY4}) vanish 
in the present situation but the essence of the problem described here is also present when $\overline{\lambda} \neq 0$ 
(see, in this respect, subsection \ref{subs54} and discussion therein).}. 
Equations (\ref{PE1})--(\ref{PB1}) {\em do not} coincide with the gauge power spectra after inflation and determine directly the spectra energy density of the gauge fields \cite{fortyseven}. The power spectra of Eqs. (\ref{PE1})--(\ref{PB1}) are both comoving 
since they are both computed in terms $\vec{E}$ and $\vec{B}$; the 
physical power spectra are instead denoted throughout in calligraphic style:
\begin{eqnarray}
{\mathcal P}_{E}(k,\tau) &=& \frac{P_{E}(k,\tau)}{\lambda(\tau)\,\, a^{4}(\tau)} = \frac{g^2(\tau)\,  P_{E}(k,\tau)}{4\,\pi \,a^4(\tau)},
\label{PE1F}\\
{\mathcal P}_{B}(k,\tau) &=& \frac{P_{B}(k,\tau)}{\lambda(\tau)\,\, a^4(\tau)} = \frac{g^2(\tau)\, P_{B}(k,\tau)}{4\,\pi \,a^4(\tau)}.
\label{PB1F}
\end{eqnarray}
The definitions of Eqs. (\ref{PE1F})--(\ref{PB1F}) do not depend on the inflationary dynamics 
but on the relation between physical and comoving fields expressed by Eq. (\ref{2four}).
Both the physical and the comoving spectra can be related to the various components 
of the energy-momentum tensor:
\begin{equation}
T_{\mu}^{\,\,\,\, \nu} = \frac{1}{g^2} \biggl[ - Y_{\mu}^{\,\,\,\, \alpha} \,\, Y_{\alpha}^{\,\,\,\,\nu} + 
\frac{1}{4} Y_{\alpha\beta}\, Y^{\alpha\beta} \, \delta_{\mu}^{\,\,\, \nu} \biggr].
\label{TENS00}
\end{equation}
From the timelike components of Eq. (\ref{TENS00}) the energy density can be written in terms of the comoving fields and, in this case, it is suppressed as $a^{-4}$. If the energy density is instead expressed in terms of the physical fields 
it scales as $g^{-2}$; In formulas we have 
\begin{equation}
\rho_{Y}(\vec{x},\tau) = \frac{\lambda}{8 \pi} \biggl[ {\mathcal e}^2 + {\mathcal b}^2] = 
\frac{1}{ 8 \pi\, a^4} \biggl[ E^2 + B^2 \biggr],
\label{TENS00a}
\end{equation}
and similar considerations  hold for all the other components of the energy momentum tensor.
In Eq. (\ref{TENS00a})  ($\vec{{\mathcal e}}$, $\vec{{\mathcal b}}$) are 
the physical fields while ($\vec{E}$, $\vec{B}$) denote the corresponding comoving 
fields. If the energy density is averaged with the help 
of the two-point functions of Eqs. (\ref{EIGHTEEN})--(\ref{NINETEEN}):
\begin{equation}
\langle \mathrm{vac} | \hat{\rho}_{Y} | \mathrm{vac} \rangle = \frac{1}{4\pi a^4} \int \frac{d k}{k} \biggl[ 
P_{E}(k,\tau) + P_{B}(k,\tau) \biggr].
\label{EN1}
\end{equation}
The integral appearing in Eq. (\ref{EN1}) is defined, in practice over all the modes 
that are amplified during the inflationary stage and, for this reason, it is practical to divide 
Eq. (\ref{EN1}) by the critical energy density with the purpose of 
introducing the spectral energy density in critical units, i.e. $\Omega_{Y}(k,\, a) = (d \langle 
\hat{\rho}_{Y} \rangle/d \ln{k})/\rho_{crit}$ where $\rho_{crit} = 3 \, H^2 \, \overline{M}_{P}^2$.
Therefore, from the explicit form of Eq. (\ref{EN1}) and from Eqs. (\ref{PE1})--(\ref{PB1}) we 
obtain
\begin{equation}
\Omega_{Y}(k,a) = \frac{2}{3} \frac{H^2}{M_{P}^2} \biggl[ {\mathcal C}(|\gamma -1/2|) \, 
\biggl( \frac{k}{a \, H} \biggr)^{m_{E}} + {\mathcal C}(\gamma +1/2) \, \biggl( \frac{k}{a \, H} \biggr)^{m_{B}}\biggr].
\label{OM1}
\end{equation}
Equation (\ref{OM1}) estimates the spectral energy density {\em during the inflationary stage}
and it is only meaningful when $a \leq a_{1}$.
Since the spectral energy density must always be subcritical  we must have $\Omega_{Y}(k,\, a) \ll 1$
and this only happens when $\gamma \leq 2$. If we now go back to Eqs. (\ref{PE1})--(\ref{PB1}) we see 
that when $\gamma \to 2$ the electric spectrum is flat while the magnetic spectrum is steeply 
increasing (i.e. $m_{E} \to 0$ while $m_{B} \to 2$ for $\gamma\to 2$). This 
conclusion does not forbid that {\em after inflation the magnetic power spectrum 
is flat} since, as stressed above, the previous expressions are only valid during the inflationary stage.

The previous results based on Eq. (\ref{ggg1}) can be easily translated to the case of {\em decreasing gauge coupling} where 
\begin{equation}
\frac{g^{\prime}}{g} = - \,\overline{\gamma} \,\,{\mathcal H}, \qquad \mathrm{ with} \qquad \overline{\gamma}>0,
\end{equation}
now measures the rate of decrease of the gauge coupling in units of the inflationary expansion rate. The (comoving) gauge power spectra become therefore
\begin{eqnarray}
\overline{P}_{E}(k, \tau) &=& {\mathcal H}^4 {\mathcal C}(\overline{\gamma} +1/2) \, \biggl( \frac{k}{{\mathcal H}} \biggr)^{\overline{m}_{E}}, \qquad \qquad \overline{m}_{E} = 4 - 2 \overline{\gamma},
\label{PE2}\\
\overline{P}_{B}(k, \tau) &=& {\mathcal H}^{4} {\mathcal C}(|\overline{\gamma} -1/2|) \, \biggl( \frac{k}{{\mathcal H}} \biggr)^{\overline{m}_{B}}, \qquad \qquad \overline{m}_{B} = 5 - |2\overline{\gamma} - 1|.
\label{PB2}
\end{eqnarray}
The results of Eqs. (\ref{PE2})--(\ref{PB2}) are related to the ones of Eqs. (\ref{PE1})--(\ref{PB1}) 
by duality transformations since, under inversion of the gauge coupling (i.e. $g \to 1/g$),
$\gamma \to \overline{\gamma}$ and 
\begin{equation}
P_{E}(k,\tau) \to \overline{P}_{B}(k, \tau), \qquad\qquad P_{B}(k,\tau) \to \overline{P}_{E}(k, \tau).
\end{equation}
As a consequence the spectral energy density in critical units now takes the form:
\begin{equation}
\overline{\Omega}_{Y}(k,a) = \frac{2}{3} \frac{H^2}{M_{P}^2} \biggl[ {\mathcal C}(\overline{\gamma} +1/2) \, 
\biggl( \frac{k}{a \, H} \biggr)^{\overline{m}_{E}} + {\mathcal C}(|\overline{\gamma} -1/2|) \, \biggl( \frac{k}{a \, H} \biggr)^{\overline{m}_{B}}\biggr].
\label{OM2}
\end{equation}
In this case we see that the conditions imposed by the critical density bound imply that $\overline{\gamma} \leq 2$ 
and this means that, during inflation, it is possible to have a flat hyperelectic spectrum. Conversely 
the hypermagnetic spectrum can be flat during inflation provided $\overline{\gamma}\to 2$.
According to Ref. \cite{CC3} it is mandatory to have a decreasing gauge coupling if we want, for some reason, 
a flat hypermagnetic spectrum at the present time. This argument is however 
incorrect since the power spectra during inflation and after inflation are physically and mathematically 
different, as we shall see in a moment.

\subsection{General scaling after inflation}
\label{subs24}
After inflation both the evolution of the gauge coupling and of the 
expansion rate are modified but duality still remains the relevant symmetry of the problem as long as the evolution takes place for $ - \tau_{1} \leq \tau < \tau_{k}$.
We are generally interested in the case where, after inflation, the gauge coupling flattens out and this 
dynamics depends upon the specific inflationary dynamics. In general terms, however, the 
mode functions after inflation are related, by continuity, to their values at the end of inflation 
by an appropriate transition matrix:
\begin{eqnarray} 
f_{k}(\tau) &=& {\mathcal T}_{f \, f}(k, \tau_{1}, \tau) \, f_{k}(-\tau_{1}) + {\mathcal T}_{f \, g}(k, \tau_{1}, \tau) \, g_{k}(-\tau_{1})/k,
\nonumber\\
g_{k}(\tau) &=& {\mathcal T}_{g \, f}(k, \tau_{1}, \tau) \,k\, f_{k}(-\tau_{1}) + {\mathcal T}_{g \, g}(k, \tau_{1}, \tau) \, g_{k}(-\tau_{1}),
\label{TRANS1}
\end{eqnarray}
where $-\tau_{1}$ is the transition time between two successive stages of the evolution of the gauge 
coupling. Since the various ${\mathcal T}_{X,\, Y}$ (with $X= f,\, g$ and $Y= f,\, g$) are all real functions 
of their respective arguments we get the explicit form of  Eq. (\ref{TRANS1}) where the mode 
functions are evaluated, by definition, at the end of the inflationary stage (i.e. for $\tau= - \tau_{1}$).
The Wronskian normalization for $|\tau| > \tau_{1}$ also implies a relation among 
the elements of the transition matrix:
\begin{equation}
{\mathcal T}_{f\, f}(k, \tau_{1}, \tau)\,\,{\mathcal T}_{g\, g}(k, \tau_{1}, \tau) - {\mathcal T}_{f\, g}(k, \tau_{1}, \tau) \,\, {\mathcal T}_{g\,\, f}(k, \tau_{1}, \tau) = 1.
\label{TRANS2}
\end{equation}
Equation (\ref{TRANS1}) only assumes the continuity of the background geometry 
and of the gauge coupling across the inflationary transition and it implies that the 
comoving power spectra {\em after inflation} are given by:
\begin{eqnarray}
P_{E}(k, \tau) &=& \frac{k^3}{2\pi^2} \biggl| {\mathcal T}_{g\, f}(k, \tau_{1}, \tau) \, k\, f_{k} + {\mathcal T}_{g\, g}(k, \tau_{1}, \tau) \, g_{k} \biggr|^2,
\label{TRANS3}\\
P_{B}(k,\tau) &=&  \frac{k^5}{2\pi^2} \biggl| {\mathcal T}_{f\, f}(k, \tau_{1}, \tau) \, f_{k} + {\mathcal T}_{f\, g}(k, \tau_{1}, \tau) \, \frac{g_{k}}{k} \biggr|^2,
\label{TRANS4}
\end{eqnarray}
where, for the sake of conciseness, all the arguments of the corresponding 
mode functions have been dropped. Under a duality transformation  the mode functions transform as:
\begin{equation}
g \to 1/g, \qquad\qquad f_{k} \to \overline{g}_{k}/k, \qquad \qquad g_{k} \to - k \, \overline{f}_{k},
\label{TRANS5}
\end{equation}
and the elements of the transition matrix transform instead as 
\begin{eqnarray}
&& {\mathcal T}_{f\, f}(k, \tau_{1}, \tau) \to \overline{{\mathcal T}}_{g\,g}(k, \tau_{1}, \tau), \qquad\qquad {\mathcal T}_{g\, g}(k, \tau_{1}, \tau) \to \overline{{\mathcal T}}_{f\,f}(k, \tau_{1}, \tau),
\nonumber\\
&& {\mathcal T}_{f\, g}(k, \tau_{1}, \tau) \to - \, \overline{{\mathcal T}}_{g\,f}(k, \tau_{1}, \tau), \qquad\qquad {\mathcal T}_{g\, f}(k, \tau_{1}, \tau) \to -\,\overline{{\mathcal T}}_{f\,g}(k, \tau_{1}, \tau).
\label{TRANS6}
\end{eqnarray}
The gauge power spectra computed from the evolution with inverted gauge coupling are then given by:
\begin{eqnarray}
\overline{P}_{E}(k, \tau) &=& \frac{k^3}{2\pi^2} \biggl| \overline{{\mathcal T}}_{g\, f}(k, \tau_{1}, \tau) \, k\, \overline{f}_{k} + \overline{{\mathcal T}}_{g\, g}(k, \tau_{1}, \tau) \, \overline{g}_{k} \biggr|^2,
\label{TRANS7}\\
\overline{P}_{B}(k,\tau) &=&  \frac{k^5}{2\pi^2} \biggl| \overline{{\mathcal T}}_{f\, f}(k, \tau_{1}, \tau) \, \overline{f}_{k} + \overline{{\mathcal T}}_{f\, g}(k, \tau_{1}, \tau) \, \frac{\overline{g}_{k}}{k} \biggr|^2.
\label{TRANS8}
\end{eqnarray}
Recalling now Eqs. (\ref{TRANS3})--(\ref{TRANS4}), the transformations of Eqs. (\ref{TRANS5})--(\ref{TRANS6}) imply 
that, under duality, 
\begin{equation}
P_{E}(k,\tau) \to \overline{P}_{B}(k,\tau), \qquad \qquad P_{B}(k,\tau) \to \overline{P}_{E}(k,\tau).
\label{TRANS9}
\end{equation}
The results of Eqs. (\ref{TRANS7})--(\ref{TRANS8}) and (\ref{TRANS9}) apply generically 
after the end of inflation. The different dynamical evolutions of the gauge coupling can then be translated one into the other by using the duality properties of the related power spectra. 

\renewcommand{\theequation}{3.\arabic{equation}}
\setcounter{equation}{0}
\section{The gauge power spectra after inflation}
\label{sec3}
The bottom line of the previous section is that
the general scaling of the power spectra at late times (i.e. well after inflation) is 
determined by the properties of the transition matrix and the physical evolution
suggests that the gauge coupling must vary after inflation but it progressively 
flattens out as the post-inflationary phase develops. This is consistent 
with the idea that the gauge coupling should freeze somewhere 
between the end of inflation and the beginning of radiation since 
today the variation of the gauge couplings is severely 
constrained \cite{elevend,elevene,elevenf}.

With these two premises, we now intend to show that 
after the end of inflation the gauge power spectra necessarily exhibit standing 
oscillations that are the gauge analog of the Sakharov 
phases. The power spectra computed in this section are {\em only} valid for typical 
wavelengths larger than the Hubble radius: it would be actually incorrect to forget that
the asymptotic values of the gauge fields {\em after Hubble crossing} is fixed by
the finite value of the conductivity and not simply by the asymptotic evolution of the gauge 
coupling. As soon as the modes of the 
hypermagnetic and hyperectric fields get of the order of the comoving Hubble radius 
they are overdamped and the standing oscillations never develop. In view of its relevance 
this point is separately analyzed in section \ref{sec4}.

\subsection{The Sakharov phases in the case of increasing gauge coupling} 
\label{subs31}
After the end of inflation, 
the gauge coupling flattens out and gradually freezes. More precisely, while for $\tau< - \tau_{1}$, $g^{\prime}/g = \gamma \, {\mathcal H}$ (with $\gamma > 0$, see Eq. (\ref{ggg1}))
for $\tau \geq - \tau_{1}$ we have instead
\begin{equation}
\frac{g^{\prime}}{g} = \delta \, {\mathcal H}, \qquad\qquad 0 < \delta \ll \gamma.
\label{POST1}
\end{equation}
Both $\gamma$ and $\delta$ are positive semi-definite but the case of decreasing gauge coupling is separately discussed in the following subsection 
and it is swiftly obtained by using the a duality transformations (\ref{TRANS5})--(\ref{TRANS6}).
As before, in Eq. (\ref{POST1}) $\delta$ measures the rate of variation of the gauge coupling in units of the comoving Hubble rate. By continuity we can then determine the elements of the transition matrix and the diagonal elements are given by:
\begin{equation}
{\mathcal T}_{f\, f} = - {\mathcal M}_{\nu} \,\,{\mathcal N}_{\nu-1} \,\, \sin{(\theta_{\nu} - \phi_{\nu-1})}, \qquad \qquad{\mathcal T}_{f\, g} = {\mathcal M}_{\nu} \,\,{\mathcal N}_{\nu} \,\,\sin{(\theta_{\nu} - \phi_{\nu})},
\label{DIAG1}
\end{equation}
while the off-diagonal elements  are instead:
\begin{equation}
{\mathcal T}_{f\, g} = - {\mathcal M}_{\nu-1} \,\,{\mathcal N}_{\nu-1} \,\, \sin{(\theta_{\nu-1} - \phi_{\nu-1})}, \qquad \qquad{\mathcal T}_{g\, g} = {\mathcal M}_{\nu-1} \,\,{\mathcal N}_{\nu} \,\,\sin{(\theta_{\nu-1} - \phi_{\nu})}.
\label{DIAG2}
\end{equation}
 Equations (\ref{DIAG1})--(\ref{DIAG2}) take a simple form if we choose to represent the products 
 of Bessel functions in terms of their moduli and phases \cite{abr1,abr2}. Concerning 
  Eqs. (\ref{DIAG1})--(\ref{DIAG2}) the following comments are in order: 
 \begin{itemize}
\item{} the value of the Bessel index follows from Eq. (\ref{POST1}) and it is given by 
 $\nu = \delta +1/2$; in more general terms $\nu$ is determined by the rate of variation of the gauge coupling {\em after inflation}; 
\item{} ${\mathcal M}_{\nu}$  and ${\mathcal N}_{\nu}$ are related to the moduli of the Bessel functions and while $\theta_{\nu}$ and $\phi_{\nu}$ are the corresponding phases. 
\item{} according 
 to the standard conventions the modulus $M_{\nu}$ of the Bessel function is introduced as \cite{abr2}:
 \begin{equation}
J_{\nu}(k \,y) = M_{\nu} \cos{\theta_{\nu}}, \qquad \qquad Y_{\nu}(k \,y) = M_{\nu} \sin{\theta_{\nu}},
 \label{POST3}
 \end{equation}
 where, in our notations $M_{\nu} = M_{\nu}(k y)$ and $\theta_{\nu} = \theta_{\nu}(k y)$; note 
 that $k y \to k \tau$ when $\tau \gg -\tau_{1}$;
\item{} the continuity of the mode functions fixes their arguments so that, for instance, 
 $y(\tau) = \tau + \tau_{1} (1 + q)$  with $q = \gamma/\delta$ is the 
 ratio between the rates of variation  of the gauge coupling during and after inflation.  
\end{itemize} 
 With the same notations we introduce also the amplitudes and phases for the Bessel functions with argument $q x_{1}$ where $x_{1} = k \tau_{1}$:
 \begin{equation}
J_{\nu}(q \,x_{1}) = N_{\nu} \,\cos{\phi_{\nu}}, \qquad \qquad Y_{\nu}(q \,x_{1}) = N_{\nu} \,\sin{\phi_{\nu}},
 \label{POST4}
 \end{equation}
where, this time, $N_{\nu}= N_{\nu}(q \,x_{1})$ and $\phi_{\nu} = \phi_{\nu}(x_{1})$. The variables ${\mathcal M}_{\nu}(k\, y)$ and ${\mathcal N}_{\nu}( q \,x_{1})$ appearing in Eqs. (\ref{DIAG1})--(\ref{DIAG2}) are related to $M_{\nu}(k y)$ and $N_{\nu}( q x_{1})$ as 
 \begin{equation}
 {\mathcal M}_{\nu}(k \,y) = \sqrt{\frac{\pi}{2}} \,\, \sqrt{k\,y} \,\,M_{\nu}(k\, y), \qquad \qquad 
{\mathcal N}_{\nu}(q \,x_{1}) = \sqrt{\frac{\pi}{2}} \,\, \sqrt{q x_{1}} \,\,N_{\nu}(q\, x_{1}).
\label{POST5}
\end{equation}
Since the continuity of the mode functions is explicitly implemented
the limit $\delta \to 0$ is well defined and it corresponds to the case where $\delta$ is arbitrarily small but finite in all the intermediate steps of the calculation. The second small parameter 
of the problem is $k\tau_{1}$ since it is always true that the comoving wavenumbers 
must not exceed the maximal frequency of the spectrum:
\begin{equation}
x_{1} = k \,\tau_{1} = \frac{k}{a_{1} \, H_{1}} \ll  1, \qquad \qquad \biggl(\frac{H_{1}}{M_{P}}\biggr) = \frac{\sqrt{\pi \, r_{T}\, {\mathcal A}_{{\mathcal R}}}}{4},
\label{POST6}
\end{equation}
where ${\mathcal A}_{{\mathcal R}}$ denotes the amplitude of the scalar power spectrum at the pivot scale $k_{p}=0.002\,\mathrm{Mpc}^{-1}$ and $r_{T}$ is the tensor to scalar ratio\footnote{Equation  (\ref{POST6}) has actually the same content of Eq. (\ref{DEFes}) with the difference that we now traded $\epsilon$ for $r_{T}$ by using the consistency relations.}. In practice, the limit of Eq. (\ref{POST6}) is always realized for all the modes that are substantially amplified so that, inserting  Eqs. (\ref{DIAG1})--(\ref{DIAG2}) into Eq. (\ref{TRANS7}) the explicit 
expressions for the comoving power spectra take a factored form where the first term 
depends on $k \tau_{1}$ while the second is only a function of $k\, y$:
\begin{equation}
P_{E}(k,\,\tau_{1},\, \tau) = a_{1}^4 \, H_{1}^4\, p_{E}(k\, \tau_{1}, \gamma, \delta) \,\, f_{E}(k\,y, \delta).
\label{POST7}
\end{equation}
In the limit $k \tau_{1} \ll 1$ the exact expressions of $p_{E}(k\, \tau_{1}, \gamma, \delta)$ and $f_{E}(k \,y, \delta)$ take a particularly simple form that is often used in the forthcoming considerations:
\begin{eqnarray} 
\lim_{|k \tau_{1}| < 1} \,\, p_{E}(k\, \tau_{1},\, \gamma,\, \delta) &=& {\mathcal C}(\gamma +1/2)  \,\, 
\biggl(\frac{k}{a_{1} \, H_{1}}\biggr)^{ 4 - 2 \gamma - 2 \delta},
\label{POST7b}\\
\lim_{|k \tau_{1}| < 1} \,\, f_{E}(k \, y, \delta) &=& \biggl(\frac{q}{2}\biggr)^{-2 \delta} \,\, \biggl(\frac{k \, y}{2}\biggr)\, \,J_{\delta- 1/2}^{2}(k\, y).
\label{POST7c}
\end{eqnarray}
Recalling that $y(\tau) = \tau + \tau_{1} (1 + q)$, it follows that $k \, y \to k\tau$ in the limit $k \tau_{1} < 1$ so that, in practice, $f_{E}(k \, y, \delta) \to f_{E}(k\tau, \delta)$. By going through the same steps can be in the case of the comoving hypermagnetic spectrum we obtain 
a similar result but with different phases:
\begin{equation}
P_{B}(k,\,\tau_{1},\, \tau) = a_{1}^4 \, H_{1}^4 \,\,p_{B}(k\, \tau_{1}, \gamma, \delta) \,\, f_{B}(k\,y, \delta).
\label{POST8}
\end{equation}
Again, in the limit $k \tau_{1} < 1$,  $p_{B}(k\, \tau_{1}, \gamma, \delta)$ and $f_{B}(k\, y, \delta)$ 
can be written as:
\begin{eqnarray} 
\lim_{|k \tau_{1}| < 1} \,\, p_{B}(k\, \tau_{1},\, \gamma,\, \delta) &=& {\mathcal C}(\gamma +1/2)  \,\, 
\biggl(\frac{k}{a_{1} \, H_{1}}\biggr)^{ 4 - 2 \gamma - 2 \delta},
\label{POST8b}\\
\lim_{|k \tau_{1}| < 1} \,\, f_{B}(k\, y, \delta) &=& \biggl(\frac{q}{2}\biggr)^{-2 \delta} \,\, \biggl(\frac{k \, y}{2}\biggr)\, \,J_{\delta+1/2}^{2}(k\, y).
\label{POST8c}
\end{eqnarray}
The explicit expression appearing in Eqs. (\ref{POST7c})--(\ref{POST8c}) depend on the 
range of $\delta$. We are chiefly interested in the range $0\leq \delta < 1/2$ since
the relevant physical situation, for the present ends is the one where the gauge coupling eventually 
flattens out and $\delta \ll 1$. Equations (\ref{POST7}) and (\ref{POST8}) are also meaningful 
in the limit $\delta \to 0$ where\footnote{ Note that since $y(\tau) = \tau + \tau_{1} (1 + q)$, in the limit 
$\delta \to 0$ we have $k \, y \to (x + x_{1})$ (recall, in this respect, that $ q = \delta/\gamma$ 
and it goes to $0$ when $\delta\to 0$).}
\begin{equation}
\lim_{\delta \to 0} \,f_{E}(k \,y,\,\delta) = \cos^2{(x+ x_{1})}, \qquad\qquad \lim_{\delta \to 0} \,f_{B}(k \, y,\,\delta) = \sin^2{(x + x_{1})},
\label{POST8a0}
\end{equation}
where $x = k\tau$.  The finite limit of  ${\mathcal f}_{E}(x, \delta)$ and ${\mathcal f}_{B}(x, \delta)$ for $\delta \to 0$ follows from the continuity of the mode functions 
across the transition regime: the lack of continuity could lead to 
spurious divergences (for instance in $q = \gamma/\delta$); these divergences do not 
arise since the typical combinations involving $q$ are of the type of $q^{-2 \delta}$ (see Eqs. (\ref{POST7c})--(\ref{POST8c})) and are finite in the limit $\delta \to 0$. Since the power spectra for $\delta \to 0$ are well defined, we could invert the strategy and take first the limit $\delta \to 0$ and then expand for $x_{1} \ll 1$. This complementary discussion, for the sake of accuracy, reported in appendix \ref{APPA} and it is relevant for the determination of the final asymptotic value of the hypermagnetic power spectra. All in all Eq. (\ref{POST8a0}) justifies the result already mentioned 
in Eq. (\ref{FOURb}) so that we can also write
\begin{equation}
P_{E}(k,\tau) = a_{1}^4 \, H_{1}^4 \,\,p_{E}(k\, \tau_{1}, \gamma) \cos^2{k\tau}, \qquad\qquad
P_{B}(k,\tau) = a_{1}^4 \, H_{1}^4 \,\,p_{B}(k\, \tau_{1}, \gamma) \sin^2{k\tau}.
\label{DUAL1aaa}
\end{equation}
where $p_{B}(k\, \tau_{1}, \gamma)$ and $p_{E}(k\, \tau_{1}, \gamma)$ must be 
understood as the limits of $p_{B}(k\, \tau_{1}, \gamma, \delta)$ and $p_{E}(k\, \tau_{1}, \gamma, \delta)$
for $\delta \to 0$. These limits are always finite is the transition if the gauge coupling and its derivative are continuous (see, in this respect, also the discussion of appendix 
\ref{APPA}). The important aspect of Eq. (\ref{DUAL1aaa}) can be 
ultimately understood from Eqs. (\ref{POST7b})--(\ref{POST7c}) and 
(\ref{POST8b})--(\ref{POST8c}) in the limit $\delta\to 0$ where, 
in fact, $p_{B}(k \tau_{1}, \gamma) = p_{E}(k \tau_{1}, \gamma)$.
This means that when the gauge coupling flattens out the magnetic power 
spectrum after inflation matches with the {\em electric power 
spectrum during inflation}. This is the reason why the consideration 
following from Eq. (\ref{OM1}) is not conclusive: according to that 
argument the hypermagnetic and the hyperelectric power 
spectra {\em after} inflation coincide with the ones obtained {\em during}
inflation. Equation (\ref{DUAL1aaa}) shows actually that this conclusion 
is not correct since, as we shall see, for the magnetogenesis 
considerations what matters is the power spectrum for at the crossing time 
$\tau_{k}$.

\subsection{Decreasing gauge coupling and duality}
\label{subs32}
As in the case of Eqs. (\ref{POST7}) and (\ref{POST8}), 
when the gauge coupling decreases the comoving power spectra have a factored expression and the explicit results follow from the duality transformations 
of Eqs. (\ref{TRANS6}) and (\ref{TRANS7})--(\ref{TRANS8}).
We are assuming here that the gauge coupling decreases during inflation and flattens out later on;
this means that for $\tau< - \tau_{1}$, $g^{\prime}/g = - \overline{\gamma} \, {\mathcal H}$ 
while for $\tau \geq - \tau_{1}$ we have 
\begin{equation}
\frac{g^{\prime}}{g} = - \overline{\delta} \, {\mathcal H}, \qquad\qquad 0 < \overline{\delta} \ll \overline{\gamma}, \qquad\qquad \overline{\gamma} >0.
\label{POST9}
\end{equation}
In Eq. (\ref{POST9}) $\overline{\delta}$ measures the rate of decrease of the gauge coupling in units of the (comoving) Hubble rate. From Eq. (\ref{POST9}) we get a new transition matrix and the discussion 
follows the same steps of the case where the gauge coupling increases. The 
gauge power spectra in their factored form are:
\begin{eqnarray}
\overline{\,P\,}_{E}(k,\tau) &=& a_{1}^{4} \, H_{1}^4 \, \overline{\,p\,}_{E}(k\, \tau_{1}, \overline{\gamma}, \overline{\delta}) \,\, \overline{\,f\,}_{E}(k \,y, \overline{\delta}),
\label{POST10}\\
\overline{\,P\,}_{B}(k,\tau) &=& a_{1}^{4} \, H_{1}^4 \, \overline{\,p\,}_{B}(k\, \tau_{1}, \overline{\gamma}, \overline{\delta}) \,\, \overline{\,f\,}_{B}(k \,y, \overline{\delta}),
\label{POST11}
\end{eqnarray}
and in the limit $|k \tau_{1}| < 1$ (i.e. for all the wavenumbers smaller than the maximal) the explicit 
expressions of $\overline{\,p\,}_{E}(k\, \tau_{1}, \overline{\gamma}, \overline{\delta})$ and of $\overline{\,f\,}_{E}(k \,y, \overline{\delta})$ become:
\begin{eqnarray} 
\lim_{|k \tau_{1}| < 1} \,\, \overline{\,p\,}_{E}(k\, \tau_{1},\, \overline{\gamma},\, \overline{\delta}) &=& {\mathcal C}(\overline{\gamma} +1/2)  \,\, 
\biggl(\frac{k}{a_{1} \, H_{1}}\biggr)^{ 4 - 2 \overline{\gamma} - 2 \overline{\delta}},
\label{POST10b}\\
\lim_{|k \tau_{1}| < 1} \,\, \overline{\,f\,}_{E}(k\, y, \overline{\delta}) &=& \biggl(\frac{\overline{q}}{2}\biggr)^{-2 \overline{\delta}} \,\, \biggl(\frac{k \, y}{2}\biggr)\, \,J_{\overline{\delta}+1/2}^{2}(k\, y),
\label{POST10c}
\end{eqnarray}
where now $\overline{q} = \overline{\delta}/\overline{\gamma}$ and $y(\tau) = \tau + \tau_{1} (1 + \overline{q})$.
With the same notations the analog quantities for the hypermagnetic fields are:
\begin{eqnarray} 
\lim_{|k \tau_{1}| < 1} \,\, \overline{\,p\,}_{B}(k\, \tau_{1},\, \overline{\gamma},\, \overline{\delta}) &=& {\mathcal C}(\overline{\gamma} +1/2)  \,\, 
\biggl(\frac{k}{a_{1} \, H_{1}}\biggr)^{ 4 - 2 \overline{\gamma} - 2 \overline{\delta}},
\label{POST11b}\\
\lim_{|k \tau_{1}| < 1} \,\, \overline{\,f\,}_{B}(k\, y, \overline{\delta}) &=& \biggl(\frac{\overline{q}}{2}\biggr)^{-2 \overline{\delta}} \,\, \biggl(\frac{k \, y}{2}\biggr)\, \,J_{\overline{\delta}-1/2}^{2}(k\, y).
\label{POST11c}
\end{eqnarray}
Also in this case, by continuity, the limit $\overline{\delta}\to 0$ is well defined and, in this limit, 
\begin{equation}
\lim_{\overline{\delta} \to 0} \overline{\,f\,}_{E}(x,\overline{\delta}) = \sin^2{(x+ x_{1})},\qquad \qquad 
\lim_{\overline{\delta} \to 0} \overline{\,f\,}_{B}(x,\overline{\delta}) = \cos^2{(x + x_{1})}.
\end{equation}
The duality properties of the obtained results are
evident by comparing the final power spectra in the limit 
$\delta \to 0$ and $\overline{\delta}\to 0$. In this case 
the inversion of the gauge coupling implies, in the notations of Eqs. (\ref{POST1}) and (\ref{POST9}) 
\begin{equation}
g \to 1/g, \qquad \qquad \gamma \to \overline{\gamma}, \qquad\qquad \delta \to \overline{\delta}.
\label{DUAL1a}
\end{equation}
If we now consider the explicit expressions of the comoving power spectra we can see that, under 
the duality transformation (\ref{DUAL1a}) the gauge power spectra are interchanged as
\begin{eqnarray}
P_{B}(k,\tau) &\to& \overline{\,P\,}_{E}(k,\tau) = a_{1}^4 \, H_{1}^4 \,\,\overline{p}_{E}(k\, \tau_{1}, \overline{\gamma}) \sin^2{k\tau},
\nonumber\\
P_{E}(k,\tau) &\to& \overline{\,P\,}_{B}(k,\tau) = a_{1}^4 \, H_{1}^4 \,\,\overline{p}_{B}(k\, \tau_{1}, \overline{\gamma}) \cos^2{k\tau}.
\label{DUAL2a}
\end{eqnarray}
Equation (\ref{DUAL2a}) is the counterpart of Eq. (\ref{FOURc}) and it tells that duality exchanges the phases of Sakharov oscillations. In particular 
if the gauge coupling increases the hyperelectric and the hypermagnetic 
spectra oscillate, respectively, as 
cosines and sines squared (see Eq. (\ref{DUAL1aaa})). Thanks to duality the result of Eq. (\ref{DUAL1aaa}) also implies that the situation is reversed when the gauge coupling decreases; in this case Eq. (\ref{DUAL2a}) applies and the hyperelectric and the hypermagnetic spectra 
oscillate, respectively, as sines and cosines squared. 

\subsection{The physical power spectra}
\label{subs33}
The comoving power spectra do not scale with the expansion of the Universe 
and with the gauge coupling so that, for the phenomenological applications 
we must compute the physical power spectra whose definition coincides 
exactly with the one already introduced in Eqs. (\ref{PE1F})--(\ref{PB1F}).
Since the non-screened vector modes of the hypercharge
field project on the electromagnetic fields through the cosine of the Weinberg angle, 
the expressions of Eqs. (\ref{PE1F})--(\ref{PB1F}) must be multiplied  by $\cos^2{\theta_{W}}$. So for instance, the hypermagnetic power
spectrum\footnote{To avoid possible confusions we shall not introduce further variables and the physical power spectra are defined as in Eqs. (\ref{PE1F})--(\ref{PB1F}) by including, however, the appropriate 
dependence on the Weinberg's angle.} is related to the magnetic one as  ${\mathcal P}_{B}(k, \tau) = \cos^2{\theta_{W}}\,\, P_{B}(k,\, \tau)/[a^4(\tau)\, \lambda(\tau)]$.  Recalling now Eqs. (\ref{POST8}) and (\ref{POST8b})--(\ref{POST8c}) we can write the explicit form of the physical magnetic spectrum as
\begin{equation}
{\mathcal P}_{B}(k, \, \tau_{1}, \, \tau) = H_{1}^4 \,\, \biggl(\frac{a_{1}}{a}\biggr)^4 
Q(g_{1}, \,\cos{\theta_{W}}) \,\, {\mathcal p}_{B}(k\, \tau_{1},\, \gamma,\, \delta) 
\,\, {\mathcal f}_{B}(k\, \tau,\, \delta),
\label{PHYSPS}
\end{equation}
where now $Q(g_{1}, \, \cos{\theta_{W}})$ accounts for the dependence 
on the gauge coupling $g_{1}$ and on the Weinberg angle
\begin{equation}
Q(g_{1}, \, \cos{\theta_{W}}) = \cos^2{\theta_{W}} \, \frac{g_{1}^2}{4\,\pi}, \qquad \qquad \lambda_{1} = \, 
\frac{4 \, \pi}{g_{1}^2},
\label{PHYSPSQ}
\end{equation}
while ${\mathcal p}_{B}(k\, \tau_{1},\, \gamma,\, \delta)$ and ${\mathcal f}_{B}(k\, \tau,\, \delta)$ are defined in terms of the corresponding quantities already introduced for the 
comoving power spectra: 
\begin{eqnarray}
{\mathcal p}_{B}(k \tau_{1}, \, \gamma, \, \delta) &=& \biggl(\frac{k}{a_{1}\, H_{1}}\biggr)^{ - 2\delta} \, p_{B}(k\tau_{1}, \gamma, \delta) = \biggl(\frac{k}{a_{1} \, H_{1}}\biggr)^{ 4 - 2 \gamma - 4 \delta}, 
\nonumber\\
{\mathcal f}_{B}(k\,\tau, \delta) &=& q^{2 \delta} \,\, (k \tau)^{ 2 \delta} \,\, f_{B}(k\tau, \delta) = (k\, \tau)^{ 2 \delta + 1} \,\, 2^{2 \delta -1}\,\,
J_{\delta+1/2}^2(k \tau),
\label{PHYSPSQ2}
\end{eqnarray}
where we wrote directly the final expressions valid in the limit $k \tau_{1}< 1$ and $\tau > |\tau_{1}|$.
To obtain the correct form of the physical power spectra it is essential to use of the continuity
of the gauge coupling and of its time derivative in $\tau= -\tau_{1}$; this means, 
in particular, that 
\begin{equation}
g(\tau) = g_{1}\, [q (1 + \tau/\tau_{1}) +1]^{\delta}, \qquad  \mathrm{for} \qquad \tau\geq - \tau_{1} \qquad \mathrm{and} \qquad  0\leq \delta \ll 1.
\end{equation}
When $\tau \gg |\tau_{1}|$ we then have, approximately, that $g(\tau) = g_{1} [ q (\tau/\tau_{1})]^{ \delta}$.

In the case of decreasing gauge coupling the final expression of the physical power 
spectrum is obtained from the same steps outlined above with the difference that 
for $\tau \geq - \tau_{1}$ we have $\sqrt{\overline{\lambda}} = \sqrt{\overline{\lambda}_{1}} [ \overline{q} ( 1 + \tau/\tau_{1}) + 1]^{\overline{\delta}}$. In this case $\overline{{\mathcal P}}_{B}(k,\tau) = 
\cos^2{\theta_{W}} \,\, P_{B}(k,\, \tau)/[ \overline{\lambda}(\tau) \, \, a^4(\tau)]$ which becomes, in more 
explicit terms, 
\begin{equation}
\overline{{\mathcal P}}_{B}(k, \, \tau_{1}, \, \tau) = H_{1}^4 \,\, \biggl(\frac{a_{1}}{a}\biggr)^4 
Q(\overline{g}_{1}, \,\cos{\theta_{W}}) \,\, \overline{{\mathcal p}}_{B}(k\, \tau_{1},\, \gamma,\, \delta) 
\,\, \overline{{\mathcal f}}_{B}(k\, \tau,\, \delta),
\label{PHYSPS3}
\end{equation}
where $\overline{g}_{1}^2 = 4\pi/\overline{\lambda}_{1}$ while $\overline{{\mathcal p}}_{B}(k\, \tau_{1},\, \gamma,\, \delta)$ and  $\overline{{\mathcal f}}_{B}(k\, \tau,\, \delta)$ are now given by:
\begin{eqnarray}
\overline{{\mathcal p}}_{B}(k \tau_{1}, \, \gamma, \, \delta) &=& \biggl(\frac{k}{a_{1}\, H_{1}}\biggr)^{ 2\overline{\delta}} \, \overline{p}_{B}(k\tau_{1}, \overline{\gamma}, \overline{\delta}) 
\nonumber\\
\overline{{\mathcal f}}_{B}(k\,\tau, \overline{\delta}) &=& q^{-2 \overline{\delta}} \,\, (k \tau)^{ - 2 \overline{\delta}} \,\, \overline{f}_{B}(k\tau, \overline{\delta}),
\label{PHYSPSQ3}
\end{eqnarray}
where $\overline{p}_{B}(k\tau_{1}, \overline{\gamma}, \overline{\delta})$ and $\overline{f}_{B}(k\tau, \overline{\delta})$ have been already introduced in their physical limits (see Eqs. (\ref{POST11b})--(\ref{POST11c})).  By duality it follows that the values of the gauge couplings for $\tau = -\tau_{1}$ 
are related as $\overline{g}_{1} = 1/g_{1}$. For instance if $g_{1} \leq {\mathcal O}(0.1)$, 
 $g(\tau)$ was much smaller at the onset of inflation and it is still perturbative 
when the gauge coupling flattens out. Conversely if $\overline{g}_{1} \leq {\mathcal O}(0.1)$ 
the gauge coupling was much larger at the beginning of the inflationary stage of expansion. 
Recalling Eq. (\ref{POST9}) we then have that\footnote{In Eq. (\ref{PPP1})  $g_{i}$ is the value of the gauge coupling at a reference time $-\tau_{i}$ close 
to the onset of inflation; $N_{g}$ is the total number of $e$-folds associated with the variation 
of $g$ and, by definition, $g_{i}= g(-\tau_{i})$.}
\begin{equation}
g_{i} = \overline{g}_{1} \,\, \biggl( \frac{a_{i}}{a_{1}}\biggr)^{- \overline{\gamma}} \to \overline{g}_{1} 
e^{\overline{\gamma}\,\,N_{g}}.
\label{PPP1}
\end{equation}
If $N_{g}$ is estimated from the critical number $e$-folds (i.e. $N_{g} = N_{crit}$) we have  
\begin{equation}
N_{crit}  = 61.88 - \ln{\biggl(\frac{h_{0}}{0.7}\biggr)} + \frac{1}{4} \ln{\biggl(\frac{r_{T}}{0.06}\biggr)} +\frac{1}{4} \ln{\biggl(\frac{{\mathcal A}_{{\mathcal R}}}{2.41 \times 10^{-9}}\biggr)}
+   \frac{1}{4} \ln{\biggl(\frac{h_{0}^2 \, \Omega_{R0}}{4.15 \times 10^{-5}}\biggr)},
\label{Ncrit}
\end{equation}
where $N_{crit}$ is obtained by requiring the the present value of the Hubble radius 
is of the order of the (redshifted) event horizon. The total number of $e$-folds 
$N_{t}$ does not have to coincide with $N_{crit}$ but, in any case, $N_{t} \geq N_{crit}$. Even if 
 $N_{g}$ can be either 
larger or smaller than $N_{crit}$,  $g_{i} \gg 1$ whenever $\overline{g}_{1} 
= {\mathcal O}(0.01)$. This means that, at the onset of inflation, the gauge coupling should be 
extremely large and, for this reason, we shall rather focus on the case 
where the gauge coupling increases as then flattens out\footnote{ A possibility 
suggested in Ref. \cite{fortyseven} has been that the $g(\tau)$ decreases during inflation, increases sharply during reheating, and then flattens out again. This suggestion, often assumed by various authors without justifications, has some drawbacks that are however 
not central to the present discussion.}.

\renewcommand{\theequation}{4.\arabic{equation}}
\setcounter{equation}{0}
\section{The physical power spectra at the crossing time}
\label{sec4}
\subsection{The crossing time $\tau_{k}$}
\label{subs41}
Unlike the case of matter density inhomogeneities \cite{fortytwo,fortythree,SAK2} the standing oscillations appearing, for instance, in Eqs. (\ref{POST8}) and (\ref{POST11}) never develop since they are suppressed 
by the finite value of the conductivity that operates as soon as the corresponding wavelengths get of the order of the Hubble radius.
In the present section we analyze the standard thermal history and focus the attention on the bunch of wavenumber that are relevant for the large-scale magnetism (i.e. $k = {\mathcal O}(\mathrm{Mpc}^{-1})$) with the purpose of showing that the relevant scale 
of the problem is provided by the crossing time $\tau_{k}$ where the Sakharov phases are generically ${\mathcal O}(1)$.
 \begin{figure}[!ht]
\centering
\includegraphics[height=8cm]{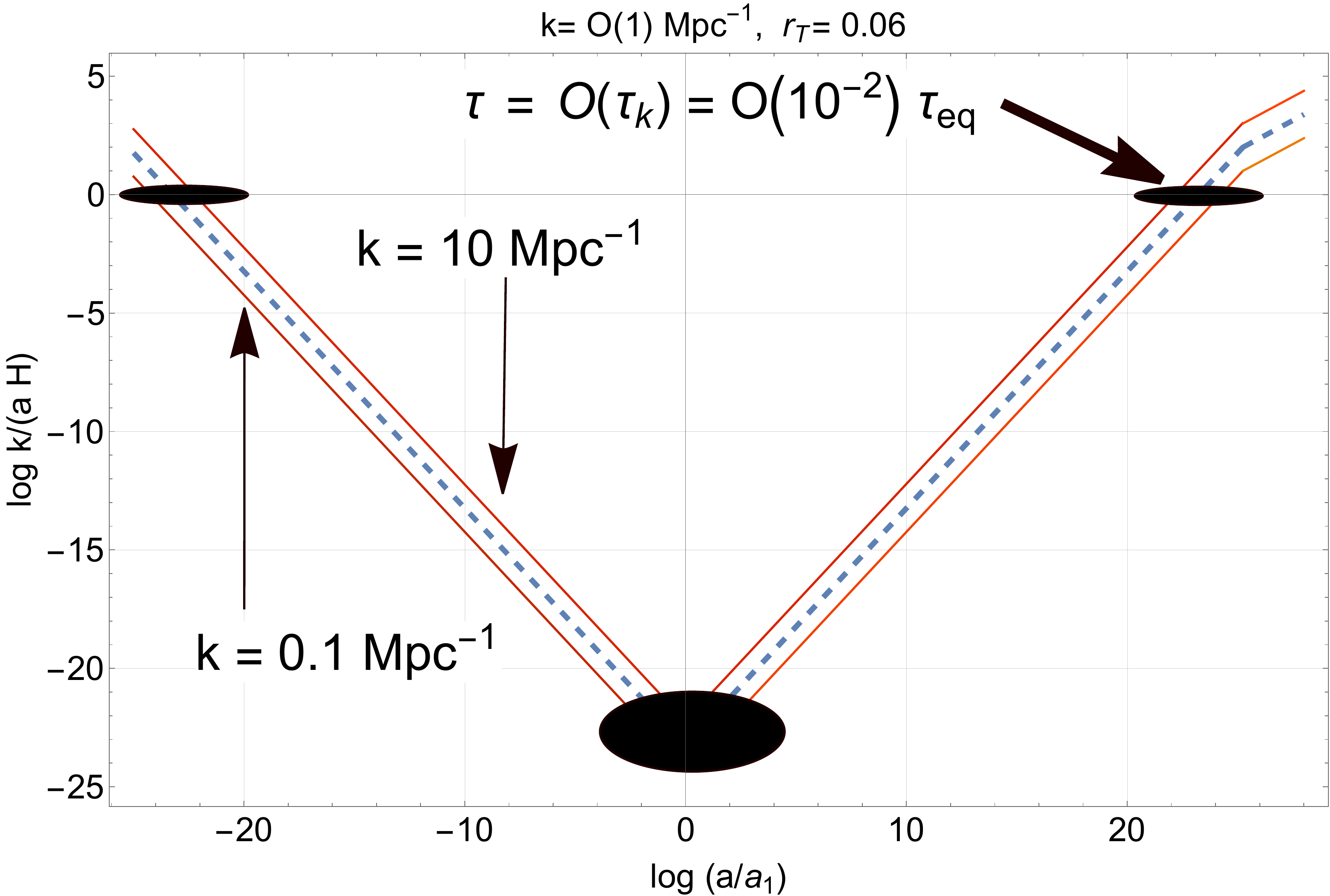}
\caption[a]{On the vertical axis  the (common) logarithm of $k/(a H)$ is illustrated as a function of the (common) logarithm of the scale factor. The inflationary stage is followed by the radiation phase and the 
three different lines illustrate a selectrion of wavenumbers that are all ${\mathcal O}(\mathrm{Mpc}^{-1})$. These wavenumbers become of the order of the comoving Hubble radius when $k/(a\, H) = {\mathcal O}(1)$ close to the elliptic disk that is present for $a>a_{1}$. This range typically
corresponds to a conformal time coordinate ${\mathcal O}(10^{-2})\, \,\tau_{eq}$.}
\label{FIGU1}      
\end{figure}
 In Fig. \ref{FIGU1} we illustrate the common logarithm of $k/(a H)$  as a function 
of the common logarithm of the scale factor and the three different lines appearing in the plot
correspond to slightly different values of $k$ while the black 
blobs indicate the time-scale $\tau_{k} = {\mathcal O}(1/k)$. The black blob 
for $a < a_{1}$ represents the inflationary crossing time
but the second blob close to matter-radiation equality 
is the relevant one for our discussion and it corresponds to  
\begin{equation}
k = {\mathcal O}(\mathrm{Mpc}^{-1}) = a H = {\mathcal H}, \qquad \qquad \tau_{k} < {\mathcal O}(\tau_{eq}).
\end{equation}
The crossing time is always slightly smaller than the equality time (i.e. $\tau_{k}= {\mathcal O}(10^{-2}) \, \tau_{eq}$) and the ratio $\tau_{k}/\tau_{eq}$ can be more accurately estimated as:
\begin{equation}
\frac{\tau_{k}}{\tau_{eq}} = 1.01 \times 10^{-2} \biggl(\frac{k}{\mathrm{Mpc}}\biggr)^{-1} 
\biggl(\frac{h_{0}^2 \Omega_{M0}}{0.1386}\biggr) \biggl(\frac{h_{0}^2 \Omega_{R0}}{4.15\times 10^{-5}}\biggr)^{-1/2}.
\label{PS1}
\end{equation}
The central elliptic disk in Fig. \ref{FIGU1} denotes the minimum of $k/(a\, H)$ at the transition regime between inflation and the radiation stage\footnote{ The transition 
regime can modify the final values of the power spectra and ultimately affect the magnetogenesis 
requirement as originally suggested in \cite{twelve}. The estimate of Eq. (\ref{PS1}) is however {\em not} affected by the transition regime, as we shall see below.}. According to Eq. (\ref{PS1}) we therefore have that for $\tau< \tau_{k}$ the gauge power spectra do not oscillate but as soon as  $\tau = {\mathcal O}(\tau_{k})$ the standing oscillations should start and eventually persist for $\tau > \tau_{k}$ but this sequence of events is never realized since  the oscillations are overdamped by the finite value of the conductivity when $k \tau \geq 1$. The different time ranges are now separately analyzed.

\subsection{The regime $\tau< \tau_{k}$}
\label{subs42}
When $\tau < \tau_{k}$ the wavenumber is always smaller than the (comoving) Hubble radius (i.e. $k < a H$) and the minimum of $k/(a H)$ (illustrated in Fig. \ref{FIGU1}) is ${\mathcal O}(10^{-23})$ for typical wavenumbers $k = {\mathcal O}(\mathrm{Mpc}^{-1})$:
\begin{equation}
\frac{k}{a_{1}\, H_{1}} = \frac{k}{ H_{0}} \biggl(2 \,\pi\, \epsilon\, \Omega_{R0} {\mathcal A}_{{\mathcal R}}\biggr)^{-1/4}\, \sqrt{\frac{H_{0}}{M_{P}}}.
\label{PS2}
\end{equation}
The estimate of Eq. (\ref{PS2}) refers explicitly to the timeline of Fig. \ref{FIGU1} but
 when the expansion history is modified the value of $k/(a\, H)$ at the minimum takes a different 
 form\footnote{See, in this respect, the discussion of appendix \ref{APPB} and, in particular, Eqs. (\ref{APPB2})--(\ref{APPB3}).}. We remind that in Eq. (\ref{PS2}) $\Omega_{R0}$ is the (present) critical fraction of massless species (in the concordance paradigm $h_{0}^2 \Omega_{R0} = 4.15 \times 10^{-5}$).  In terms of the typical values of the various parameters, Eq. (\ref{PS2}) becomes as expected 
\begin{equation}
\frac{k}{a_{1}\, H_{1}} = 10^{-23.24}\,\, \biggl(\frac{k}{\mathrm{Mpc}^{-1}}\biggr)\, \biggl(\frac{r_{T}}{0.06}\biggr)^{-1/4}\,\,\biggl(\frac{h_{0}^2 \Omega_{R0}}{4.15\times 10^{-5}}\biggr)^{-1/4} \,\,\biggl(\frac{{\mathcal A}_{{\mathcal R}}}{2.41\times10^{-9}}\biggr)^{-1/4}.
\label{PS3}
\end{equation}
Let us now consider, for example, the physical power spectra  associated with Eqs. (\ref{POST7})--(\ref{POST8}) 
for  $\tau \leq \tau_{k}$ and in the limit $\delta \to 0$:
 \begin{eqnarray}
{\mathcal P}_{E}(k,\tau) &=& H_{1}^4 \,\, Q(g_{1}, \cos{\theta_{W}}) \,\,\biggl(\frac{a_{1}}{a} \biggr)^4 \,\, 
p_{E}(k\tau_{1}, \, \gamma) \,\, \cos^2{k\tau},
\label{PS4}\\
{\mathcal P}_{B}(k,\tau) &=&  H_{1}^4 \,\, Q(g_{1}, \cos{\theta_{W}}) \,\,\biggl(\frac{a_{1}}{a} \biggr)^4 \,\, 
p_{B}(k\tau_{1}, \, \gamma) \,\, \sin^2{k\tau}.
\label{PS5}
\end{eqnarray}
As already stressed the limit $\delta \to 0$ indicates that the gauge coupling freezes after inflation and, in this case,
 Eqs. (\ref{PHYSPSQ})--(\ref{PHYSPSQ2}) imply
\begin{equation}
\lim_{\delta \to 0} \,\, {\mathcal p}_{E}(k\tau_{1}, \, \gamma, \delta) \to p_{E}(k\tau_{1}, \, \gamma), \qquad \mathrm{and}   \qquad \lim_{\delta \to 0} \,\, {\mathcal p}_{E}(k\tau_{1}, \, \gamma, \delta) {\mathcal p}_{B}(k\tau_{1}, \, \gamma, \delta) \to p_{B}(k\tau_{1}, \, \gamma).
\end{equation}
In Eqs. (\ref{PS4})--(\ref{PS5}) the limit $k \tau< 1$ has not been imposed but we simply 
applied the definition of the physical power spectra and took the limit $\delta \to 0$. 
For $k \tau < 1$ we instead have that Eqs. (\ref{PS4})--(\ref{PS5}) become:
\begin{eqnarray}
\lim_{k\,\tau< 1} \,\,{\mathcal P}_{E}(k,\tau) &=& H_{1}^4\,\,Q(g_{1}, \cos{\theta_{W}})\,\,\biggl(\frac{a_{1}}{a} \biggr)^4 \,\, 
p_{E}(k\tau_{1}, \, \gamma)\,\,\biggl[ 1 + {\mathcal O}(|k \tau|^2)\biggr],  
\label{PS4a}\\
\lim_{k\, \tau < 1} \,\,{\mathcal P}_{B}(k,\tau) &=&  H_{1}^4\,\,Q(g_{1}, \cos{\theta_{W}})\,\,\biggl(\frac{a_{1}}{a} \biggr)^4 \,\, 
p_{B}(k\tau_{1}, \, \gamma)\,\, |k\,\tau|^2 \biggl[ 1 - {\mathcal O}(|k \tau|^2)\biggr].
\label{PS5a}  
\end{eqnarray}
 Equations (\ref{PS4a})--(\ref{PS5a}) demonstrate that for 
$ \tau < \tau_{k}$ and  $k \tau <1$ {\em there is no gain in the amplitude} of the hypermagnetic power spectrum.
This is particularly evident in the flat case (i.e. $\gamma \to 2$) where 
the physical power spectra are:
\begin{eqnarray}
{\mathcal P}_{E}(k,\tau) &=& \frac{9 H_{1}^4}{4\pi^2}\,\,Q(g_{1}, \cos{\theta_{W}})\,\,\biggl(\frac{a_{1}}{a} \biggr)^4  \cos^2{k\tau} \to \frac{9 H_{1}^4}{4\pi^2}\,\,Q(g_{1}, \cos{\theta_{W}})\,\,\biggl(\frac{a_{1}}{a} \biggr)^4, 
\label{PS6a}\\
{\mathcal P}_{B}(k,\tau) &=& \frac{9 H_{1}^4}{4\pi^2}\,\,Q(g_{1}, \cos{\theta_{W}})\,\,\biggl(\frac{a_{1}}{a} \biggr)^4  \sin^2{k\tau} \to \frac{9 H_{1}^4}{4\pi^2}\,\,Q(g_{1}, \cos{\theta_{W}})\,\,\biggl(\frac{a_{1}}{a} \biggr)^4\, \, |k\,\tau|^2.
\label{PS7a}
\end{eqnarray}
If we Eq. (\ref{PS7a}) is now evaluated in the limit $\tau \to {\mathcal O}(\tau_{1})$ we would have that 
${\mathcal P}_{B}(k, \tau_{1}) = {\mathcal O}(10^{-46}) {\mathcal P}_{E}(k, \tau_{1})$. This 
is, however, {\em much smaller} than the values of the gauge power spectra for $\tau = {\mathcal O}(\tau_{k})$.  
Equation (\ref{PS7a}) can also be expressed in various equivalent ways 
that \do not alter the evaluation of the gauge power spectra; for instance 
we may divide it and multiply it by $| k \,\tau_{1}|^2$ so that the equivalent form of Eq. (\ref{PS7a}) can be expressed as:
\begin{equation}
{\mathcal P}_{B}(k,\tau) = \frac{9 H_{1}^4}{4\pi^2} \, Q(g_{1},\, \cos{\theta_{W}})  \,\, \biggl(\frac{k}{a_{1} H_{1}} \biggr)^2 \,\,
\biggl(\frac{a_{1}}{a} \biggr)^6 \, \, \biggl(\frac{H_{1}}{H}\biggr)^2.
\label{PS8a}
\end{equation}
Even if the logic should be clear we stress that Eq. (\ref{PS8a}) follows since $\tau/\tau_{1} = {\mathcal H}_{1}/{\mathcal H}$ 
and $ {\mathcal H} = a \, H$; moreover it holds as long as 
the scale factor and its derivative are continuous across $\tau = - \tau_{1}$ i.e. 
$a_{inf}(-\tau_{1}) = a_{p}(-\tau_{1})$ and ${\mathcal H}_{inf} = {\mathcal H}_{p}$
where the subscripts denote the values of the scale factors and of its rate during inflation and in the post-inflationary stage of expansion.  Equations  (\ref{PS7a})--(\ref{PS8a}) are exactly equal and this apparent confusion stems from the
largeness of $(a_{1}/a)^6\, (H_{1}/H)^2$: this quantity is ${\mathcal O}(1)$ at $\tau_{1}$ but it seems much larger than $(a_{1}/a)^4$ for $\tau \gg \tau_{1}$. This is, however, not true since for $\tau_{k}$ 
\begin{equation}
\biggl(\frac{k}{a_{1} H_{1}} \biggr)^2 \,\,
\biggl(\frac{a_{1}}{a_{k}} \biggr)^6 \, \, \biggl(\frac{H_{1}}{H_{k}}\biggr)^2 = {\mathcal O}(1),
\label{PS9bb}
\end{equation}
and the apparent largeness of $(a_{1}/a_{k})^6 (H_{1}/H_{k})^2$ is completely compensated by the smallness of $|k \tau_{1}|^2$. In particular, if we just suppose a single (radiation-dominated) post-inflationary 
epoch as illustrated in Fig. \ref{FIGU1} we would have  that  $(a_{1}/a_{k})^3\, (H_{1}/H_{k}) = {\mathcal O}(10^{23})$ but this large ratio does not affect the final value of the magnetic fields since, as we saw before, $k \tau_{1} = {\mathcal O}(10^{-23})$ and the contribution of Eq. (\ref{PS9bb}) never gets large.

\subsection{The regime $\tau\geq \tau_{k}$}
\label{subs43}
For $\tau \geq \tau_{k}$ the mode functions do not obey the equations discussed so far but they are modified by the presence of the conductivity and as soon as $\tau = {\mathcal O}(\tau_{k})$ their evolution is given by:
\begin{equation}
g_{k}^{\prime} = - k^2 f_{k} - \sigma_{c} \, g_{k}, \qquad\qquad f_{k}^{\prime} = g_{k}.
\label{COND1}
\end{equation}
The solutions of Eq. (\ref{COND1}) can be phrased in various different ways;
 in particular, we could use an expansion in $(k/\sigma)$ and directly insert, as initial data at $\tau =\tau_{k}$, the values of the mode functions for $\tau \leq \tau_{k}$. In Eq. (\ref{COND1}) the typical time scale 
of variation of the plasma is in fact much smaller than the conductivity i.e.
\begin{equation}
\sigma_{c} \gg {\mathcal H} = {\mathcal O}(\tau^{-1}) \qquad\Rightarrow \qquad \sigma_{ph} \gg H.
\label{COND2}
\end{equation}
where $\sigma_{c}(\tau)$ and $\sigma_{ph}(\tau)$ are, respectively, the comoving and 
the physical conductivities related as $\sigma_{ph}(\tau) = \sigma_{c}(\tau)/a(\tau)$.
The condition (\ref{COND2}) is verified at $\tau = {\mathcal O}(\tau_{k})$ (where it implies $\sigma_{c} \gg k$) and also later on; this means, incidentally, that out of the two solutions of Eq. (\ref{COND1}) only one dominates. Furthermore, since at equality the conductivity is approximately given by $\sigma_{ph}(t_{eq}) =\, \sqrt{T_{eq}/m_{e}} (T_{eq}/\alpha_{em})$ we also have that $\sigma_{ph}(t_{eq})\gg H_{eq}$.  If we then take into account the previous
observations together with Eq. (\ref{COND2}) the mode functions for $\tau \geq \tau_{k}$ are determined as follows:
\begin{equation}
f_{k}(\tau) = f_{k}(\tau_{k}) \,\, e^{- \frac{k^2}{k_{\sigma}^2}}, 
\qquad\qquad 
g_{k}(\tau) = \biggl(\frac{k}{\sigma}\biggr) g_{k}(\tau_{k}) \,\,e^{- \frac{k^2}{k_{\sigma}^2}}.
\label{cond3}
\end{equation}
In Eq. (\ref{cond3}) $k_{\sigma}(\tau)$ denotes the magnetic diffusivity scale 
 \begin{equation}
 \frac{k^2}{k_{\sigma}^{2}}  = k^2 \int_{\tau_{k}}^{\tau} \, \frac{d z}{\sigma_{c}(z)}\to\frac{{\mathcal O}(10^{-26})}{ \sqrt{2 \, h_{0}^2 \Omega_{M0} (z_{\mathrm{eq}}+1)}} \, \biggl(\frac{k}{\mathrm{Mpc}^{-1}} \biggr)^2,
 \label{cond4}
 \end{equation}
 where $\Omega_{M0}$ is the present critical fraction in matter and $z_{\mathrm{eq}} + 1 = a_{0}/a_{\mathrm{eq}}\simeq {\mathcal O}(3200)$ is the redshift of matter-radiation equality. The estimate of Eq. (\ref{cond4}) follows by assuming that $\tau = {\mathcal O}(\tau_{eq})$ and it can be refined by computing the transport coefficients of the plasma in different regimes (see, for instance, \cite{MAC}). For the present purposes, however, what matters is that the ratio $(k/k_{\sigma})^2$  is negligibly small  for $k = {\mathcal O}(\mathrm{Mpc}^{-1})$ so that the negative exponentials of Eq. (\ref{cond3}) evaluate to $1$ and the physical power spectra for  $\tau \gg \tau_{k}$ are therefore given by:
\begin{eqnarray}
{\mathcal P}_{B}(k,\tau) &=& {\mathcal P}_{B}(k, \tau_{k}) \,\, \biggl(\frac{a_{k}}{a}\biggr)^4 \,\, e^{-2 \frac{k^2}{k_{\sigma}^2}} \to {\mathcal P}_{B}(k, \tau_{k}) \,\, \biggl(\frac{a_{k}}{a}\biggr)^4,
\label{cond5}\\
{\mathcal P}_{E}(k,\tau) &=& \biggl(\frac{k}{\sigma}\biggr)^2\,\,{\mathcal P}_{E}(k, \tau_{k}) \,e^{-2 \frac{k^2}{k_{\sigma}^2}} \to \biggl(\frac{k}{\sigma}\biggr)^2\,\,\,\,{\mathcal P}_{E}(k, \tau_{k}) \,\,\biggl(\frac{a_{k}}{a}\biggr)^4.
\label{cond6}
\end{eqnarray}
The limits appearing in Eqs. (\ref{cond5})--(\ref{cond6}) take into account 
the smallness of $(k/k_{\sigma})$ and the suppression of the electric 
power spectrum that is ultimately a consequence of the standard hydromagnetic evolution where the Ohmic electric 
field is given by $\vec{E} = (\vec{\nabla} \times \vec{B})/\sigma_{c}$ 
when the conductivity is large \cite{MAC2,MAC3}. In our notations the suppression of the electric power spectra is:
\begin{equation}
\biggl(\frac{k}{\sigma_{c}}\biggr)^2 = {\mathcal O}(10^{-48}) \biggl(\frac{T}{T_{eq}}\biggr)^{-3} \, \, \biggl(\frac{k}{\mathrm{Mpc}^{-1}}\biggr)^2.
\label{cond7}
\end{equation}
All in all the physical power spectrum at the onset of the gravitational collapse of the 
protogalaxy can therefore be expressed as: 
\begin{equation}
{\mathcal P}_{B}(k, \tau_{0}) = {\mathcal P}_{B}(k,\tau_{k}) \biggl(\frac{a_{k}}{a_{0}}\biggr)^4, \qquad \qquad \tau \geq \tau_{k}.
\label{cond8}
\end{equation}
The previous evolution of the gauge coupling enters 
Eq. (\ref{cond8}) through ${\mathcal P}_{B}(k,\tau_{k})$ but 
it is otherwise independent of the specific scenario.
After the conductivity sets in the 
duality symmetry is broken by the presence of Ohmic currents that do not have a 
counterpart in the magnetic sector; this is the ultimate physical meaning of Eq. (\ref{cond7}) implying that the electric power spectra ${\mathcal P}_{E}(k,\tau)$ are (at least) $50$ orders of magnitude smaller than their magnetic counterpart. 
\begin{figure}[!ht]
\centering
\includegraphics[height=6cm]{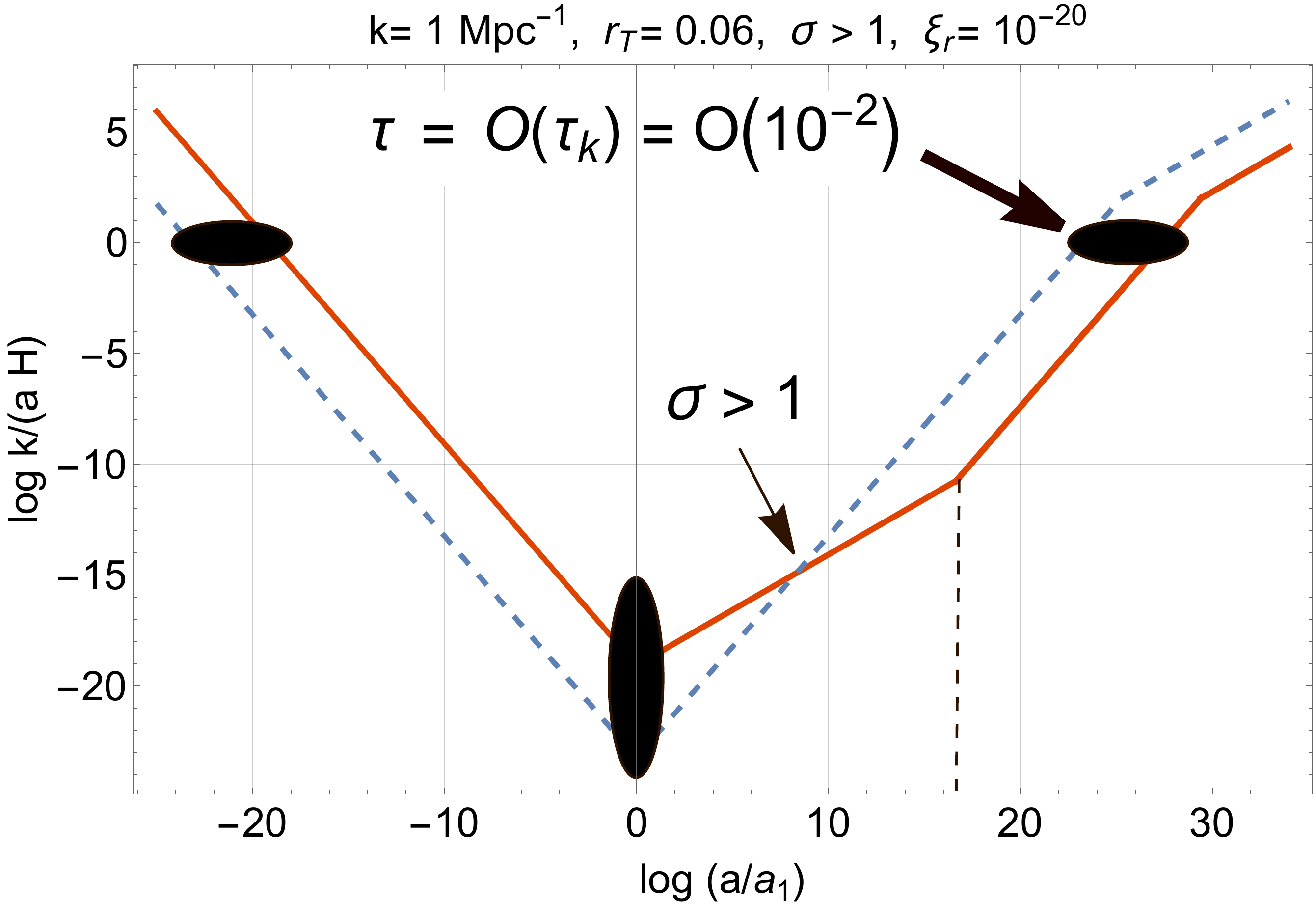}
\includegraphics[height=6cm]{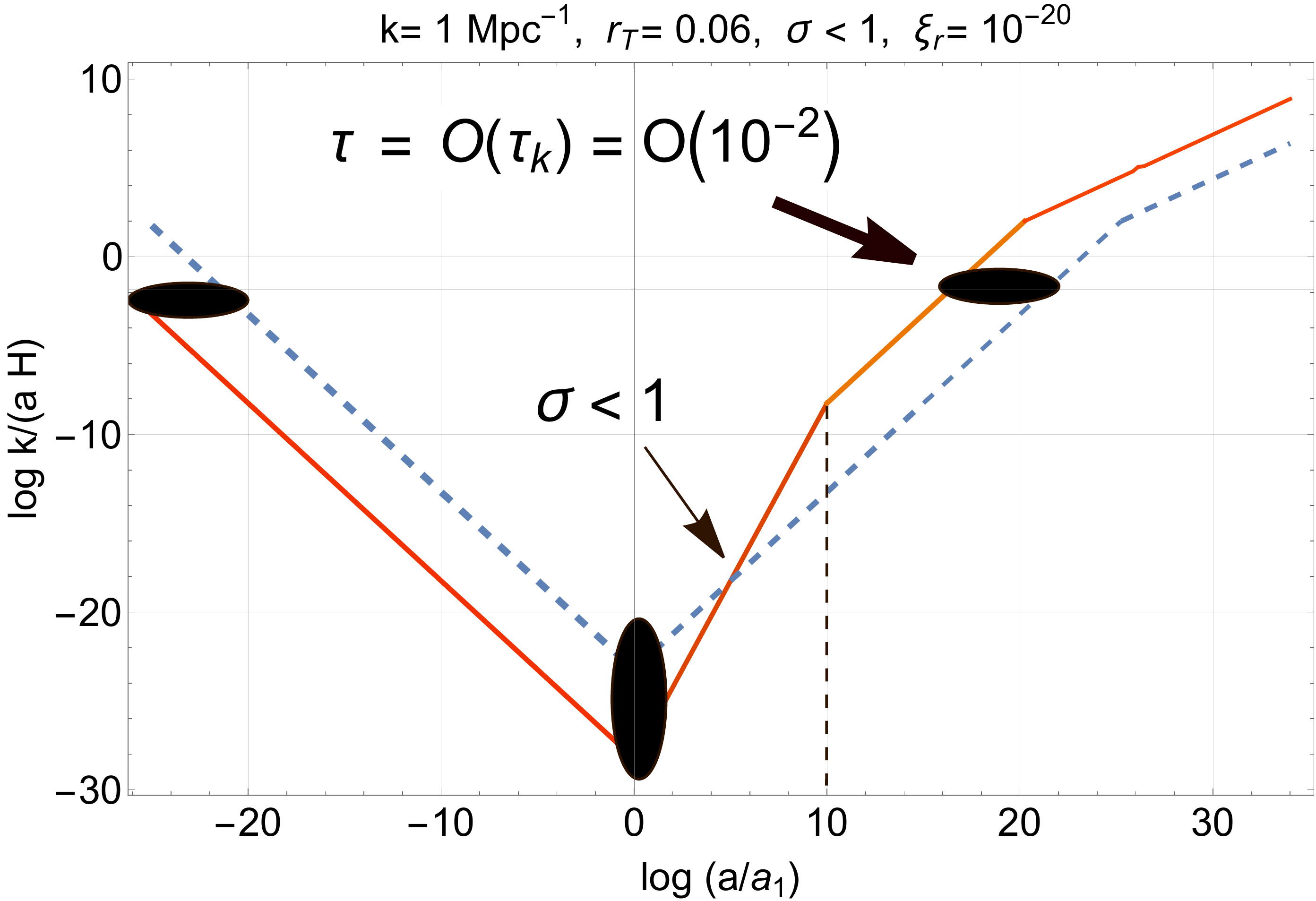}
\caption[a]{In both plots, with the dashed line, we illustrate the same situation described in Fig. \ref{FIGU1}
while the full lines account for the case where the post-inflationary expansion rate 
contains a single (supplementary) phase expanding either faster (i.e. $\sigma>1$) or slower (i.e. $\sigma < 1$) than radiation.
If the expansion rate is faster than radiation $k/(a H)$ is systematically larger than in the radiation-dominated phase. Furthermore since the Universe expands faster the total redshift 
of the post-inflationary phase is larger but the opposite is true when the expansion rate is slower than radiation; in both cases, however, $\tau_{k} = {\mathcal O}(10^{-2}) \,\, \tau_{eq}$. In these plots we have selected, for illustration, $\xi_{r} = 10^{-20}$.}
\label{FIGU2}      
\end{figure}
All the previous considerations have been derived by assuming the scenario illustrated in Fig. \ref{FIGU1} but  Eq. (\ref{cond8}) is still valid 
even if ${\mathcal P}_{B}(k,\tau_{k})$ takes a 
different value because of a modified expansion history as we are going to see in the next section and in the related appendix \ref{APPB}. Figure \ref{FIGU1} could actually be complemented by different physical situations that are specifically examined in section \ref{sec5}.  In Fig. \ref{FIGU2} we report an example where 
the radiation stage does not sets in immediately after inflation and during the 
intermediate stage the expansion rate in the conformal time coordinate 
is parametrized by\footnote{Not to be 
confused with the conductivities that have been always denoted (with appropriate subscripts) by $\sigma_{c}$ and $\sigma_{ph}$ in the comoving and physical cases.} $\sigma$ (i.e. $a(\tau) \propto \tau^{\sigma}$).
In the left plot of Fig. \ref{FIGU2} we illustrate the situation 
where $\sigma >1$ and the expansion rate is faster than radiation (corresponding 
to $\sigma \to 1$). In the right plot of the same figure we instead 
consider the case $\sigma <1$ where the expansion rate is 
slower than radiation. In both plots the dashed line corresponds, for 
comparison to the case of Fig. \ref{FIGU1} where the intermediate stage is 
absent. 

In spite of the value of $\sigma$, $\tau_{k} = {\mathcal O}(10^{-2}) \, \tau_{eq}$ even if $k/( a\, H)$ {\em overshoots the radiation result} for $\sigma >1$.
The same is true for $\sigma <1$ where $k/(a\, H)$ 
{\em undershoots the radiation result}. 
Between $H_{1}$ and $H_{r}$ the plasma expands either faster or slower 
than radiation and this implies that the decrease of $H$ is covered 
in different time-ranges; note, incidentally, that $\xi_{r} = H_{r}/H_{1} =10^{-20}$ 
in Fig. \ref{FIGU2}. This parameter is only constrained, within the present logic, 
by the nucleosynthesis considerations (see, in this respect, Eq. (\ref{WR4}) and discussion therein).
\begin{figure}[!ht]
\centering
\includegraphics[height=5.9cm]{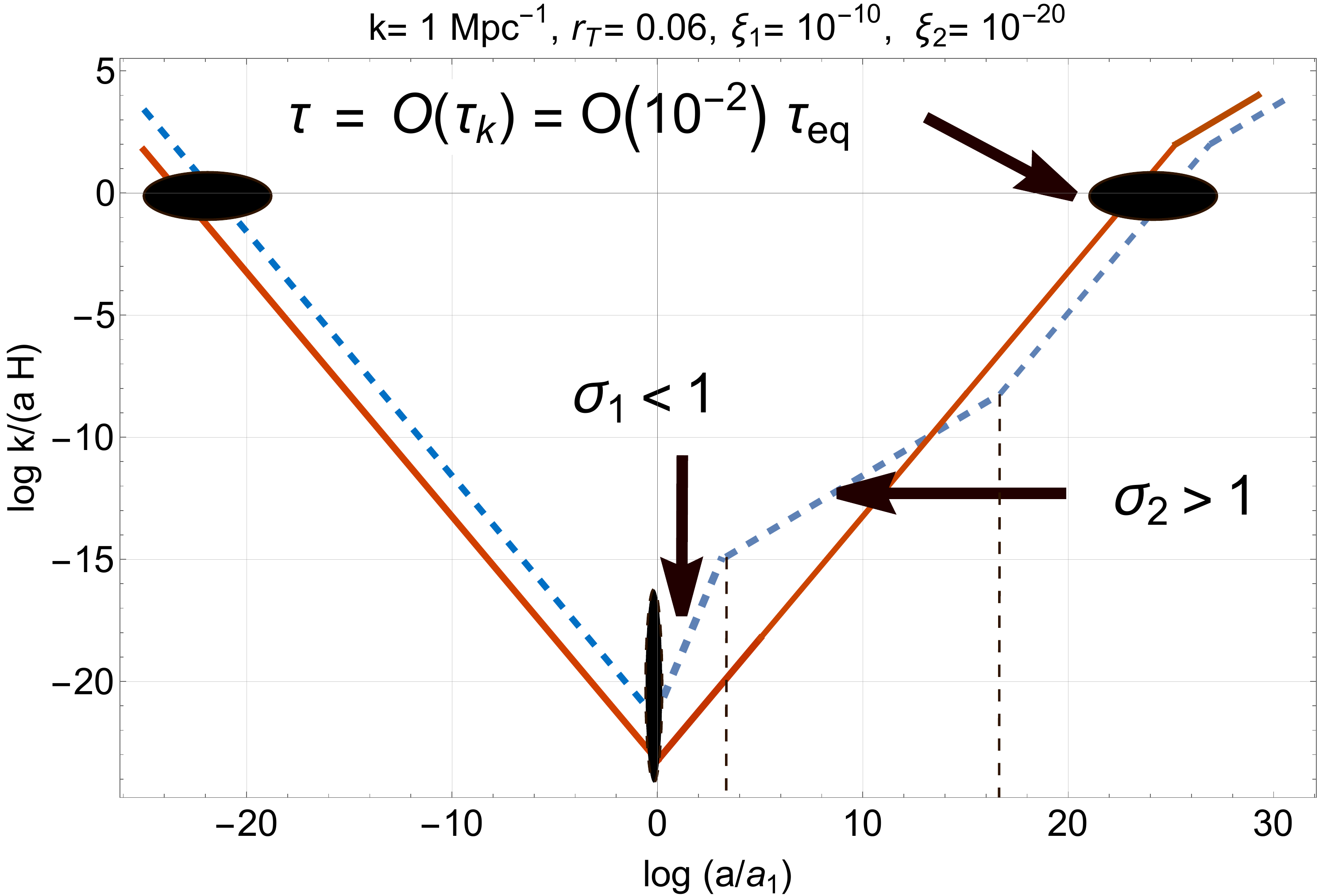}
\includegraphics[height=5.9cm]{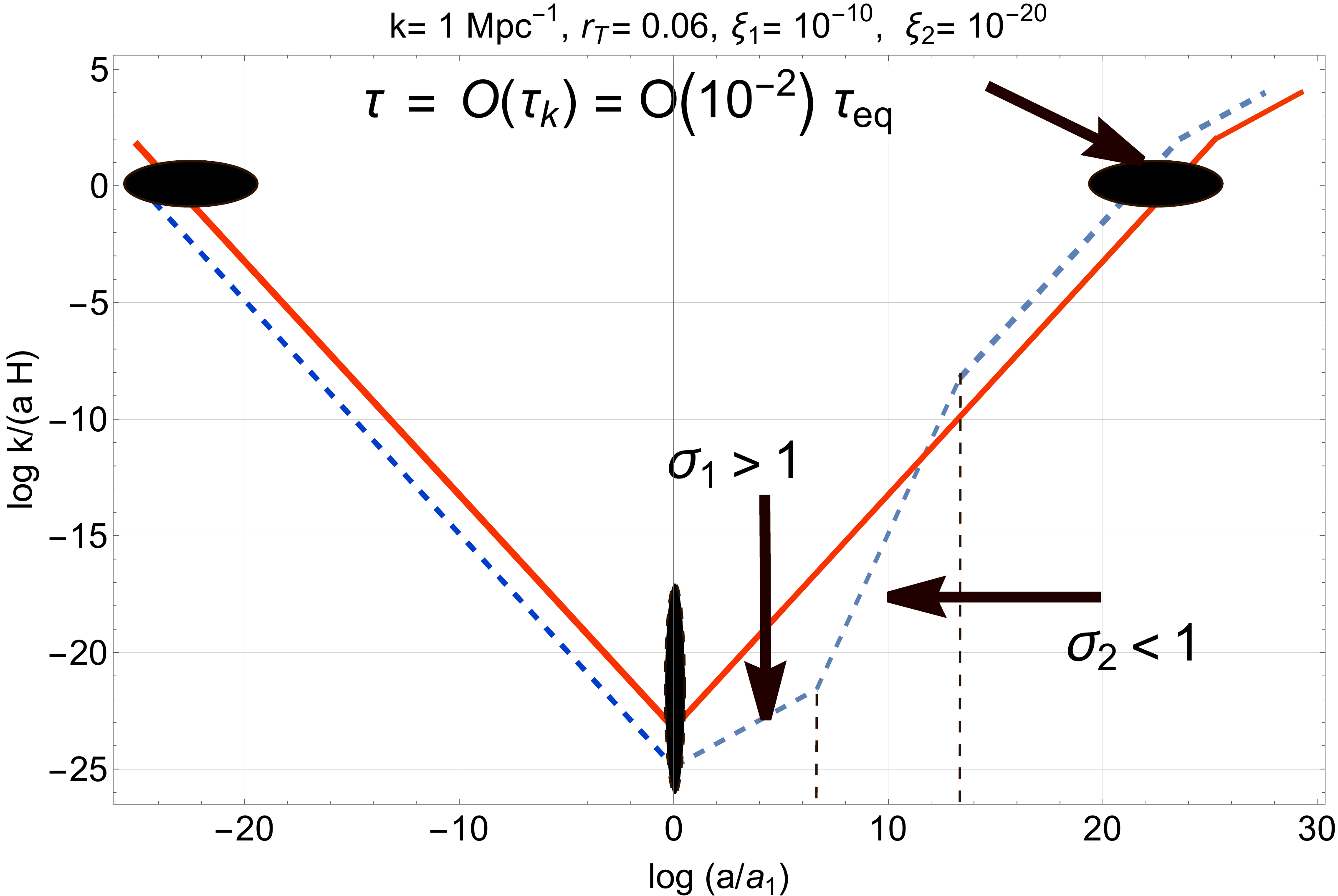}
\caption[a]{As in Fig. \ref{FIGU2} with the dashed lines we illustrate the radiation case while the full lines describe the 
situation where there are {\em two intermediate phases}. In the first case (i.e. $\sigma_{1} < 1$ and $\sigma_{2}> 1$) 
the expansion rate is first slower and then faster than radiation. In the second case  (i.e. $\sigma_{1} > 1$ and $\sigma_{2}< 1$) the expansion rate is first faster and then slower than radiation.}
\label{FIGU3}      
\end{figure}
The same logic of Fig. \ref{FIGU2} is replicated in Fig. \ref{FIGU3} 
where there are now two intermediate phases parametrized 
by two different expansion rates (i.e. $\sigma_{1}$ and $\sigma_{2}$) 
and by two intermediate durations (i.e. $\xi_{1} = H_{2}/H_{1}$ 
and $\xi_{2} = H_{r}/H_{2}$). As already mentioned the general 
situation (for an arbitrary number of intermediate phases) 
is described in appendix \ref{APPB}.

\renewcommand{\theequation}{5.\arabic{equation}}
\setcounter{equation}{0}
\section{The physical power spectra at the present time}
\label{sec5}
All in all the standing oscillations of the gauge power spectra are overdamped by the finite value of the conductivity so that the electric fields are ultimately suppressed in comparison with their magnetic counterpart and the original duality symmetry is broken down. In the current section the magnetic power spectra will 
be evaluated at the present epoch which coincides, for all the practical purposes of this analysis, with the time of the gravitational collapse of the protogalaxy. If the protogalactic matter collapsed by gravitational instability over a typical scale ${\mathcal O}(\mathrm{Mpc})$ the mean matter density before collapse was of the order of $\rho_{crit}$. 
Compressional amplification typically increases the initial values of the magnetic fields by $4$ or even $5$ orders of magnitude since, after collapse, the mean matter density got larger while the magnetic flux itself is conserved \cite{mg04,MAC,MAC2,MAC3}. The gain caused by compressional amplification is usually complemented by the one associated with the 
dynamo action: since the galaxy has a typical rotation period of 
$3 \times 10^{8}$ yrs it can happen that the kinetic energy associated with 
the global rotation is converted into magnetic energy \cite{MAC3}.
If the rotation period  is compared with the age (i.e. $\mathcal{O}(10^{10} {\rm yrs})$), the galaxy performed about $30$ rotations since the time 
of the protogalactic collapse. The achievable amplification produced by the 
dynamo instability can be at most of $10^{13}$, i.e. $e^{30}$ (see, for instance, \cite{seven,mg04}). By putting together these two elements we can estimate the typical requirements on the physical power spectra implying that 
\begin{equation}
{\mathcal P}_{B}(k, \tau_{0}) \geq 10^{-22} \, \, \mathrm{nG}^2, \qquad\qquad 
{\mathcal P}_{B}(k, \tau_{0}) \geq 10^{-32} \, \, \mathrm{nG}^2.
\label{REQ1}
\end{equation}
The requirements of Eq. (\ref{REQ1}) correspond to different dynamo efficiencies
and, in the forthcoming figures, the most constraining bound will be always 
illustrated. The physical power spectra depend on the post-inflationary 
expansion history but their specific form is also determined by the phases of the Sakharov oscillations evaluated at the crossing time $\tau_{k}$. In what follows the previous considerations are illustrated even further by explicitly comparing the results based on the analysis of section \ref{sec4}
with the ones following from the undue identification of $\tau_{k}$ with $\tau_{r}$ (or $\tau_{1}$). In what follows Eq. (\ref{REQ1}) all just be used as a useful 
reference value since, in general, we aim at larger values of the magnetic power 
spectra.

\subsection{Post-inflationary radiation stage}
\label{subs51}
 In the case of a flat spectrum (corresponding, in our notations, to the case $\gamma \to 2$ and $\delta \to 0$) the value of the physical power spectrum at the time of galaxy formation sets a useful benchmark  for all the other cases that are numerically illustrated later on. Since the general expression of the physical spectrum has been already introduced in Eq. (\ref{PS5}) we can now directly use the result of Eq. (\ref{cond8}) and deduce the final value of the physical power spectrum at $\tau_{0}$ 
\begin{equation}
{\mathcal P}_{B}(k,\tau_{0}) = \frac{9}{4 \pi^2} Q(g_{1}, \cos{\theta_{W}})  H_{r}^4 \,\, \biggl(\frac{a_{1}}{a_{k}}\biggr)^4\, \biggl(\frac{a_{k}}{a_{0}}\biggr)^{4}  \,\, \sin^2{k \tau_{k}}.
\label{RAD3}
\end{equation}
Even if $k \tau_{k} = {\mathcal O}(1)$ it is wise to include the last factor in Eq. (\ref{RAD3}) since its contribution  slightly reduces the final value of the power spectrum and, depending on $g_{1}$, the value of $Q(g_{1}, \cos{\theta_{W}})$ (see Eq. (\ref{PHYSPSQ})) may range\footnote{In what follows the typical value $g_{1} =0.1$ will be assumed throughout (see e.g. the legends of Figs. \ref{FIGU4} and \ref{FIGU5}).} between $10^{-3}$ and $10^{-2}$.  Bearing in mind these specifications, Eq. (\ref{RAD3}) can be recast in the following form:
\begin{equation}
{\mathcal P}_{B}(k,\tau_{0}) = \frac{ 9 \, H_{r}^2 \, H_{0}^2}{2 \pi^2}\,\, \Omega_{R0} \,\,
Q(g_{1}, \cos{\theta_{W}})\,\, \sin^2{k \tau_{k}},
\label{RAD3a}
\end{equation}
where $H_{r}$ now coincides with the end of the inflationary phase and, in the case of the conventional timeline of Fig. \ref{FIGU1}, we would actually have  that $H_{r} = H_{1}$ and $a_{r} = a_{1}$. For an even more explicit evaluation of Eqs. (\ref{RAD3})--(\ref{RAD3a}) it is useful to recall that, in conventional inflationary scenarios, $H_{1}/M_{P}$ is fixed by the amplitude of curvature inhomogeneities 
and that the same is true for the total redshift after the end of inflation:
\begin{equation}
 \biggl(\frac{H_{1}}{M_{P}}\biggr)^2 = \frac{\pi \, r_{T} {\mathcal A}_{{\mathcal R}}}{16}, \qquad \qquad \biggl(\frac{a_{1}}{a_{0}} \biggr) = \biggl(\frac{32 \Omega_{R0}}{\pi r_{T} {{\mathcal A}}_{{\mathcal R}}}\biggr)^{1/4}\,\, \sqrt{ \frac{H_{0}}{M_{P}}}.
 \label{RAD3b}
 \end{equation}
All in all inserting Eq. (\ref{RAD3b}) into Eqs. (\ref{RAD3})--(\ref{RAD3a}) 
we have, after simple algebra:
\begin{equation}
{\mathcal P}_{B}(k,\tau_{0}) =  \frac{9\,\, \Omega_{R\,0}}{32 \, \pi} \, H_{0}^2 \, M_{P}^2\, {\mathcal A}_{{\mathcal R}}\, \,r_{T} \,  \,\,Q(g_{1}, \cos{\theta_{W}})\,\,\sin^2{k \tau_{k}}.
 \label{RAD4}
 \end{equation}
By taking into account all the fiducial values of the physical quantities appearing in Eq. (\ref{RAD4}),  the explicit expression of the magnetic power spectrum becomes:
 \begin{equation}
\frac{ {\mathcal P}_{B}(k,\tau_{0}) }{ \mathrm{nG}^2} \,= \ 10^{- 4.261}\,\, Q(g_{1}, \cos{\theta_{W}})\,\,\biggl(\frac{r_{T}}{0.06}\biggr)\,\, 
 \biggl(\frac{{\mathcal A}_{{\mathcal R}}}{2.41 \times 10^{-9}}\biggr)\,\, \biggl(\frac{h_{0}^2 \, \Omega_{R\,0}}{4.15 \times 10^{-5}}\biggr).
 \label{RAD6}
 \end{equation}
As we can see the value of $Q(g_{1}, \cos{\theta_{W}}) = {\mathcal O}(10^{-3})$ is essential to determine the overall amplitude of the spectrum; however this 
complication is simply disregarded in various analyses by assuming that the electromagnetic field was already present during a de Sitter stage of expansion
even if inflation typically takes place before the electroweak symmetry is broken. Some other authors even avoid the notion of the gauge couplings and this means, in our language, that $g_{1} \to 1$.
 The two previous observations imply that while in our estimates $\sqrt{{\mathcal P}_{B}(k, \tau_{0})} = 
{\mathcal O}(10^{-3.63}) \,\, \mathrm{nG}$ for $Q(g_{1}, \cos{\theta_{W}}) = {\mathcal O}(10^{-3})$ there are some who would dub Eq. (\ref{RAD6}) by saying that $\sqrt{{\mathcal P}_{B}(k, \tau_{0}) }= 
{\mathcal O}(10^{-2.13}) \,\, \mathrm{nG}$ for $Q(g_{1}, \cos{\theta_{W}}) \to 1$.

\subsection{A possibly deceptive strategy}
\label{tauktaur}
The results of Eqs. (\ref{RAD3})--(\ref{RAD6}) are a direct consequence of Eqs. (\ref{COND1})--(\ref{cond8}) and of the analysis reported in \ref{sec4}. If we now assume,  by fiat,  that the crossing time $\tau_{k}$ is immaterial and that 
the conductivity dos not play any role it is tempting to replace $\tau_{k}$
by the time of radiation dominance $\tau_{r}$; this is, incidentally, the argument of Ref. \cite{FFF}. This means that instead of evaluating the power spectrum at $\tau_{k}$ (i.e. close to the elliptic disk in the right half of Fig. \ref{FIGU1}) we should instead posit that the current value of the physical power spectrum is simply determined at $\tau_{r}$. The analog of Eq. (\ref{RAD3}) becomes therefore:
 \begin{eqnarray}
 {\mathcal P}^{(r)}_{B}(k,\tau_{0}) &=& {\mathcal P}_{B}(k,\tau_{r}) \biggl(\frac{a_{r}}{a_{0}}\biggr)^4
 \label{WR1}\\
 &=& \frac{9}{4 \pi^2} Q(g_{1}, \cos{\theta_{W}})  H_{1}^4 \,\, \biggl(\frac{a_{1}}{a_{r}}\biggr)^4\, \biggl(\frac{a_{r}}{a_{0}}\biggr)^{4}  \,\, \sin^2{k \tau_{r}}.
\label{WR1a}
 \end{eqnarray}
where $\tau_{r}$ now denotes the time of radiation dominance that 
coincides with $\tau_{1}$ within the expansion history of Fig. \ref{FIGU1}.
We stress that  the superscript at the left hand side of Eq. (\ref{WR1}) indicates that the late-time power spectrum is computed from $\tau_{r}$ and not 
from $\tau_{k}$, as deduced in Eq. (\ref{cond8}). 
 The meaning of $\tau_{r}$ is different when the 
expansion history is given by Figs. \ref{FIGU2} and \ref{FIGU3} and the relevant changes arising in this situation are separately discussed in the subsection \ref{subs52}. 

If we now compare the logic of Eqs. (\ref{COND1})--(\ref{cond8}) with the (arbitrary) prescriptions of Eqs. (\ref{WR1})--(\ref{WR1a}) the expected orders of magnitude of the two evaluations are hugely different because the Sakharov phases are evaluated at $\tau_{r}$ and not at $\tau_{k}$, as implied by the late dominance of the conductivity. More specifically, since $ \tau_{r} = {\mathcal O}(\tau_{1}) \ll \tau_{k}$ denotes the moment at which the plasma is dominated by radiation, the physical power spectrum is in fact evaluated when the relevant wavelengths are still much larger than the Hubble radius:
 \begin{eqnarray}
 {\mathcal P}^{(r)}_{B}(k,\tau_{0}) &=&  \,\,\frac{9 H_{1}^2 \, H_{0}^2 }{4 \pi^2} \,\Omega_{R\, 0}\,Q(g_{1}, \cos{\theta_{W}})  \,\, \sin^2{k \tau_{r}}.
  \label{WR2}
 \end{eqnarray}
Equation (\ref{WR2}) suggests that if the standing waves predominantly oscillate like a sine, the result of Eq. (\ref{RAD6}) is suppressed by a factor that  goes as $|k \, \tau_{r} |^2\, \simeq |k\, \tau_{1}|^2 \ll 1$: 
\begin{equation}
  |k \, \tau_{r}|^2 = 10^{-46.48}  \biggl( \frac{k}{\mathrm{Mpc}^{-1}}\biggr)^2 \,\, \biggl(\frac{r_{T}}{0.06}\biggr)^{-1/2} \,\, \biggl(\frac{{\mathcal A}_{{\mathcal R}}}{2.41\times 10^{-9}}\biggr)^{-1/2} \,\, \biggl(\frac{h_{0}^2 \Omega_{R0}}{4.15 \times 10^{-5}}\biggr)^{-1/2}.
 \label{WR3}
 \end{equation}
This means that the estimates of Eqs. (\ref{WR2}) and (\ref{RAD4})
differ by more than ${\mathcal O}(45)$ orders of magnitude:
\begin{equation}
\frac{ {\mathcal P}^{(r)}_{B}(k,\tau_{0})}{\mathrm{nG}^2} = 10^{-50.59}\,\,Q(g_{1}, \cos{\theta_{W}})\,\,\biggl(\frac{r_{T}}{0.06}\biggr)^{1/2}\,\, \biggl( \frac{k}{\mathrm{Mpc}^{-1}}\biggr)^2\,\,
\biggl(\frac{{\mathcal A}_{{\mathcal R}}}{2.41 \times 10^{-9}}\biggr)^{1/2}\,\, \biggl(\frac{h_{0}^2 \, \Omega_{R\,0}}{4.15 \times 10^{-5}}\biggr)^{1/2}.
\label{WR5}
\end{equation}
If we now want to be purposely inaccurate and fix $Q(g_{1}, \cos{\theta_{W}})\to 1$ 
Eq. (\ref{WR5}) implies that $\sqrt{{\mathcal P}^{(r)}_{B}(k,\tau_{0})} = {\mathcal O}(10^{-34.29})\,\, \mathrm{G}$ which coincides with the estimate of Ref. \cite{FFF} suggesting, for the same quantity, a value ${\mathcal O}(10^{-33})\,\, \mathrm{G}$.
 \begin{figure}[!ht]
\centering
\includegraphics[height=7.5cm]{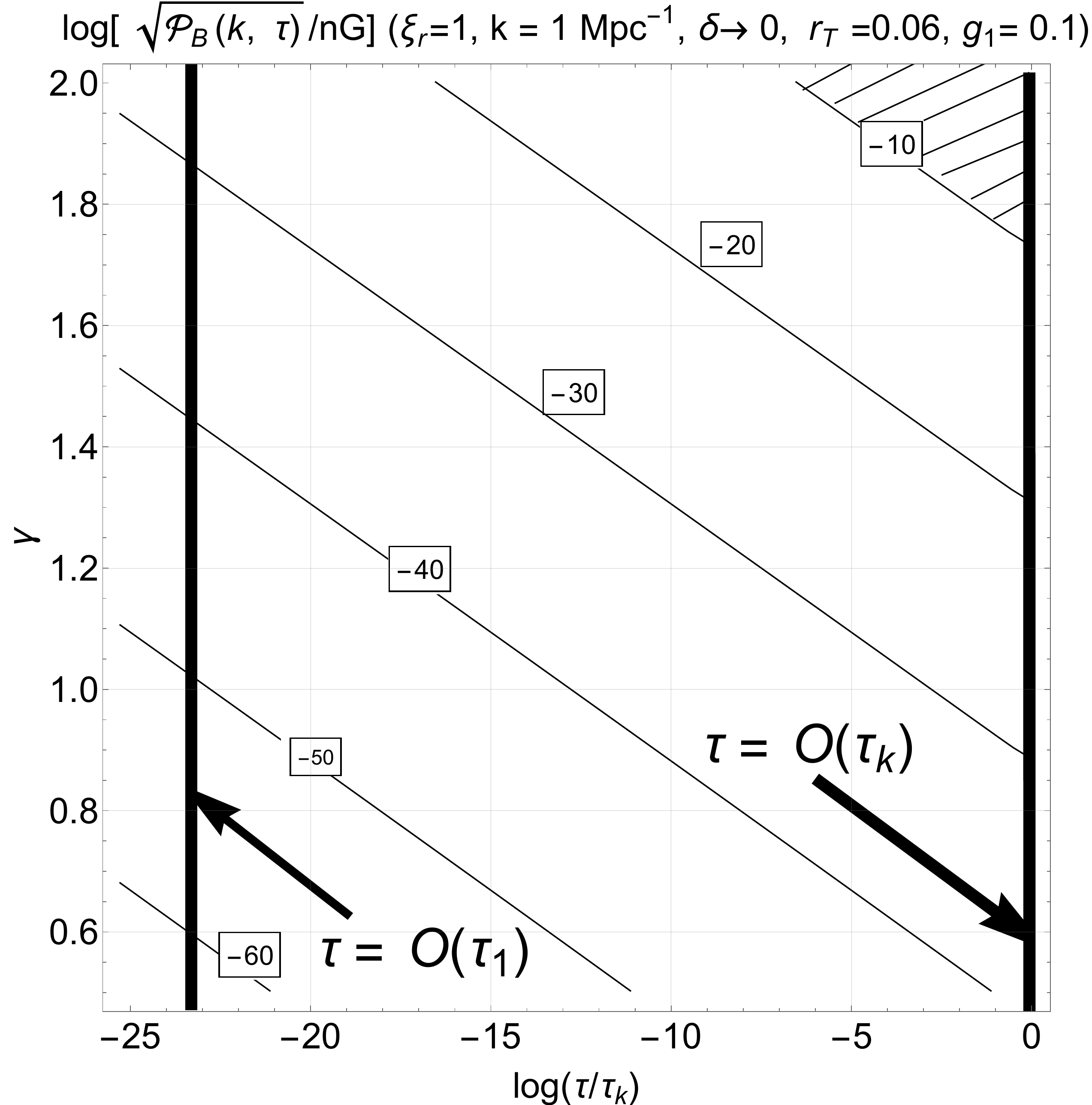}
\includegraphics[height=7.5cm]{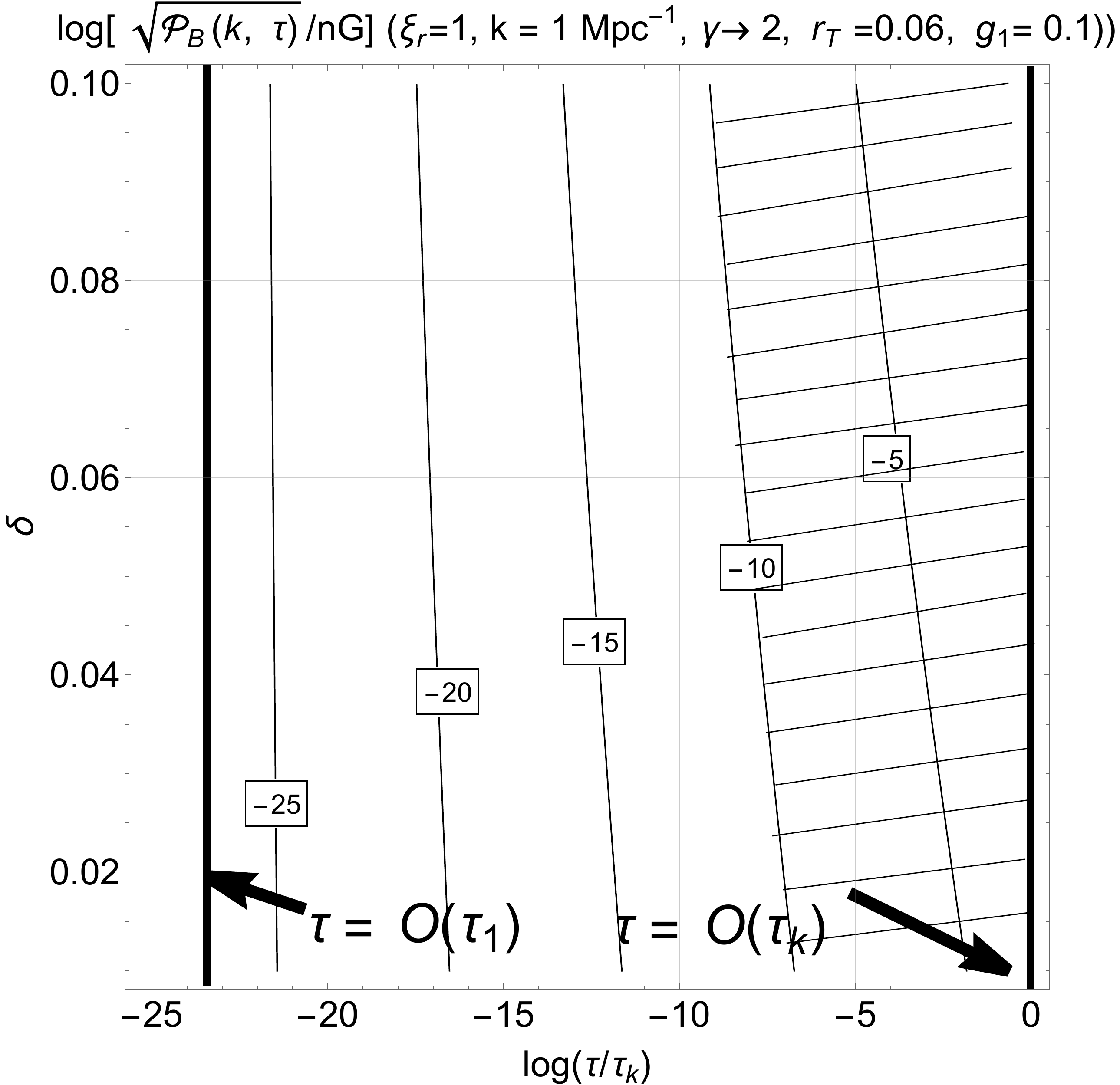}
\caption[a]{In both plots the physical power spectrum is illustrated as a function of $\tau/\tau_{k}$ and for different values of $\gamma$. The leftmost (thick) line corresponds to $\tau = {\mathcal O}(\tau_{1})$ while the rightmost (thick) line denotes $\tau = {\mathcal O}(\tau_{k})$. The labels appearing on the various contours are the common logarithms of $\sqrt{{\mathcal P}_{B}(k,\tau_{0})}$ in nG units
and are the same o each curve. The dashed regions correspond to 
the largest values of the physical power spectra. In the plot at the left $\delta\to 0$ implying that the gauge coupling freezes right after the end of inflation. In the plot at the right $\gamma$ is instead fixed 
while $\delta$ is very small but it does not vanish. The two plots are compatible 
in the case $\gamma\to 2$ and this observation justifies, once more, the limit $\delta \to 0$ that has been always assumed in the analytical estimates (see, in this respect, also appendix \ref{APPA}).}
\label{FIGU4}      
\end{figure}

All in all  the result of Eq. (\ref{WR5}) arises because the phases of the standing oscillation are  evaluated for $ \tau= {\mathcal O}(\tau_{1})$ so that we ultimately have the following hierarchies:
\begin{equation}
{\mathcal P}^{(r)}_{B}(k, \tau_{0}) = {\mathcal O}(10^{-46}) \, {\mathcal P}_{B}(k, \tau_{0})\ll {\mathcal P}_{B}(k, \tau_{0}),
\label{WR5a}
\end{equation}
where, we remind, ${\mathcal P}^{(r)}_{B}(k, \tau_{0})$ has been defined 
in Eq. (\ref{WR1}). If the phases of the standing oscillations are different Eq. (\ref{WR5a}) gets modified and, for this purpose we may suppose, for instance, that the Sakharov oscillations predominantly evolve as a cosine; in this case we may even have that ${\mathcal P}^{(r)}_{B}(k, \tau_{0}) \simeq  {\mathcal P}_{B}(k, \tau_{0})$. This result is in fact a consequence of duality and is discussed in subsection \ref{subs53}. In any case the punchline of this discussion is that the value of the physical power spectrum is not correctly estimated by ${\mathcal P}^{(r)}_{B}(k, \tau_{0})$ defined in Eq. (\ref{WR1}).
\begin{figure}[!ht]
\centering
\includegraphics[height=7.3cm]{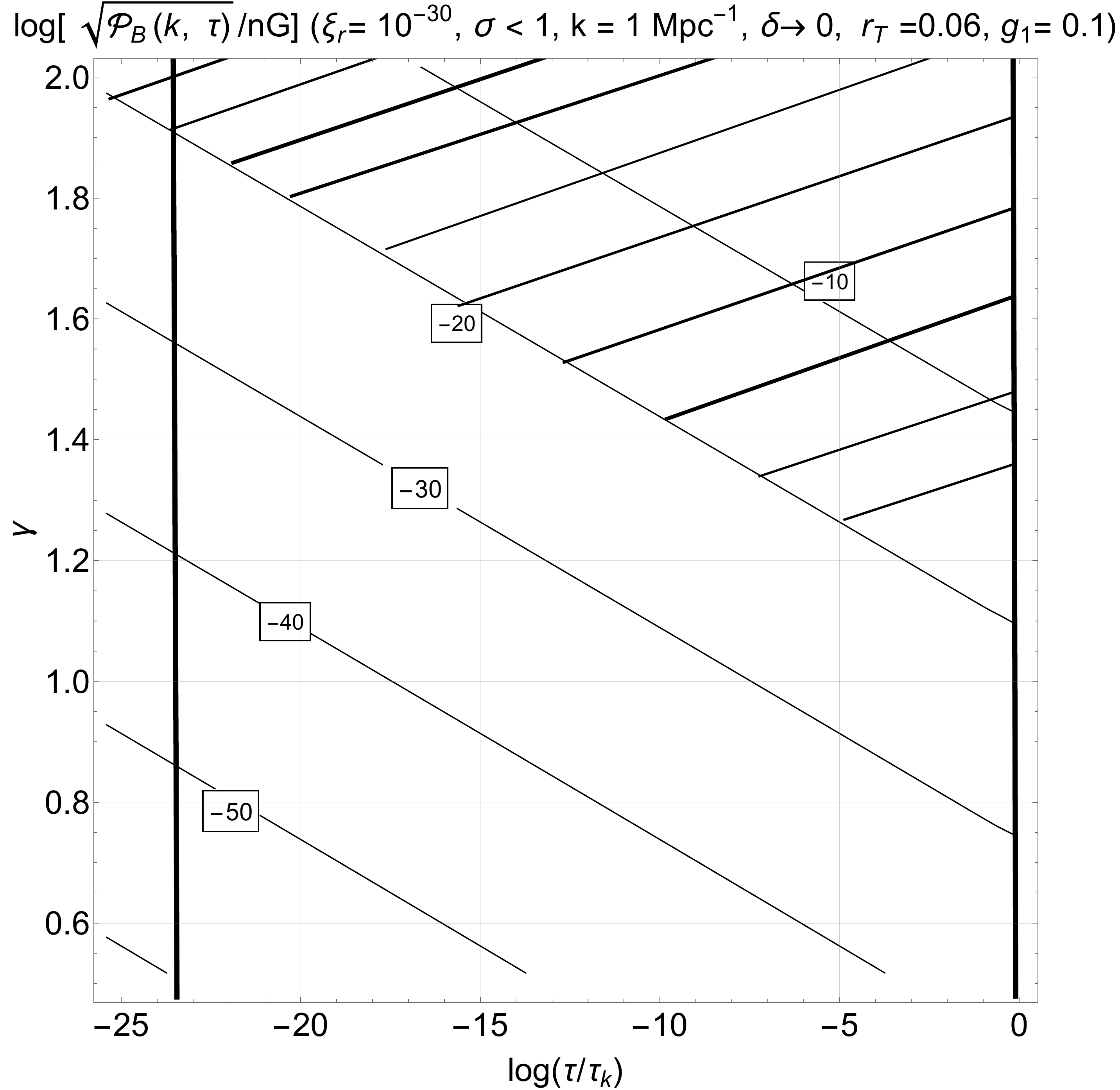}
\includegraphics[height=7.3cm]{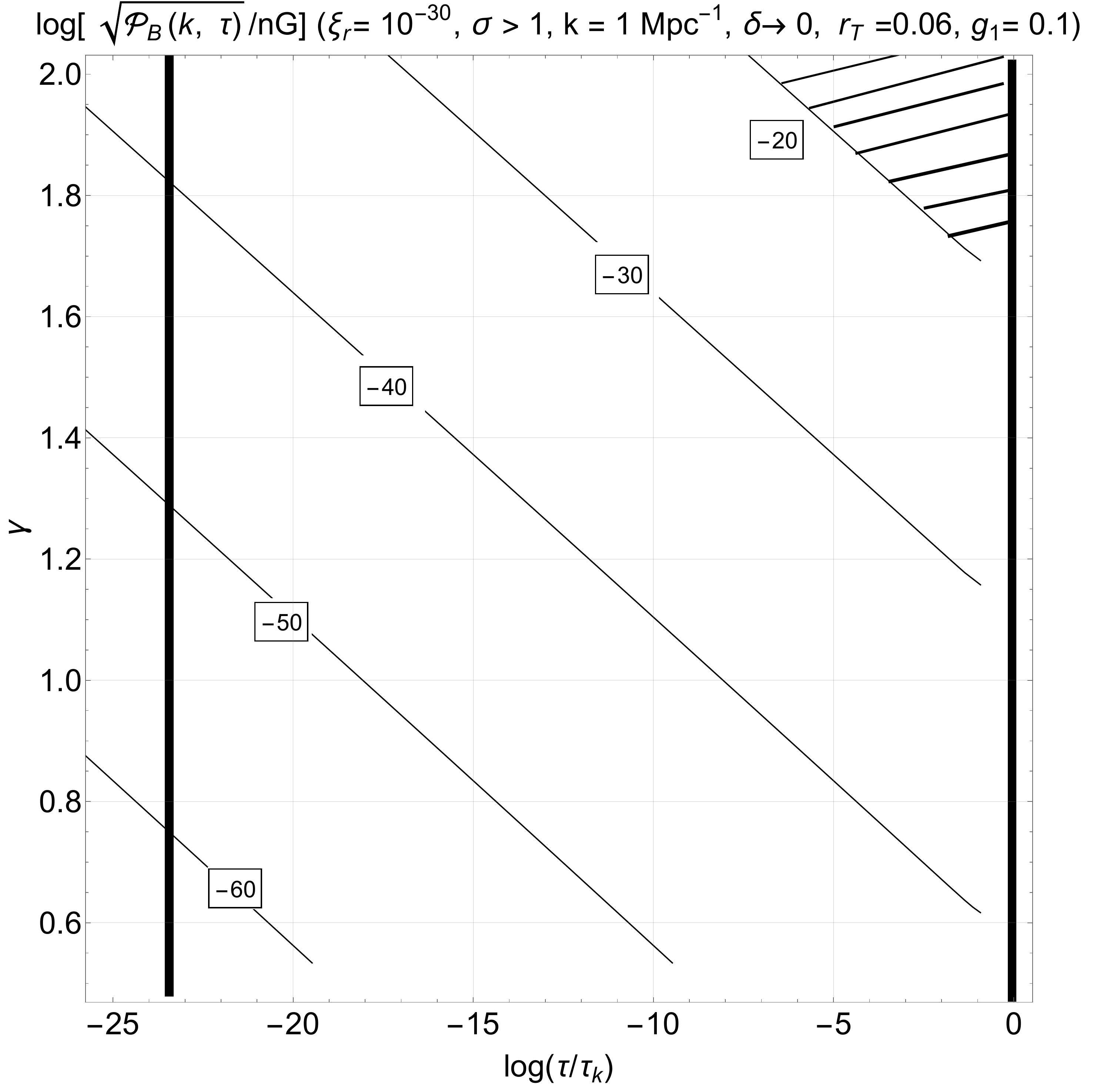}
\caption[a]{The physical power spectrum of the magnetic field is further illustrated with the same notations 
already established in Fig. \ref{FIGU4}. In both plots $\delta \to 0$ (i.e. the gauge coupling freezes 
after inflation) and in both cases we consider a long (i.e. $\xi_{r} = 10^{-30}$) post-inflationary expansion rate that 
differs from radiation. In the plot at the left $\sigma < 1$ (in particular $\sigma =1/2$) while in the plot at the right $\sigma > 1$ (i.e. $\sigma = 2$). }
\label{FIGU5}      
\end{figure}

\begin{figure}[!ht]
\centering
\includegraphics[height=7.5cm]{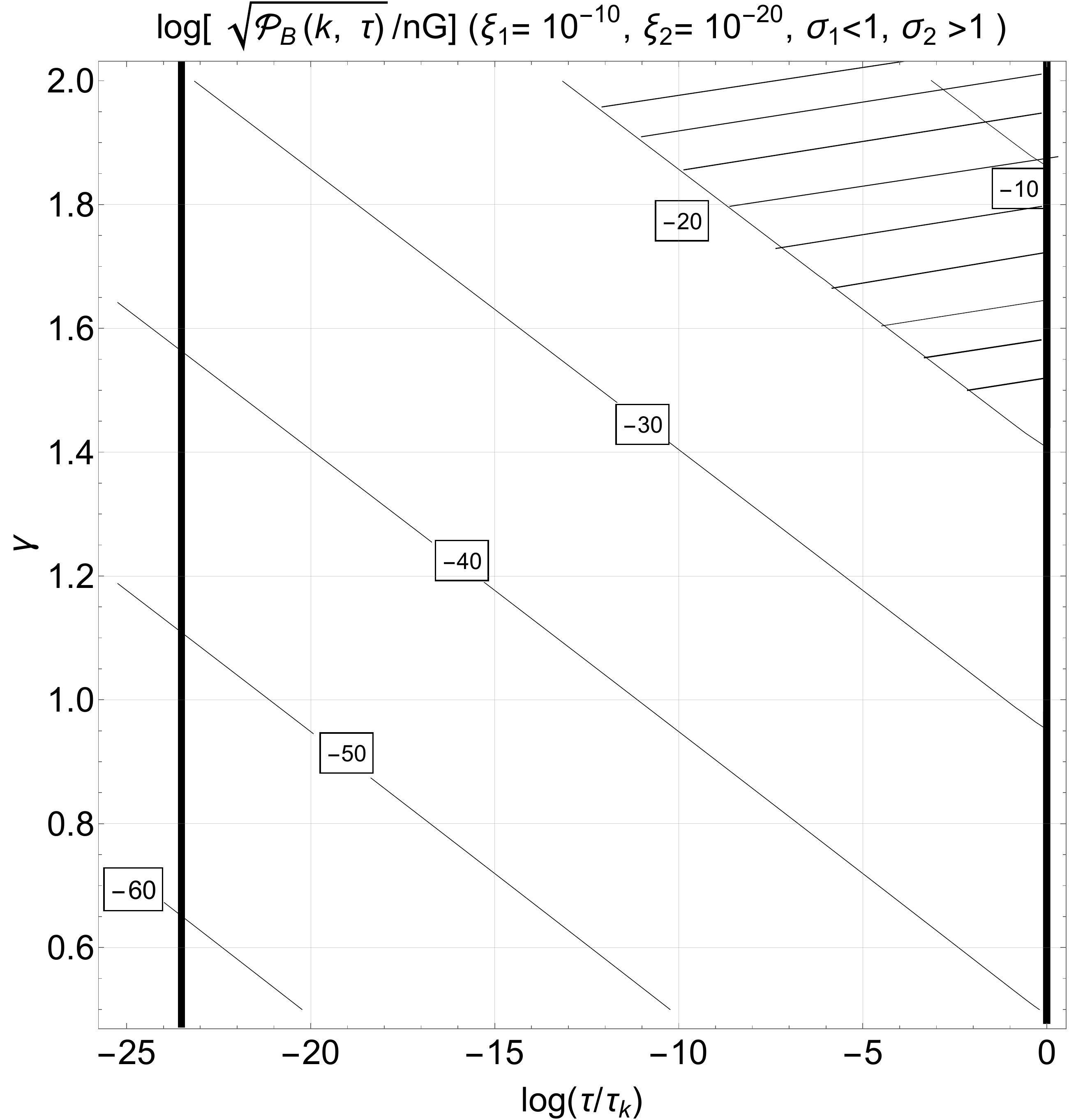}
\includegraphics[height=7.5cm]{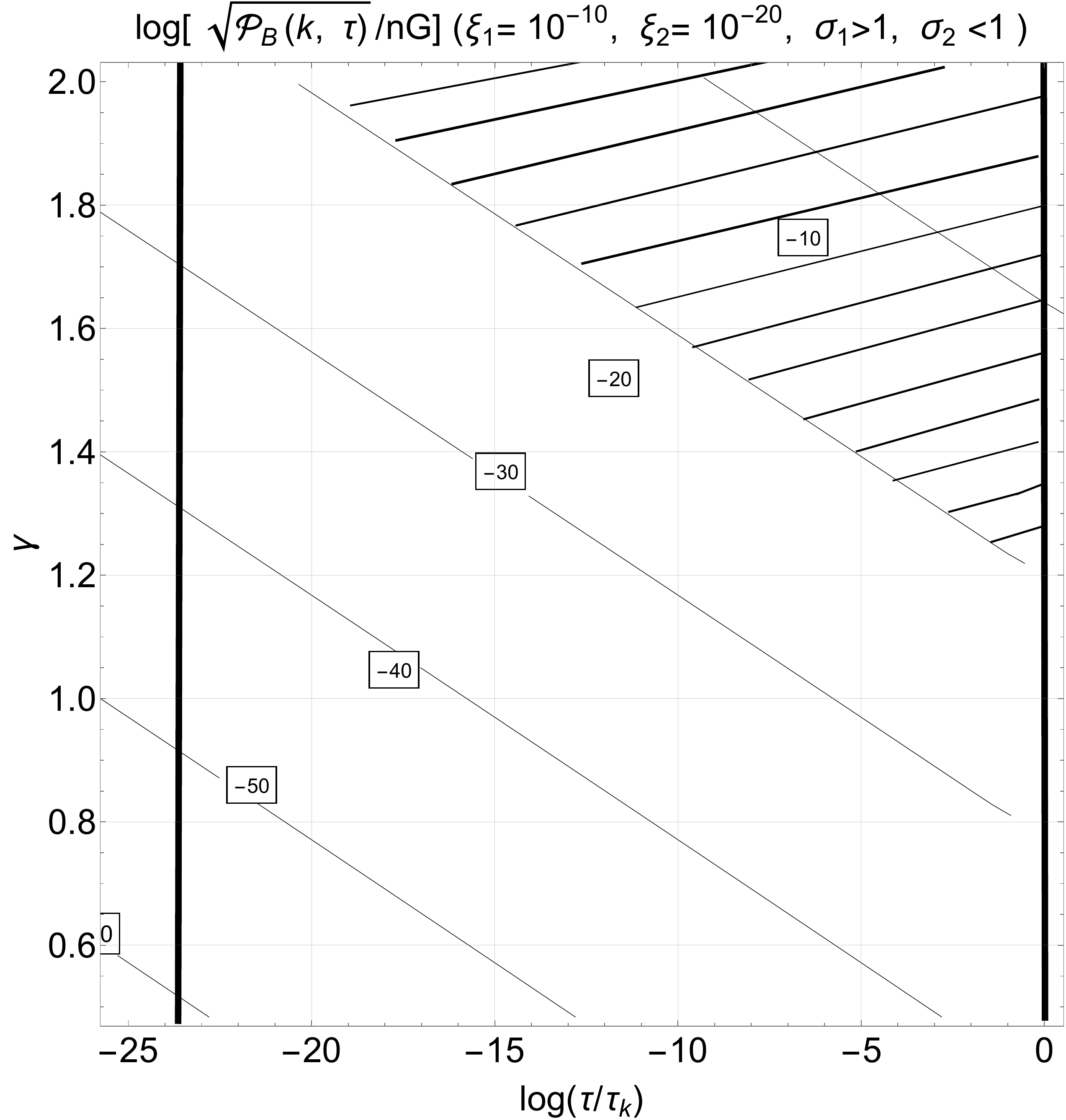}
\caption[a]{As in Fig. \ref{FIGU5} we now discuss the possibility of two separate phases 
both different from radiation. In both plots the first phase is shorter than the second 
(i.e. $\xi_{1} = 10^{-10}$ and $\xi_{2} = 10^{-20}$). The global duration of the post-inflationary stage coincides with the one analyzed in Fig. \ref{FIGU5} since $\xi_{1} \xi_{2} = \xi_{r} = 10^{-30}$. In the left plot the relevant region with large values of the magnetic power spectrum is comparatively smaller and this is because the overall duration of the phase expanding slower that radiation is smaller.}
\label{FIGU6}      
\end{figure}

At a qualitative level the results of Eqs. (\ref{RAD3a})--(\ref{RAD6}) also hold for different values of $\gamma$ and $\delta$ (see Fig. \ref{FIGU4}). In particular the hierarchy between Eqs. (\ref{RAD3a})--(\ref{RAD6}) and  Eqs. (\ref{WR1})--(\ref{WR5}) is illustrated by the two vertical (thick) lines of Figs. \ref{FIGU4}, \ref{FIGU5} and \ref{FIGU6}. These two vertical lines correspond to the two  choices described, respectively, in Eqs. (\ref{cond8}) and (\ref{WR1}).
The vertical line at the left of each plot corresponds in fact to  $\tau = {\mathcal O}(\tau_{1})$ while the vertical line at the right denotes $\tau = {\mathcal O}(\tau_{k})$.  The dashed areas of Figs. \ref{FIGU4}, \ref{FIGU5} and \ref{FIGU6} describe the regions where the magnetogenesis requirements of Eq. (\ref{REQ1}) are approximately satisfied. 

In the left plot of Fig. \ref{FIGU4} the values of $\gamma$ 
vary between $0$ and $2$ so that the results of  Eqs. (\ref{RAD3a})--(\ref{RAD6}) and (\ref{WR1})--(\ref{WR5}) correspond to the {\em upper regions of the plot}. For $ \tau= {\mathcal O}(\tau_{1})$ the power spectrum in the upper left corner of the (left) plot can be estimated from Eq. (\ref{WR5}). Indeed for  $g_{1} =0.1$ (as assumed in Fig. \ref{FIGU4}) $Q(g_{1}, \cos{\theta_{W}}) = 10^{-3.2}$ so that, from Eq. (\ref{WR5}), $\sqrt{{\mathcal P}_{B}(k,\tau_{0})} = {\mathcal O}(10^{-28}) \, \, \mathrm{nG}$ which is consistent with the results of\footnote{It is relevant to bear in mind  that in Fig. \ref{FIGU4} and in the remaining 
figures the labels indicate the common logarithm of $\sqrt{{\mathcal P}_{B}(k,\tau_{0})}$ in nG units. } Fig. \ref{FIGU4}. In the right plot of Fig. \ref{FIGU4} 
the value of $\gamma$ has been fixed (i.e. $\gamma \to 2$) but we reported the final value of the power spectrum for different values of $\delta \leq 0.1$. 
As clarified in Eq. (\ref{PHYSPSQ2}) the different values of $\delta$ slightly break the scale invariance of the result. Even in this case, however, $\sqrt{{\mathcal P}_{B}(k,\tau_{0})}$ evaluated for $\tau = {\mathcal O}(\tau_{1})$ is roughly $23$ orders of magnitude smaller 
than the result obtained when $\tau = {\mathcal O}(\tau_{k})$, as anticipated in Eq. (\ref{WR5a}). The results of Fig. \ref{FIGU4} confirm the conclusions derived above and extends them to the case where the comoving spectra are not necessarily flat but they are instead charaterized by different values of $\gamma$ and $\delta$ compatible with the inflationary constraints.

\subsection{The role of intermediate phases different from radiation}
\label{subs52}
Figure \ref{FIGU4} refers to the timeline of Fig. \ref{FIGU1} but 
the same qualitative features persist for the different post-inflationary evolutions 
of Figs. \ref{FIGU2} and \ref{FIGU3} that are specifically 
examined in Figs. \ref{FIGU5} and \ref{FIGU6}. The main difference 
between the case of Fig. \ref{FIGU4} and the timelines of Figs. \ref{FIGU5} \ref{FIGU6} is that the two time-scales $\tau_{1}$ and $\tau_{r}$ do not coincide.
Thus, the results of the two previous subsections are modified for two complementary reasons:
\begin{itemize}
\item{} the value of $k/(a_{1}\,\,H_{1})$ (or of $k/(a_{r} \, H_{r})$) 
depends of the whole expansion history of the plasma in the presence of multiple phases (see, in this respect, Eqs. (\ref{APPB1})--(\ref{APPB2}) and the related discussions in appendix \ref{APPB});
\item{} the expansion history also affects the amplitude of the physical  power spectrum via the modification of the redshift factors. 
\end{itemize}
Both differences ultimately determine the asymptotic values of the power spectrum as originally suggested long ago \cite{CC0} where the insertion 
of a post-inflationary stage has been considered by mainly focussing 
on the spectral energy density. If between  the end of inflation 
and the onset of the radiation stage there is a single post-inflationary phase (see Fig. \ref{FIGU2}) the overall profile of $k/(a H)$ depends on the value of $\sigma$ (see also, in this respect, the discussion of appendix \ref{APPB}). The absolute minimum of $k/(a \,\,H)$ gets also modified and, in particular, the analog of Eq. (\ref{PS2}) is\footnote{Equation (\ref{NRAD1}) follows from Eqs. (\ref{APPB1})--(\ref{APPB2}) in the 
case $N=2$ where $\sigma_{1}= \sigma$ and $\sigma_{2}$ coincides 
with the standard expansion rate of radiation. }:
\begin{equation} 
\frac{k}{a_{1} \, H_{1}} = 10^{-23.24} \biggl(\frac{k}{\mathrm{Mpc}^{-1}}\biggr)\,\, \biggl(\frac{r_{T}}{0.06}\biggr)^{-1/4} \,  
\biggl(\frac{{\mathcal A}_{{\mathcal R}}}{2.41\times 10^{-9}}\biggr)^{-1/4} \, \biggl(\frac{h_{0}^2 \, \Omega_{R\,0}}{4.15 \times 10^{-5}}\biggr)^{-1/4}\, \xi_{r}^{\frac{(1-\sigma)}{2(1 + \sigma)}}.
\label{NRAD1}
\end{equation}
In Eq. (\ref{NRAD1}) $\xi_{r}$ is the ratio between the curvature scales in $H_{r}$ and $H_{1}$ 
\begin{equation}
 \xi_{r} = \frac{H_{r}}{H_{1}} < 1, \qquad \qquad H_{r} \geq 10^{-38} \, M_{P}.
\label{WR4}
 \end{equation}
 The limit indicated in Eq. (\ref{WR4}) follows by requiring that $H_{r}$ always 
 exceeds the approximate curvature scale $10^{-44}\,\, M_{P}$ which roughly corresponds to a temperature of the plasma $T= {\mathcal O}(\mathrm{MeV})$. Equation (\ref{NRAD1}) eventually follows by observing that:
\begin{equation}
\frac{a_{0} \, H_{0}}{a_{1} \, H_{1}} = \frac{\sqrt{H_{0}/M_{P}}}{[2 \,\pi \,\epsilon \, {\mathcal A}_{{\mathcal R}} \, \Omega_{R0}]^{1/4}} \, \, \xi_{r}^{\frac{(1-\sigma)}{2(1 + \sigma)}}.
\label{NRAD2}
\end{equation}
By combining together the results of Eqs. (\ref{NRAD1})--(\ref{NRAD2}) we obtain that  $| k \, \tau_{r}|^2$ is now modified as: 
\begin{equation}
 |k \, \tau_{r}|^2 = \xi_{r}^{-1} \,10^{-46.48}  \biggl( \frac{k}{\mathrm{Mpc}^{-1}}\biggr)^2 \,\, \biggl(\frac{r_{T}}{0.06}\biggr)^{-1/2} \,\, \biggl(\frac{{\mathcal A}_{{\mathcal R}}}{2.41\times 10^{-9}}\biggr)^{-1/2} \,\, \biggl(\frac{h_{0}^2 \Omega_{R0}}{4.15 \times 10^{-5}}\biggr)^{-1/2}.
\label{NRAD3}
\end{equation}
According to Eq. (\ref{NRAD3}) the contribution of Eqs. (\ref{NRAD1})--(\ref{NRAD2}) combine so that the net addition to $|k \, \tau_{r}|^2$ is given by $\xi_{r}^{-1}$. 
We could consider next a series of $N$ expanding stages characterized by different rates $\sigma_{i}$ (see appendix 
\ref{APPB} and discussion therein); the analog of Eq. (\ref{RAD4}) is then modified, in this case, as 
\begin{equation}
{\mathcal P}_{B}(k,\tau_{0}) =  \frac{9\,\, \Omega_{R\,0}}{32 \, \pi} \, H_{0}^2 \, M_{P}^2\, {\mathcal A}_{{\mathcal R}}\, \,r_{T} \,  \,\,Q(g_{1}, \cos{\theta_{W}})\,\,\sin^2{k \tau_{k}} \,\, \prod_{i =1}^{N-1} \, \, \xi_{i}^{2\frac{(\sigma_{i} -1)}{(\sigma_{i} +1)}},
 \label{NRAD4}
 \end{equation}
 where now the physical power spectrum is computed by assuming the presence  of an intermediate stage of expansion preceding the radiation phase.
\begin{figure}[!ht]
\centering
\includegraphics[height=7.5cm]{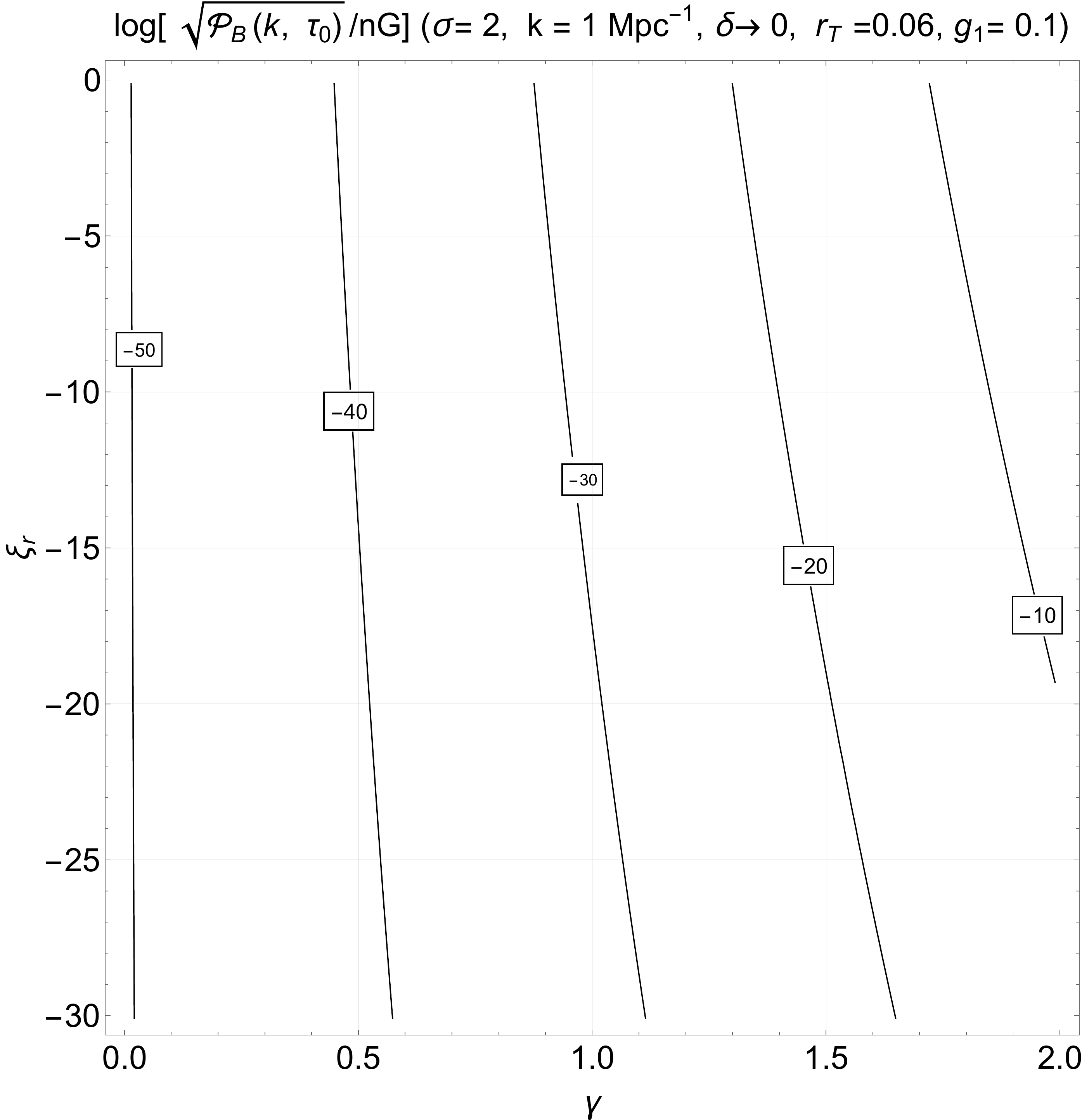}
\includegraphics[height=7.5cm]{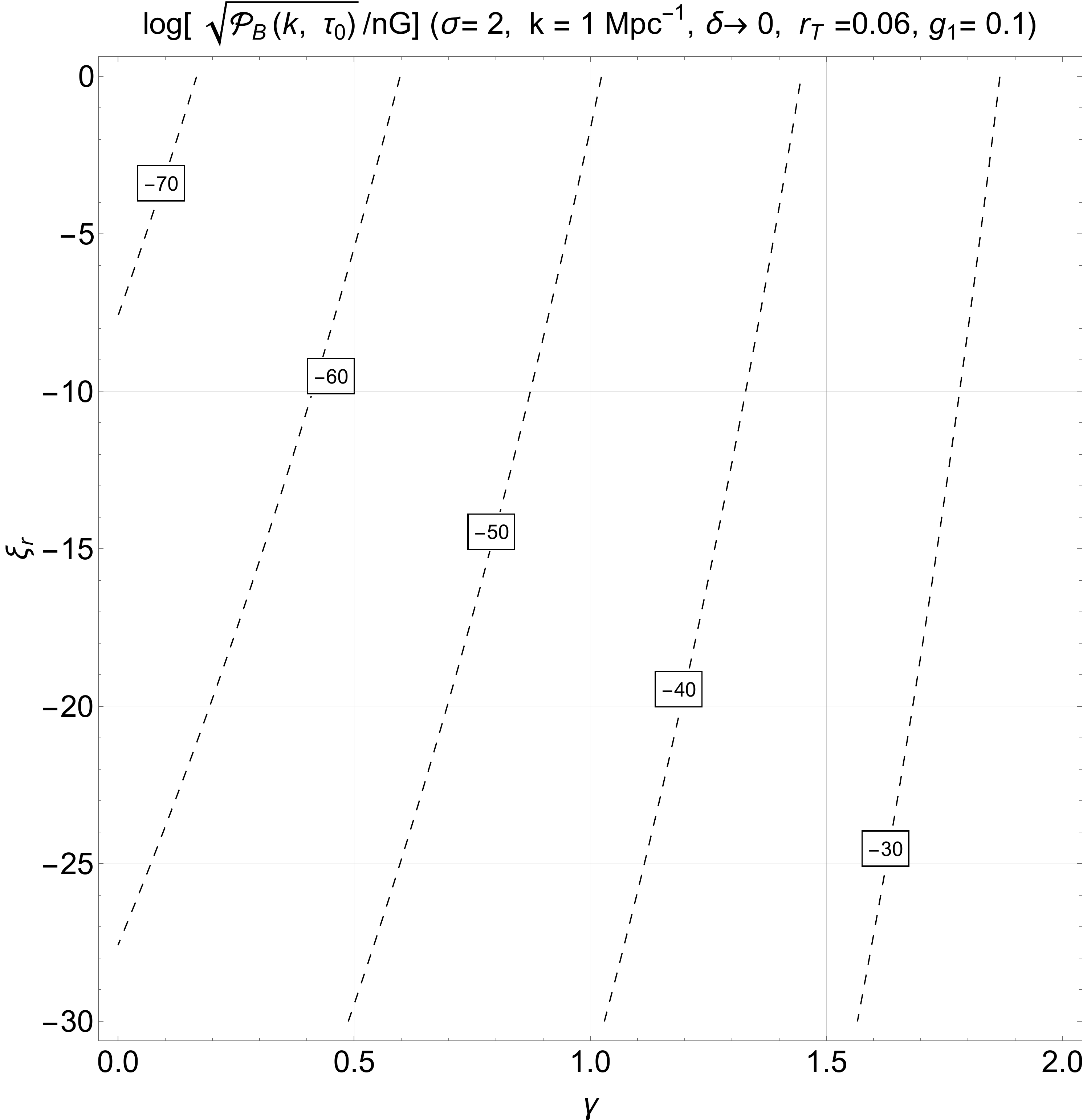}
\caption[a]{If the physical power spectrum before the Hubble crossing 
is evaluated for $\tau= {\mathcal O}(\tau_{k})$ we obtain the plot at the left. Conversely if the power spectrum is evaluated for $ \tau = {\mathcal O}(\tau_{r}) \ll \tau_{k}$ the result is obviously much smaller and it is illustrated with the dashed contours. In both plots of this figure $\sigma > 1$. }
\label{FIGU7}      
\end{figure}
Using the fiducial values of the physical parameters 
Eq. (\ref{NRAD4}) can be made even more explicit so that the final result is:
 \begin{equation}
\frac{ {\mathcal P}_{B}(k,\tau_{0}) }{ \mathrm{nG}^2} \,= \ 10^{- 4.261}\,\, Q(g_{1}, \cos{\theta_{W}})\,\,\biggl(\frac{r_{T}}{0.06}\biggr)\,\, 
 \biggl(\frac{{\mathcal A}_{{\mathcal R}}}{2.41 \times 10^{-9}}\biggr)\,\, \biggl(\frac{h_{0}^2 \, \Omega_{R\,0}}{4.15 \times 10^{-5}}\biggr) \, \prod_{i =1}^{N-1} \, \, \xi_{i}^{2\frac{(\sigma_{i} -1)}{(\sigma_{i} +1)}}.
 \label{NRAD6}
 \end{equation}
Equation (\ref{NRAD6}) suggests that, for a given number of post-inflationary stages different from radiation, the final value of the physical power spectrum gets larger as long as the number of phases expanding slower than radiation is maximized. This is realized by increasing the duration of the stages where 
$\sigma_{i} < 1$. Conversely when the expansion rate is larger than radiation 
the final value of the physical power spectrum gets comparatively 
smaller.

The results of Eq. (\ref{NRAD4}) hold for a flat spectral slope but $\gamma$ may span the whole allowed range and the outcome of this analysis appears in Figs. \ref{FIGU5} and \ref{FIGU6}. In Fig. \ref{FIGU5} the case of {\em a single intermediate phase} with rate $\sigma$ is specifically examined while in Fig. \ref{FIGU6} we considered two intermediate stages (characterized by the rates $\sigma_{1}$ and $\sigma_{2}$). In Fig. \ref{FIGU5} the duration of the intermediate stage has been fixed to $\xi_{r} = 10^{-30}$ and we considered the cases $\sigma <1$ (plot at the left) and $\sigma > 1$ (plot at the right). The two plots of Fig. \ref{FIGU5} must be compared with the {\em left plot} of Fig. \ref{FIGU4}. From these two complementary situations and from the areas of the the shaded regions it turns out that the phenomenologically interesting region increases when $\sigma < 1$ while it shrinks when $ \sigma > 1$. This result confirms the trend already pointed out by when analyzing Eq. (\ref{RAD6}) and for a single intermediate phase the spectrum of Eq. (\ref{RAD4}) (originally obtained on the basis of the timeline of Fig. \ref{FIGU1}) is therefore multiplied by a single factor containing $\xi_{r}$, i.e. 
\begin{equation}
 {\mathcal P}^{(\sigma)}_{B}(k,\tau_{0}) = \xi_{r}^{\frac{2 (\sigma -1)}{\sigma+1}} \,\,  {\mathcal P}_{B}(k,\tau_{0}) =  \xi_{r}^{\frac{2 (1 - 3 \,w)}{3 \,(1 + w)}} \,\,  {\mathcal P}_{B}(k,\tau_{0}),
 \label{NRAD7}
 \end{equation}
 where  the superscript $\sigma$ has been included at the left hand side since the corresponding result holds in the presence of an intermediate $\sigma$-phase\footnote{The second equality in Eq. (\ref{NRAD7}) follows by assuming that the sources during the single intermediate stage are parametrized in terms of a perfect fluid with barotropic index $w$; this means that Eq. (\ref{NRAD7}) refers, in practice, to the timeline of Fig. \ref{FIGU2}.}. 
 \begin{figure}[!ht]
\centering
\includegraphics[height=7.5cm]{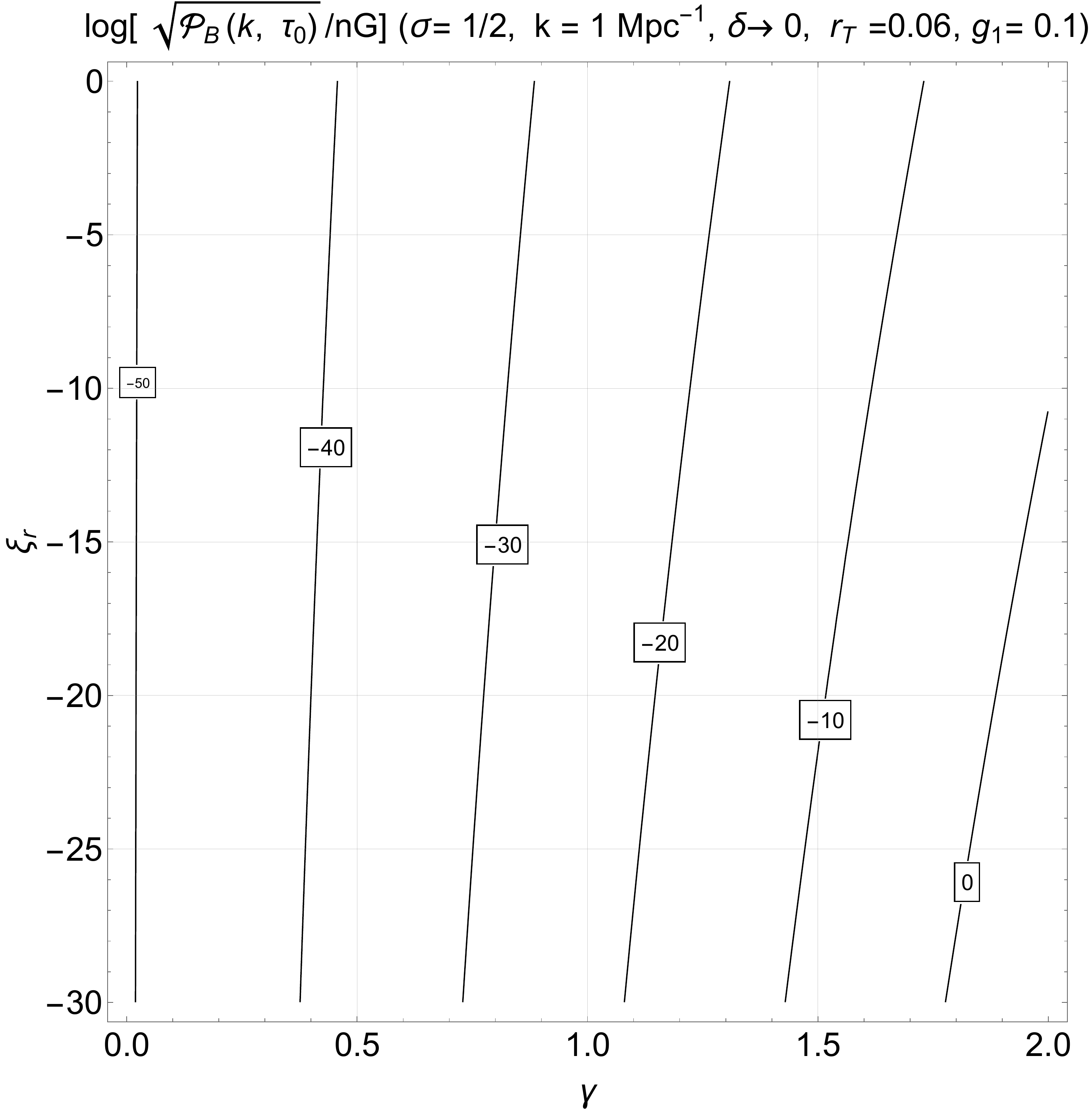}
\includegraphics[height=7.5cm]{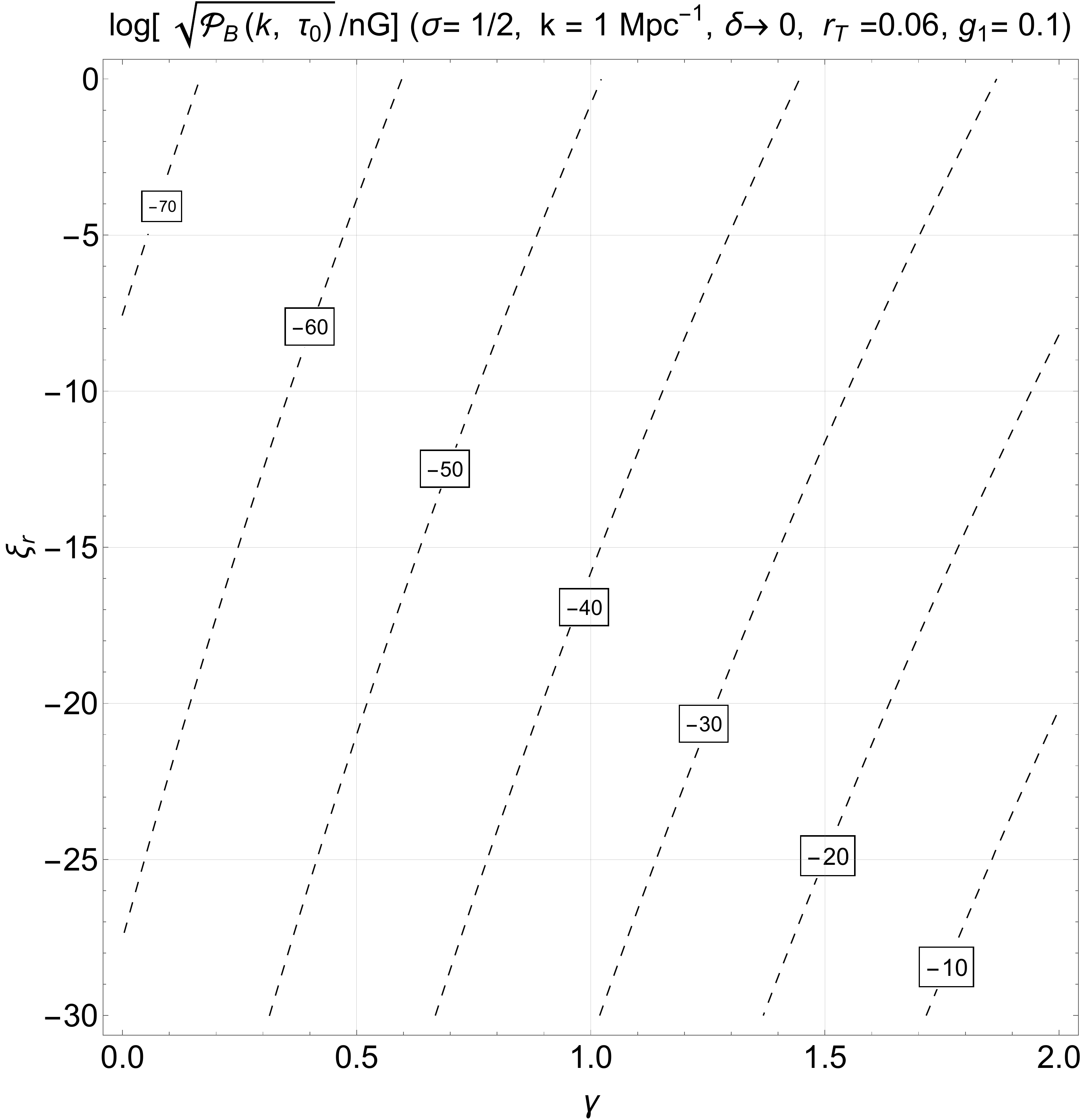}
\caption[a]{The same analysis illustrated in Fig. \ref{FIGU7} is now discussed 
for $\sigma = 2$ and it is representative of all the cases $\sigma >1$. 
The results with what we regard as the correct normalization are illustrated with the full lines while with the dashed lines we consider the case where the value of the physical power 
spectrum before Hubble crossing is set for $\tau= \tau_{r} \ll {\mathcal O}(\tau_{r})$. }
\label{FIGU8}      
\end{figure}

 The results of Fig. \ref{FIGU5} are complemented by Fig. \ref{FIGU6} 
where we actually considered the next level of complication represented by {\em two} intermediate phases with rates $\sigma_{1}$ and $\sigma_{2}$. According to Fig. \ref{FIGU6} first stage (with rate $\sigma_{1}$) is characterized by $\xi_{1}= H_{2}/H_{1} = 10^{-10}$ while the second stage (with rate $\sigma_{2}$) is characterized by $\xi_{2} = H_{r}/H_{2}$. Since $ \xi_{1} \, \xi_{2} = \xi_{r} = 10^{-30}$ (see also Eq. (\ref{APPB1})) the global duration of the intermediate stage coincides with the one of Fig. \ref{FIGU5}.  In the plot at the left of Fig. \ref{FIGU6} we took $\sigma_{1} <1$ and $\sigma_{2} >1$ (and this 
means that the phase expanding slower than radiation is {\em shorter} than the one expanding faster than radiation). In the plot at the right the values of $\sigma_{1}$ and $\sigma_{2}$ have been exchanged (i.e. $\sigma_{1} > 1$ and $\sigma_{2} <1$): this means that the phase expanding slower than radiation is comparatively {\em longer} than the one expanding 
faster than radiation. The two cases are not symmetric since when the slow phase is longer (as in the right plot) the phenomenologically 
interesting region (illustrated by the shaded area) gets wider.  This is the same 
trend already observed in Fig. \ref{FIGU5}.

\begin{figure}[!ht]
\centering
\includegraphics[height=7.5cm]{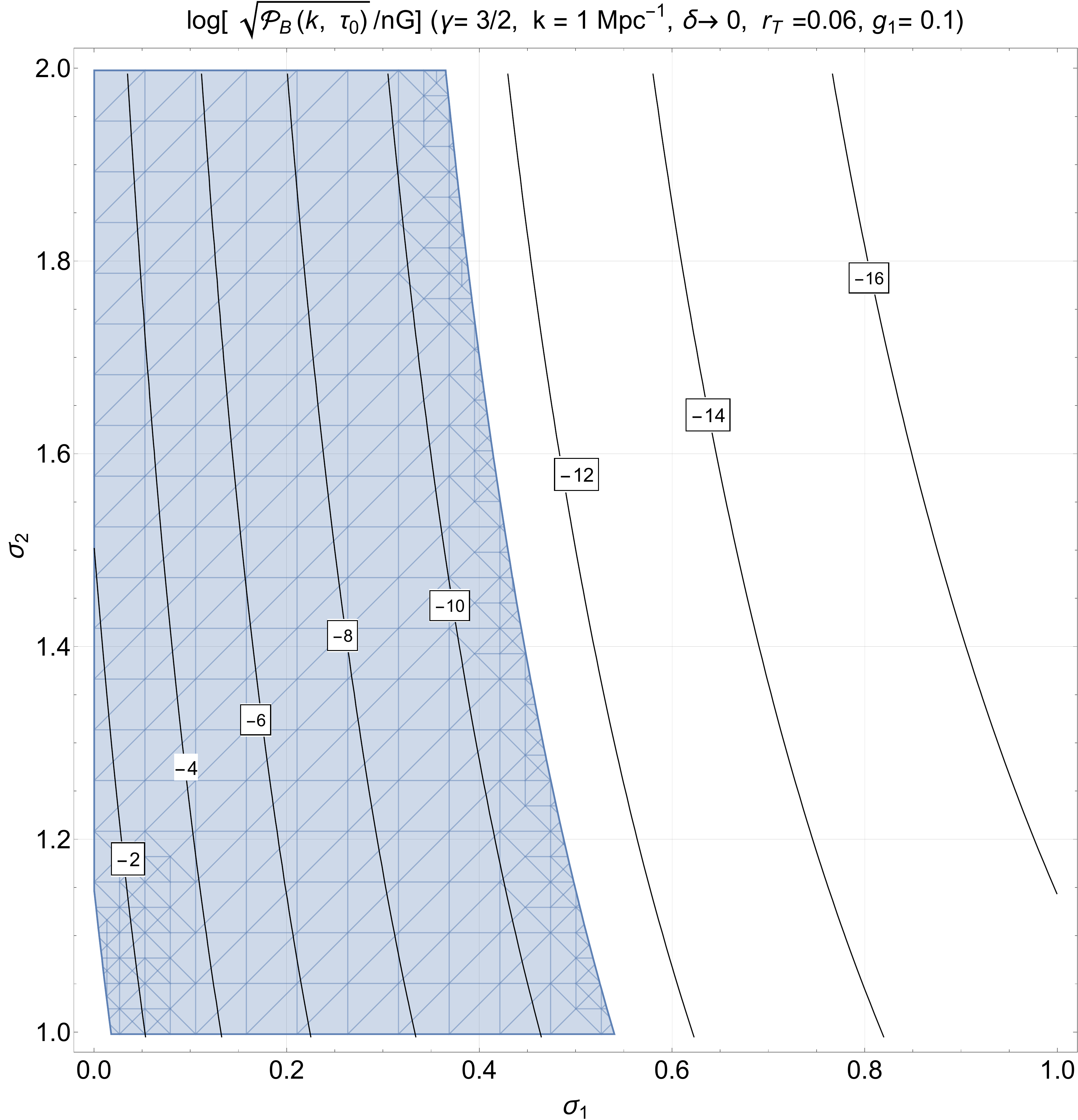}
\includegraphics[height=7.5cm]{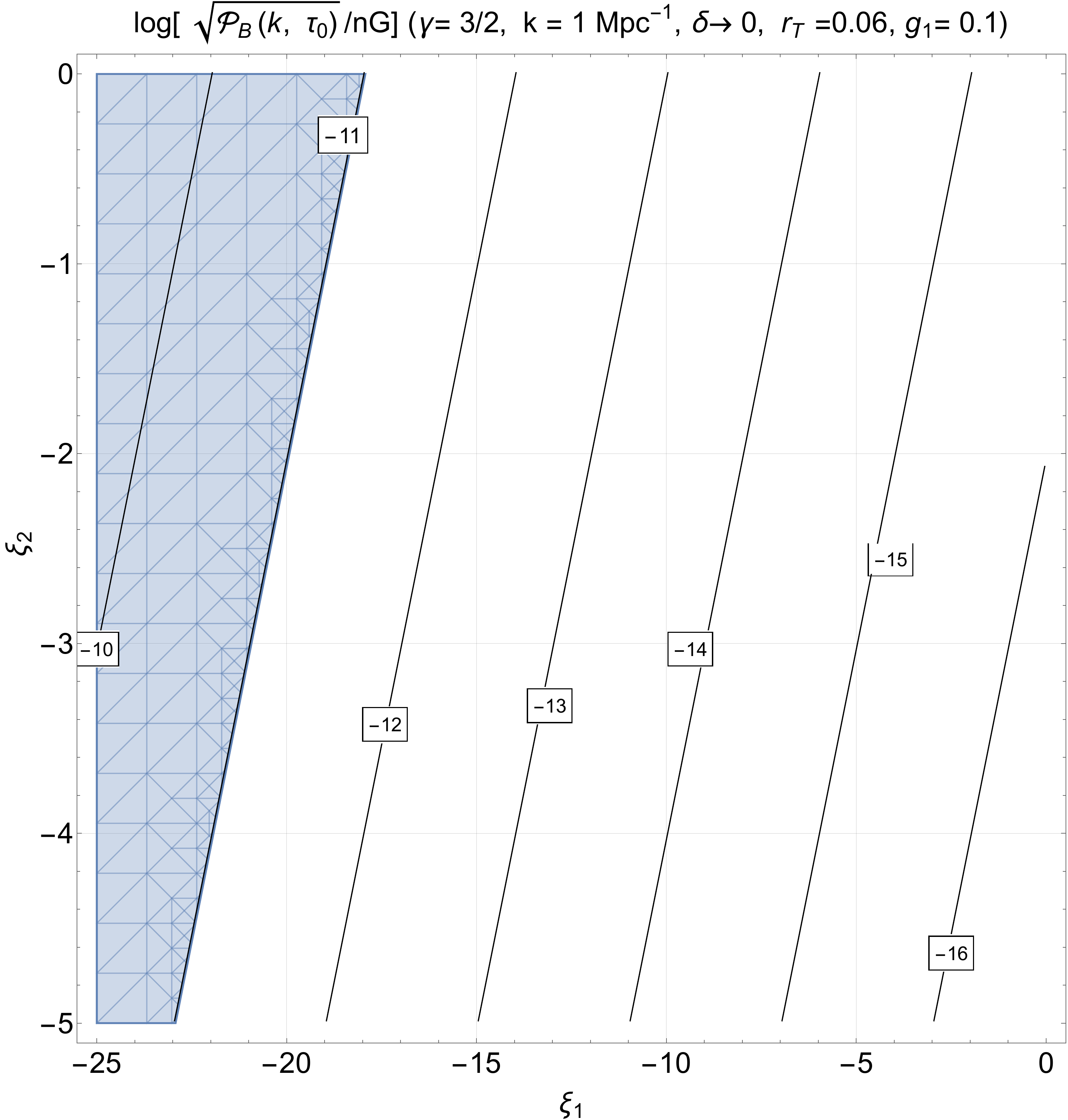}
\caption[a]{We illustrate here the possibility of two post-inflationary phases parametrized by the expansion rates $\sigma_{1}$ and $\sigma_{2}$. In the plot at the left we assumed $\xi_{1}= 10^{-20}$ and $\xi_{2} = 10^{-10}$. In the plot at the right $\sigma_{1} = 1/2$ and $\sigma_{2} = 2$. As before the shaded area denotes the region where the magnetogenesis 
requirements are satisfied and the spectral energy density is smaller than $10^{-6}$. }
\label{FIGU11}      
\end{figure}

As established before the physical power spectrum must be evaluated at $\tau_{k}$ whose value is always ${\mathcal O}(10^{-2})\,\, \tau_{eq}$ but, for the record, the prescription based on Eqs. (\ref{WR1})--(\ref{WR2}) 
may also be complemented by an intermediate stage of expansion. If the phases 
are then evaluated in $\tau_{r}$  Eq.  (\ref{NRAD4}) becomes:
\begin{equation}
{\mathcal P}^{(r)}_{B}(k,\tau_{0}) =  \frac{9\,\, \Omega_{R\,0}}{32 \, \pi} \, H_{0}^2 \, M_{P}^2\, {\mathcal A}_{{\mathcal R}}\, \,r_{T} \,  \,\,Q(g_{1}, \cos{\theta_{W}})\,\,\sin^2{k \tau_{r}} \,\, \prod_{i =1}^{N-1} \, \, \xi_{i}^{2\frac{(\sigma_{i} -1)}{(\sigma_{i} +1)}}.
 \label{NWR1}
 \end{equation}
 \begin{figure}[!ht]
\centering
\includegraphics[height=7.5cm]{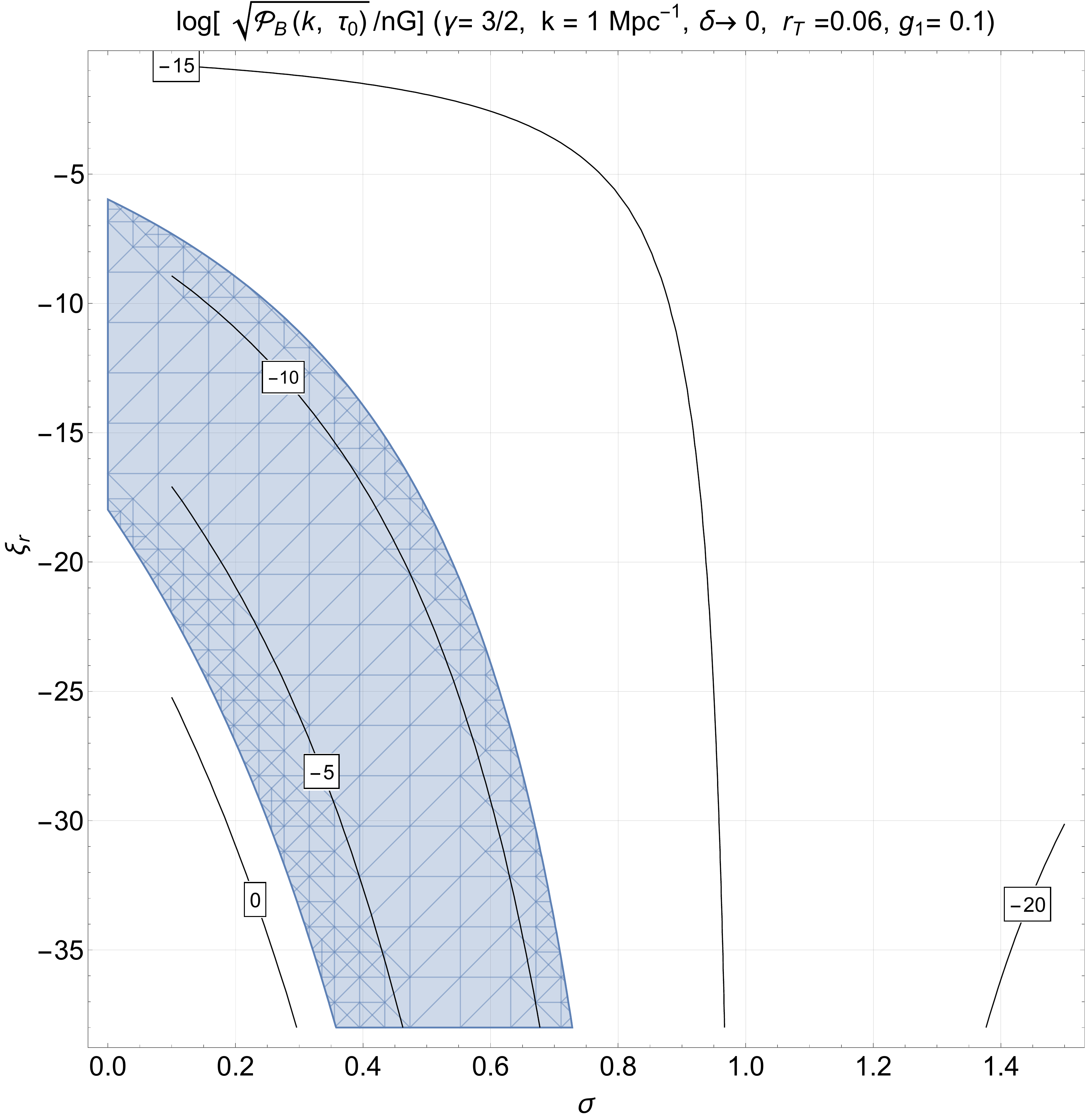}
\includegraphics[height=7.5cm]{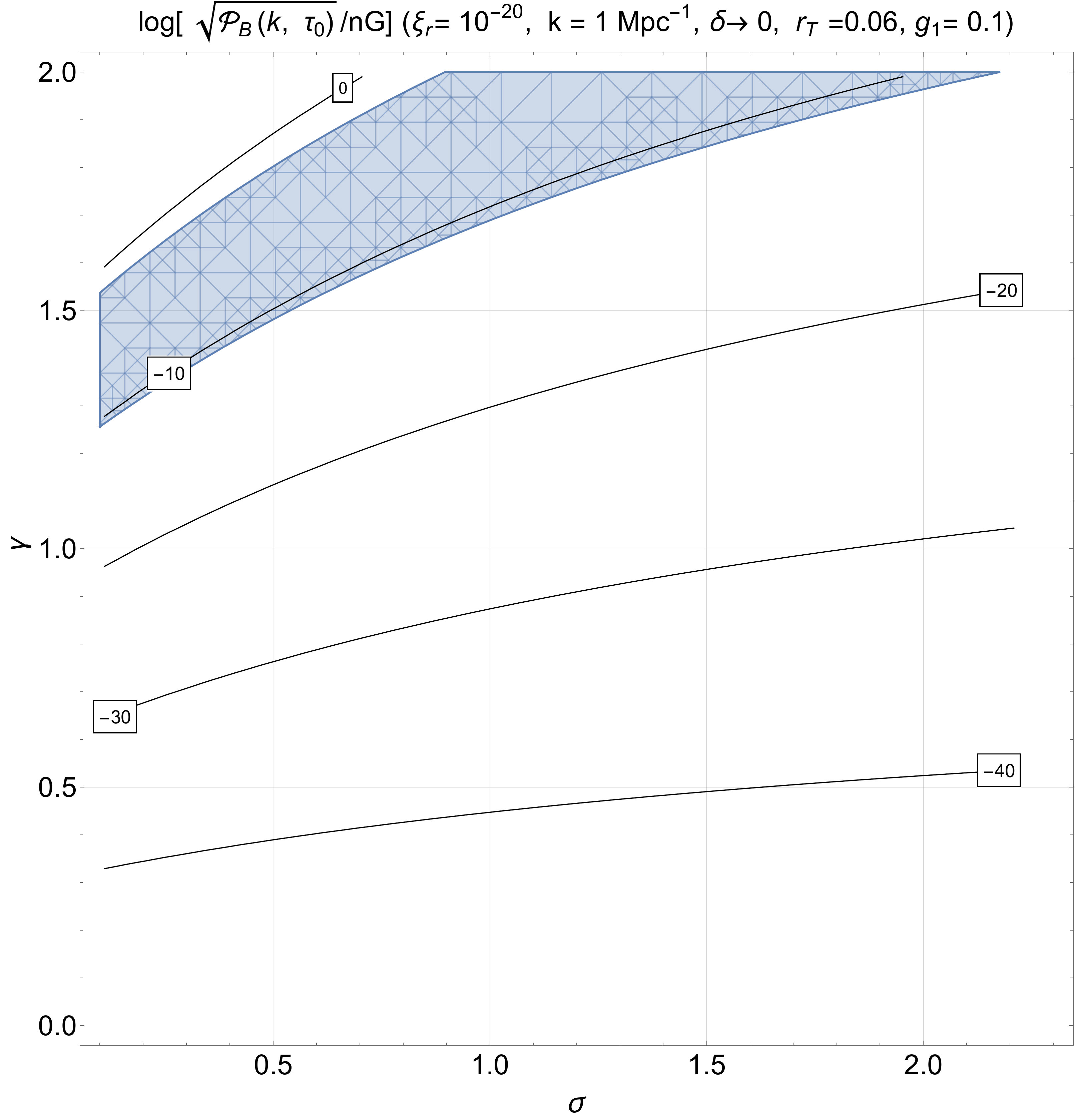}
\caption[a]{In both plots the late-time power spectrum is illustrated in the presence of an intermediate 
post-inflationary phase with $\xi_{r}=H_{r}/H_{1}< 1$; the post-inflationary expansion rate is controlled by $\sigma$. The value of the physical power spectrum is evaluated at $\tau_{k}$ and subsequently 
scaled by following the dominance of the conductivity. Th shaded area corresponds to the region 
where where the spectral energy density is smaller than $10^{-6}$ and the magnetogenesis requirements 
are satisfied in their most demanding form (i.e. $\sqrt{{\mathcal P}_{B}(k,\tau_{0})} > 10^{-22}$). In the plot at the left 
the value of $\gamma$ is fixed to $3/2$ (implying a rather small power spectrum at the magnetogenesis scale). In the 
plot at the right $\xi_{r} =10^{-30}$.}
\label{FIGU9}      
\end{figure}
According to Eq. (\ref{NWR1}) the sine contribution is extremely small since $k \tau_{r} \ll 1$ and, according to Eq. (\ref{NRAD3}), its square also depends on the intermediate stages of expansion. This implies that the overall 
 dependence of the physical power spectrum upon the different stages of expansion can now 
 be written as:
 \begin{equation}
  \frac{ {\mathcal P}^{(r)}_{B}(k,\tau_{0})}{\mathrm{nG}^2} = 10^{-50.59}\,\,Q(g_{1}, \cos{\theta_{W}}) \prod_{i =1}^{N-1} \, \, \xi_{i}^{\frac{\sigma_{i} -3}{\sigma_{i} +1}}.
 \label{NWR2}
\end{equation}
where we used the same typical values of $r_{T}$, ${\mathcal A}_{{\mathcal R}}$ and $\Omega_{R0}$ already employed in Eq. (\ref{WR5}). For a single intermediate phase we therefore have that the modified 
physical power spectrum becomes:
\begin{equation}
 {\mathcal P}^{(r,\, \sigma)}_{B}(k,\tau_{0}) = \xi_{r}^{\frac{(\sigma -3)}{\sigma+1}} \,\,  {\mathcal P}^{(r)}_{B}(k,\tau_{0}) =  \xi_{r}^{-\frac{(1 + 9 \,w)}{3 \,(1 + w)}} \,\,  {\mathcal P}^{(r)}_{B}(k,\tau_{0}),
 \label{NWR3}
 \end{equation}
  where, as in Eq. (\ref{NRAD7}) the superscript $\sigma$ has been included 
  at the left hand side just to stress that this result holds in the presence of an intermediate $\sigma$-phase.  If we now compare the expressions of Eqs. (\ref{NRAD7}) and (\ref{NWR3}) we can observe sharp differences not only in the overall amplitude (as already pointed out in Eq. (\ref{WR5a})) but also in the dependence upon the intermediate phase. According to Eq. (\ref{NRAD7}) the power spectra get comparatively larger when the plasma passes through a stage expanding at a rate 
  that is slower than radiation (i.e. $\sigma < 1$) and this is consistent with the fact that the energy density gets less diluted during a stiff stage of expansion where $w>1/3$. Equation (\ref{NWR3}) suggests instead {\em any} post-inflationary stage increases the value of the magnetic power spectrum and this is just a consequence of the fact that the power spectrum gets artificially smaller when we evaluate its phases for $k \tau_{r} \ll 1$ rather than for $k \tau_{k} = {\mathcal O}(1)$. The 
  final value of the power spectrum gets enhanced even when the evolution is dominated 
  by radiation in sharp contrast with Eq. (\ref{NRAD7}) where 
  in the limit $w\to 1/3$ the results of subsection \ref{subs51} (based on Fig. \ref{FIGU1}) 
  must be recovered; this is what has been already observed, incidentally, in Ref. \cite{CC0}.
\begin{figure}[!ht]
\centering
\includegraphics[height=7.5cm]{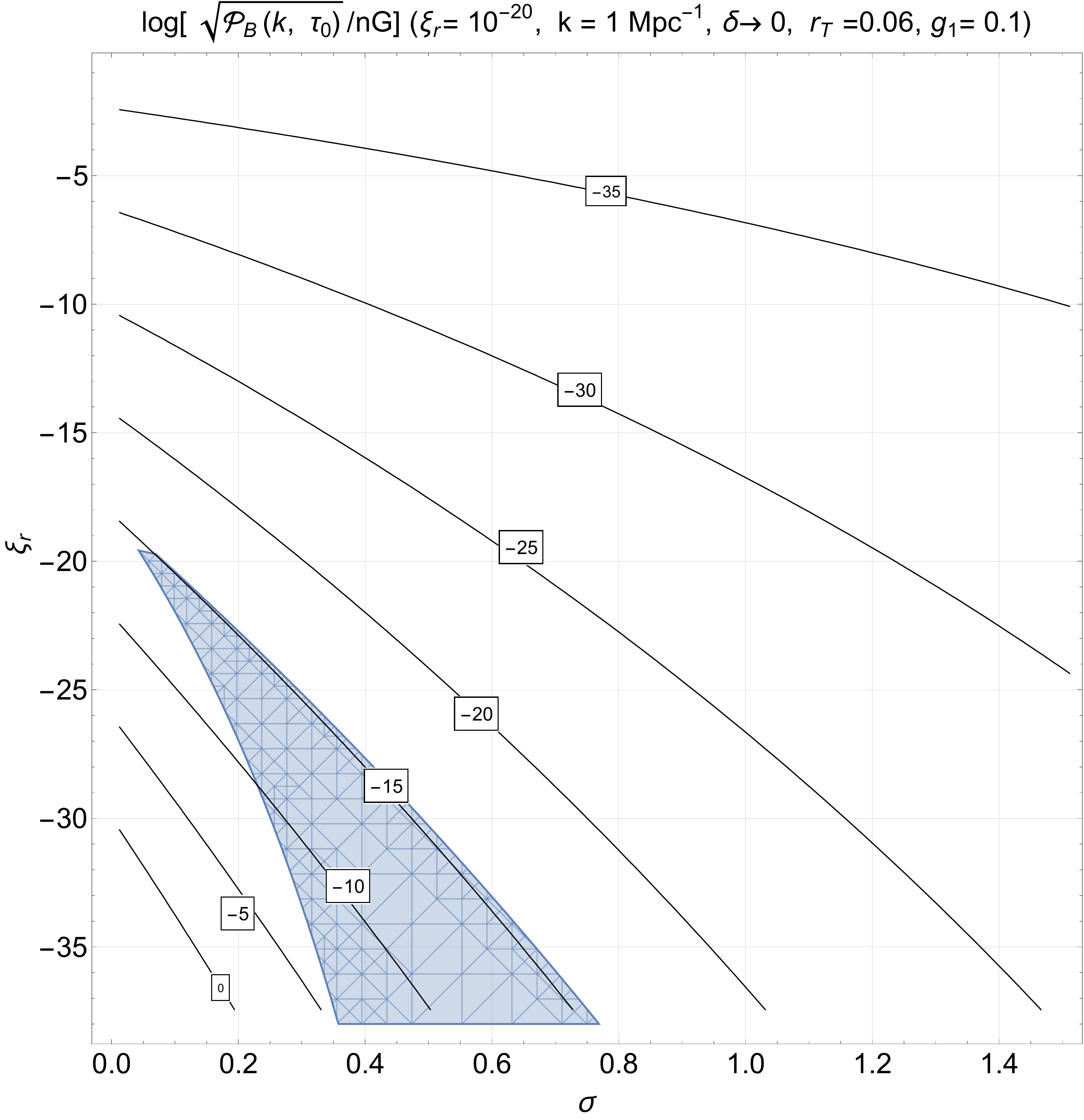}
\includegraphics[height=7.5cm]{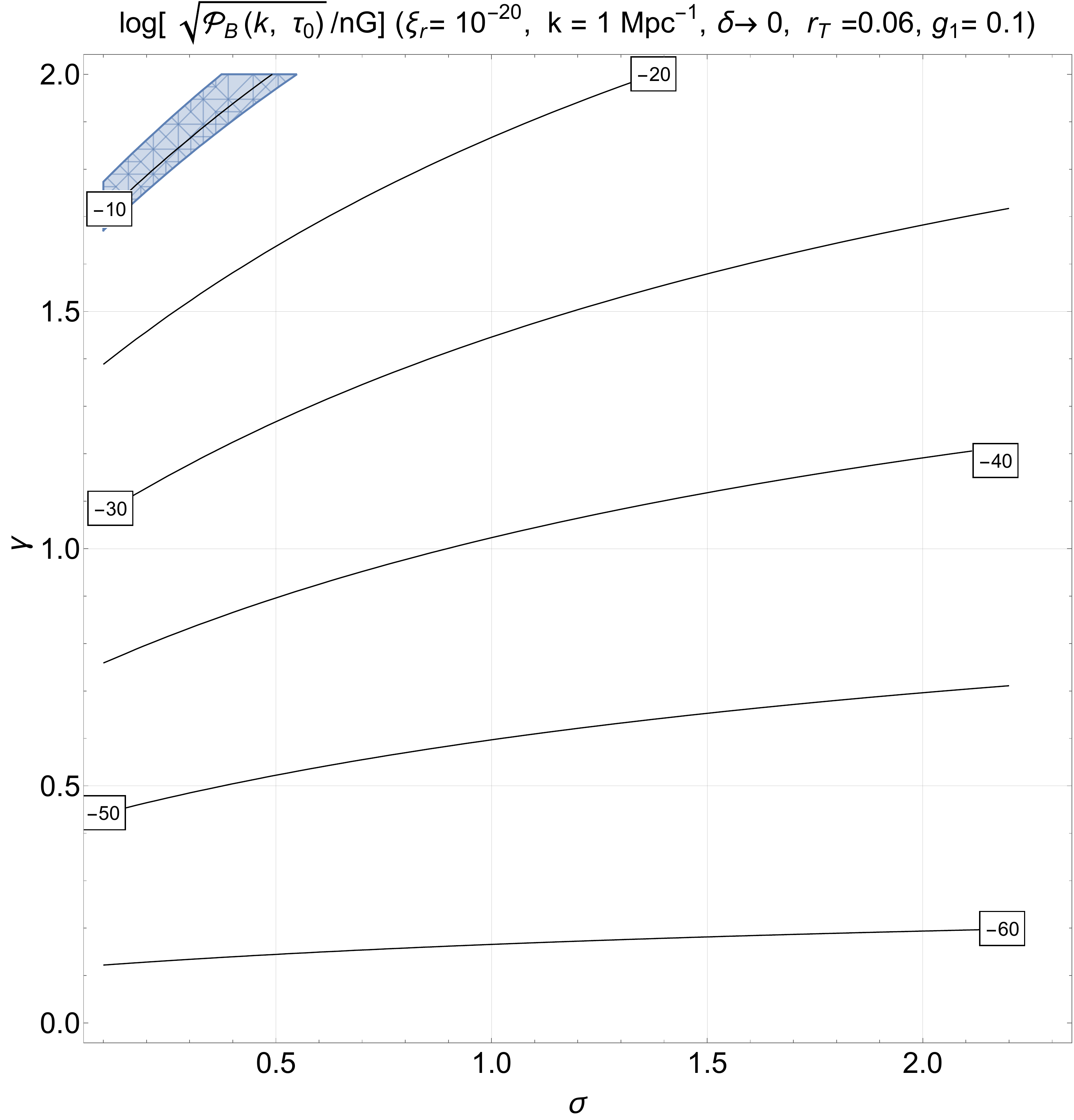}
\caption[a]{The same situation examined in Fig. \ref{FIGU9} is now considered 
when the power spectrum is evaluated according to the strategy described in 
Eqs. (\ref{NWR1})--(\ref{NWR2}) and (\ref{NWR3}). As anticipated before on the basis 
of general arguments the values of the power spectrum at the magnetogenesis scale 
are much smaller.}
\label{FIGU10}      
\end{figure}
This last point is illustrated in Figs. \ref{FIGU7} and \ref{FIGU8}. 
More specifically in  Fig. \ref{FIGU7} we consider a single phase 
expanding faster than radiation (i.e. $\sigma > 1$) while in Fig. \ref{FIGU8}
the intermediate rate is slower than radiation (i.e. $\sigma < 1$). 
In both figures the full line correspond to the case where the 
phases of the standing waves are evaluated for $\tau = {\mathcal O}(\tau_{k})$ 
as implied by the evolution of the mode functions analyzed in Eqs. (\ref{COND1})--(\ref{cond8}). The dashed lines correspond instead to the case 
where the phases have been arbitrarily fixed at the onset of the radiation-dominated 
stage. Even from a superficial comparison between Figs. \ref{FIGU7} and \ref{FIGU8} 
we can infer two complementary conclusions that confirm the ones already obtained 
before. First of all the phases evaluated at $\tau_{k}$ always lead to the larger power 
spectrum. Furthermore, when the expansion rate is slower 
than radiation the values of the physical power spectrum for fixed $\gamma$ and 
fixed $\xi_{r}$ are systematically larger than in the case when the intermediate 
expansion rate is faster than radiation. 
We can make the previous conclusions probably more transparent by charting parameter 
space whose {\em allowed region} is given by the shaded areas 
appearing, respectively, in Figs. \ref{FIGU9}, \ref{FIGU10} and \ref{FIGU11}.
More specifically in Figs. \ref{FIGU9} and \ref{FIGU10} we illustrate 
the case of a single post-inflationary stage of expansion; in the left plot of 
both figures the spectral slope is not flat (i.e. $\gamma \to 3/2$)
but nonetheless within the shaded areas the physical power 
spectra satisfy the magnetogenesis requirements 
and are compatible with all other bounds coming from large 
and short scales.
Always in Figs. \ref{FIGU9} and \ref{FIGU10} (but in the right plots) we fixed 
$ \xi_{r} \to 10^{-30}$ and charted the parameter space in the $(\gamma, \, \sigma)$ 
plane. The allowed region is much larger in  Fig. \ref{FIGU9} than in Fig. \ref{FIGU10} 
and this is just a consequence of the choice of the time at which the phases 
of the standing waves have been evaluated. The same kind of analysis 
has been finally illustrated in Fig. \ref{FIGU11} where 
instead of a single post-inflationary phase we included two phases parametrized, as before, by two different expansion rates $\sigma_{1}$ and $\sigma_{2}$.

\subsection{Typical time-scales and duality}
\label{subs53}
In subsection \ref{tauktaur} we considered the 
possibility of exchanging $\tau_{k}$ and $\tau= \tau_{r}$ and this arbitrary 
choice led to a deceiving estimate of the final gauge power spectrum. 
The obtained results suggest that it is quite dangerous to exchange 
(even by mistake) $\tau_{k}$ and $\tau_{r}$. Since the two time-scales correspond to different stages the
of the evolution, the confusion between $\tau_{k}$ and $\tau_{r}$ should not arise on a physical ground; however from a mere mathematical viewpoint there 
are situations where the actual difference between the 
gauge power spectra evaluated in $\tau_{k}$ or $\tau_{r}$ 
may effectively disappear.  These observations are now examined by referring to the timeline of Fig. \ref{FIGU1} where the post-inflationary evolution is dominated by radiation. For this purpose we first recall that, for an increasing gauge coupling during inflation, the comoving power spectra are
\begin{equation}
P_{B}(k, \, \tau) = a_{1}^4 \, H_{1}^4 \,\,p_{B}(k\,\tau_{1}, \gamma) \,\, \sin^2{k \tau}, \qquad \qquad 
P_{E}(k, \, \tau) = a_{1}^4 \, H_{1}^4 \,\, p_{E}(k\,\tau_{1}, \gamma) \,\, \cos^2{k \tau},
\label{dual1}
\end{equation}
where $p_{B}(k\,\tau_{1}, \gamma)$ and  $p_{E}(k\,\tau_{1}, \gamma)$ contain the spectral dependence and have been evaluated in the limit $\delta\to 0$ since 
we are assuming that after inflation the gauge coupling freezes. The physical power spectra associated with Eq. (\ref{dual1}) always satisfy the following hierarchy 
\begin{equation}
{\mathcal P}_{B}(k, \tau_{1}, \tau_{r}) \ll {\mathcal P}_{B}(k, \tau_{1}, \tau_{k}),
\label{dual2}
\end{equation}
that ultimately follows from the explicit analysis of the evolution of the mode 
functions. Equation (\ref{dual2}) holds since the comoving power spectrum at the onset of 
radiation dominance is ${\mathcal O}(k^2 \, \tau_{r}^2)$ {\em smaller} than the spectrum at the 
moment when the corresponding wavelength crosses the Hubble radius i.e. 
\begin{equation}
P_{B}(k, \tau_{r}) \simeq \biggl( \frac{k}{a_{r} \, H_{r}} \biggr)^2 P_{E}(k, \tau_{k}).
\label{dual3}
\end{equation}
Is the conclusion stemming from Eq. (\ref{dual3}) generic? Short answer to the 
question is: no since there exist situations where we can safely evaluate
the magnetic power spectra at $\tau_{k}$ or at $\tau_{r}$ 
without any appreciable difference. We can consider, for instance, the dual version of the comoving power spectra of Eq. (\ref{dual1}):
\begin{equation}
\overline{P}_{B}(k, \, \tau) = a_{1}^4 \, H_{1}^4 \,\,\overline{p}_{B}(k\,\tau_{1}, \overline{\gamma}) \,\, \cos^2{k \tau}, \qquad \qquad 
\overline{P}_{E}(k, \, \tau) = a_{1}^4 \, H_{1}^4 \,\, \overline{p}_{E}(k\,\tau_{1}, \overline{\gamma}) \,\, \sin^2{k \tau},
\label{dual4}
\end{equation}
By duality we have that $\overline{p}_{B}(k\,\tau_{1}, \overline{\gamma}) = p_{E}(k\,\tau_{1}, \gamma)$
and that $\overline{p}_{E}(k\,\tau_{1}, \overline{\gamma}) = p_{B}(k\,\tau_{1}, \gamma)$. Before the conductivity operates duality is still 
a good symmetry of the problem while for $\tau \geq \tau_{k}$ 
duality is broken by the presence of the Ohmic currents. 
Since now the dominant mode of oscillation 
of the power spectrum is not the sine but rather the cosine, in the case of Eq. (\ref{dual4}) the analog of Eq. (\ref{dual3}) is 
\begin{equation}
\overline{P}_{B}(k, \tau_{r}) \simeq  \overline{P}_{B}(k, \tau_{k}).
\label{dual5}
\end{equation}
Because of Eq. (\ref{dual5}) the values of the electromagnetic power spectra at the present time will be the same for the two choices. Of course, thanks to duality, we now have that 
\begin{equation}
\overline{P}_{E}(k, \tau_{r}) \simeq \biggl( \frac{k}{a_{r} \, H_{r}} \biggr)^2 \overline{P}_{B}(k, \tau_{k})
\end{equation}
but this is practically unimportant if the hyperelectric component is dissipated for $\tau > \tau_{k}$.

In summary $\tau_{k}$ and 
$\tau_{r}$ are two time-scales that are {\em physically different}
and must not be identified.  It happens, however, that 
for some special choices of the Sakharov phases the choice 
of $\tau_{k}$ or of $\tau_{r}$ is immaterial. The appearance 
of these phases has nothing to do with the initial conditions (that are always dictated by quantum mechanics) but rather with the dynamical 
evolution of the gauge coupling. If the gauge coupling 
increases during inflation and then flattens out then the magnetic power spectrum exhibits standing oscillations that predominantly go as a sine square: in this 
case if we evaluate the physical power spectra at $\tau_{r}$ 
(as opposed to $\tau_{k}$) the final result is underestimated 
by about $40$ orders of magnitude. In the dual situation (i.e. 
when the gauge coupling decreases during inflation)
the difference between $\tau_{k}$ and $\tau_{r}$ 
is immaterial from the mathematical viewpoint even if it 
is generally not advisable to identify the two scales that are 
physically diverse. 

It is interesting to remark that the dependence on the crossing time disappears from the spectral energy density defined from Eq. (\ref{EN1}) and evaluated 
after the end of inflation. This happens since since the phases of the Sakharov oscillations combine and eventually disappear as soon as the gauge coupling flattens out and, for this reason, the spectral energy density can be used to estimate jointly the hyperelectric and hypermagnetic contributions. For this purpose we recall that the spectral energy density in critical units is
\begin{equation}
\Omega_{Y}(k,\tau) = \frac{2}{3 \, H^2 \, M_{P}^2 \, a^4} \biggl[ P_{B}(k,\,\tau) + P_{E}(k,\,\tau) \biggr].
\label{expo}
\end{equation}
If we now insert Eqs. (\ref{POST7}) and (\ref{POST8}) in Eq. (\ref{expo}) we can obtain an even more explicit expression where, for the sake of generality, we kept the dependence on $\delta$:
\begin{equation}
\Omega_{Y}(k,\tau) = \frac{2\,\, H_{1}^4}{3 \, H^2 \, M_{P}^2}\,\, \biggl(\frac{a_{1}}{a}\biggr)^{4} \,\,
\biggl[ p_{B}(k \tau_{1},\, \gamma, \, \delta) f_{B}(k\,\tau, \delta) + p_{E}(k \tau_{1},\, \gamma, \, \delta) f_{E}(k\,\tau, \delta) \biggr].
\label{expo2}
\end{equation}
If $\delta \ll 1$ the oscillations of the power spectra interfere destructively 
and Eq. (\ref{expo2}) becomes 
\begin{equation}
\Omega_{Y}(k,\tau) = \frac{2\,\, H_{1}^4}{3 \, H^2 \, M_{P}^2}\,\, \biggl(\frac{a_{1}}{a}\biggr)^{4} \,\,
{\mathcal C}(\gamma +1/2) \biggl(\frac{k}{a_{1}\, H_{1}}\biggr)^{ 4 - 2 \gamma - 2 \delta}.
\label{OMY1}
\end{equation}
For $N$ intermediate phases the effect of the intermediate stages modifies 
the overall amplitude of the spectral energy density:
\begin{equation}
\frac{ H_{1}^4}{ \, H^2 \, M_{P}^2}\,\, \biggl(\frac{a_{1}}{a}\biggr)^{4} = \biggl(\frac{H_{1}}{M_{P}}\biggr)^2 \prod_{i = 1}^{N-1} 
\biggl(\frac{a_{i}}{a_{i + 1}}\biggr)^{2(1 - 1/\sigma_{i})}.
\label{OMY2}
\end{equation}
Note that the right-hand side of Eq. (\ref{OMY2}) goes correctly to a constant for $\sigma_{i}\to 1$ 
(i.e. if all the phases collapse to radiation). However if some of the $\sigma_{i} \neq 1$ the overall scaling of $\Omega_{Y}(k,\tau)$ is indeed different from the usual situation, as previously remarked in Ref. \cite{CC0}.
Within these notations the post-inflationary stage prior to radiation is in fact partitioned in a series of $N$ stages characterized by an expansion rate $\sigma_{i}$ in conformal time. By definition
the expansion rate at the end of the last stage must coincide with expansion rate of the radiation 
stage, i.e. $H_{N} = H_{r}$. With the same notations we then have 
\begin{equation}
\frac{k}{a_{1} \, H_{1}} = \biggl(\frac{8}{\pi}\biggr)^{1/4} \, \, \frac{k}{\sqrt{H_{0} \, M_{P}} \, (r_{T} \,\, \Omega_{R\, 0}\,\, {\mathcal A}_{{\mathcal R}})^{1/4}} \,\, \prod_{j= 1}^{N-1} \,\xi_{j}^{\frac{1 - \sigma_{j}}{2 ( \sigma_{j} +1 )}},
\end{equation}
where, by definition, $\xi_{j} = H_{j}/H_{j+1} < 1$ and, as usual, $ a\, H = {\mathcal H}$. All in all the 
spectral energy density for $\tau \leq \tau_{k}$ is given by:
\begin{equation}
\Omega_{Y}(k,\tau) = \frac{2}{3} \,\biggl(\frac{H_{1}}{M_{P}} \biggr)^2\, \, {\mathcal C}(\gamma +1/2)\,\,
\biggl[ \biggl(\frac{8}{\pi}\biggr)^{1/4} \, \frac{k}{\sqrt{ H_{0} \, M_{P}} \, \, (r_{T} \, \Omega_{R0}\, {\mathcal A}_{{\mathcal R}})^{1/4} }
\biggr]^{4 - 2 \gamma}.
\end{equation}

 There are some who recently considered 
magnetic fields with vector-like coupling to fermions \cite{SCH1a,SCH1b,SCH1c,SCH2} 
and argued that various magnetogenesis scenario are constrained by  
 the Schwinger effect in de Sitter space-time. This viewpoint is not entirely 
 consistent for several reasons\footnote{These estimates are typically
done by assuming a constant and uniform electric field in de Sitter space and by also postulating a vector-like coupling to fermions. These assumptions do not specifically hold in the present case since the fields obtained here are non-uniform and non-homogeneous, 
they are not constant in time and they have, in general, a chiral coupling to fermions.  The constancy and uniformity of the field configurations entering the estimate 
of the Schwinger effect is achieved by considering a class of  currents and this physical situation is markedly different  from the flat space-time case where the constancy and uniformity of the field imposes the absence of time-dependent currents \cite{SCH2}. } explained in Ref. \cite{SCH2}. In this paper we specifically considered the case of hypermagnetic fields whose coupling to fermions is chiral so that those considerations, strictly speaking, do not apply. 
Still one may ask the question of what would be the correct form of the pair production rate assuming that the fields considered here has just a vector coupling to fermions. The first observation is that for $\gamma \leq 2$ the critical density constraint is always satisfied since both the hyperelectric and the hypermagnetic power spectra are extremely minute; the relation among them can be easily deduced from the phases of the Sakharov oscillations and it is in fact a 
generalization of Eq. (\ref{dual3})
\begin{eqnarray}
P_{B}(k,\tau) &=& \biggl(\frac{k}{a H}\biggr)^2 P_{E}(k, \tau), 
\label{SCC0}
\end{eqnarray}
The same relation follows for all the relevant values of $\gamma\leq 2$ during the inflationary stage of expansion
so that $P_{B}(k,\tau)$ is always smaller than $P_{E}(k,\tau)$. We now estimate under which conditions the probability of pair creation  per unit volume per unit time is small in an averaged sense.
Denoting with $\Gamma$ the rate of pair production 
 per unit volume and per unit time a reasonable condition to consider seems to be:
\begin{equation}
\frac{\Gamma}{H^4} = \frac{\alpha_{em}}{\pi^2} \langle E^2(\tau,\vec{x}) \rangle < 1,\qquad \Rightarrow \qquad 
\frac{2}{\pi^2 \, \lambda} \int_{0}^{k_{max}}\frac{d\, k}{k}\, P_{E}(k,\tau) < H^4,
\label{SCC2}
\end{equation}
where  $\alpha_{em} = e^2/(4\pi)$ and $k_{max}$ denotes the maximally amplified wavenumber. It can be shown that this condition is always satisfied \cite{SCH3}. For all these reasons it does not seem that the probability 
of pair creation represents a relevant constraint when the gauge coupling 
increases during the inflationary stage.

\newpage
\subsection{Sakharov phases and pseudoscalar couplings}
\label{subs54}
The gauge analog of the Sakharov phases have been analyzed in the absence of parity-breaking terms whose implications are now briefly analyzed.
The main point we ought to convey is that, for the purposes of the present discussion, the conclusions remain practically unchanged if $\overline{\lambda} \neq 0$ and $\overline{\chi} \neq 0$ in Eqs. (\ref{RMODE1})--(\ref{LMODE2}). 
In general terms the presence of $\overline{\lambda}$ 
affects very little the hypermagnetic and hyperelectric fields 
but it is instead crucial for the analysis of the gyrotropic spectra; this 
means that the pseudoscalar vertex is practically irrelevant for the magnetogenesis requirements \cite{CC13}.
Indeed Eqs. (\ref{RMODE1})--(\ref{LMODE2}) are mildly affected by the 
pseudoscalar terms and, in particular, it has been shown\footnote{ As already stressed in section \ref{sec2} we are here preferentially considering the  generic models of inflation where, in practice, there is no asymmetry 
between hyperelectric and hypermagnetic susceptibilities; this means that $\chi_{E} \simeq \chi_{B}$ 
so that $n \to 1$ and $\eta\to \tau$. This analysis would be slightly different for non-generic 
models like the ones based on a derivative coupling of the inflaton as in the case of the relativistic theory of Van der Waals (or Casimir-Polder) interactions \cite{fiftythree,fiftyfour,fiftyfive}.}
that the whole effect depends on the time-evolution 
of the ratio $\overline{\lambda}^{\prime}/\lambda$ whose 
time evolution can be parametrized as $\tau^{- \beta-1}$ \cite{CC13}. 
If $\beta > 1$ the same hierarchies  
of the case $\overline{\lambda} \to 0$ are directly translated 
in the pseudoscalar case. If $ 0<\beta < 1$ the slopes of the hypermagnetic and hyperelectric power spectra are still not affected but the overall amplitude gets modified depending on the value of $\beta$. The dependence of the amplitude on $\overline{\lambda}$ is maximal in the case $\beta\to 0$ where $\overline{\lambda} = \lambda_{0} \lambda$.  Following the results of Ref. \cite{CC13} we have that the phases of oscillations of $P_{B}(k,\tau)$ and $P_{E}(k,\tau)$ coincide exactly with the ones already mentioned above. More specifically, if we assume the timeline of Fig. \ref{FIGU1} the hyperelectric and hypermagnetic power spectra can be directly computed by solving Eqs. (\ref{RMODE1})--(\ref{LMODE2}) in the case of increasing gauge coupling and by subsequently computing the gauge power spectra. This 
specific analysis is partially reported in \cite{CC13} and in what follows 
we just focus on the relevant limit where $\delta\to 0$ and the gauge coupling increases. 
When the gauge modes evolve after the end of inflation but before $\tau_{k}$ the analog of Eq. (\ref{DUAL1aaa}) are obtained from the 
same general expression of the power spectra given in Eq. (\ref{TWENTY3}):
\begin{eqnarray}
P_{B}(k,\tau) &=& a_{1}^4 \, H_{1}^4 \,\,p_{B}(k\, \tau_{1}, \gamma) \,\,{\mathcal D}(|\zeta|, \gamma) \,\,\sin^2{k\tau},
\nonumber\\
P_{E}(k,\tau) &=& a_{1}^4 \, H_{1}^4 \,\,p_{E}(k\, \tau_{1}, \gamma) \,\,{\mathcal D}(|\zeta|, \gamma) \,\,\cos^2{k\tau},
\nonumber\\
{\mathcal D}(|\zeta|,\gamma) &=& \frac{\Gamma^2(\gamma)}{|\Gamma( \gamma - i |\zeta|)|^2} \cosh{\pi |\zeta|},
\label{CS1}
\end{eqnarray}
where $|\zeta| = 2 \lambda_{0} \gamma$. It is relevant to mention that the same phases 
also arise in the gyrotropic contributions following from Eq. (\ref{TWENTY4}):
\begin{eqnarray}
P_{B}^{(G)}(k,\tau) &=& a_{1}^4 \, H_{1}^4 \,\,p_{B}(k\, \tau_{1}, \gamma) \,\,{\mathcal D}^{(G)}(|\zeta|, \gamma) \,\,\sin^2{k\tau},
\nonumber\\
P_{E}^{(G)}(k,\tau) &=& a_{1}^4 \, H_{1}^4 \,\,p_{E}(k\, \tau_{1}, \gamma) \,\,{\mathcal D}^{(G)}(|\zeta|, \gamma) \,\,\cos^2{k\tau},
\nonumber\\
{\mathcal D}^{(G)}(|\zeta|,\gamma) &=& \frac{\Gamma^2(\gamma)}{|\Gamma( \gamma - i |\zeta|)|^2} \sinh{\pi |\zeta|}.
\label{CS2}
\end{eqnarray}
Equations (\ref{CS1})--(\ref{CS2}) have various interesting implications in the context 
of baryogenesis \cite{SCH4}. For the problem at hand, however,
nothing changes if the pseudoscalar interactions are turned on. It may superficially seem that ${\mathcal D}(|\zeta|,\gamma)$ introduces a further (exponential) amplification of the hypermagnetic field. This is however not true and an accurate analysis shows that whenever the amplification induced by this factor is phenomenologically relevant the critical density bound is violated and this also happens in the case of decreasing gauge coupling \cite{SCH4}. We finally remark that also the result of Eq. (\ref{SCC0}) holds in the case of the power spectra given in Eq. (\ref{CS1}). This means that similar considerations may be developed in the case of pseudo-scalar coupling.

\newpage
\renewcommand{\theequation}{6.\arabic{equation}}
\setcounter{equation}{0}
\section{Concluding remarks}
\label{sec6}
If the inflationary stage is sufficiently long to dissipate any preexisting classical 
inhomogeneity, the fields of various spin are copiously produced from their quantum mechanical fluctuations provided Weyl invariance is broken and since the evolution of the gauge couplings is not directly constrained before radiation dominance it is natural to consider its dynamical implications.  For the sake of concreteness we focused on the description of generic single field inflationary scenarios where the leading correction to the tree-level effective action consists of all possible terms containing four space-time derivatives. While during a quasi-de Sitter stage the corrections always lead to different hypermagnetic and the hyperelectric susceptibilities, in the generic case (which is the one suggested by the current large-scale observations) the potential asymmetry of the gauge couplings can be neglected. 

If the quantum mechanical initial conditions are imposed the gauge mode functions initially oscillate as travelling waves and then turn asymptotically into standing waves that are the analog of the Sakharov oscillations. The related phases are determined by the rate of variation of the gauge coupling that may either increase or decrease during inflation. As a result the Sakharov oscillations appear directly in the hypermagnetic and hyperelectric power spectra that generically arise as Bessel functions whose index is fixed by the rate of variation of the gauge coupling after inflation. These patterns of oscillation are exchanged by duality and unlike the case of the density contrasts in a relativistic plasma (where this phenomenon has been primarily identified), the gauge analog of Sakharov phases do not develop and  they are  overdamped by the finite value of the conductivity as soon as the modes become of the order of the comoving Hubble radius. The late-time value of the magnetic field is therefore not simply determined by radiation dominance (and in spite of the value of the wavenumber) but  it depends on the moment when the given wavelength (comparable with the Mpc) get feel the effect of the conductivity before equality. More specifically there are two physically different time-scales $\tau_{k}$ and $\tau_{r}$. If we evaluate the physical power spectra at $\tau_{r}$  (as opposed to $\tau_{k}$) the final result is underestimated by about $40$ orders of magnitude. The turning point in the evolution of the mode functions corresponds in fact to $\tau_{k}$ and since for $\tau > \tau_{k}$ duality is broken, in this regime the electric fields are suppressed while the magnetic fields persist over typical scales larger than the magnetic diffusivity scale.

Under some specific conditions the Sakharov phases lead to a modified scaling of the hypermagnetic power spectrum that might superficially appear as a substantial advantage for the amplification of large-scale magnetic fields. This  bogus effect  is always weighted by terms that are negligibly small 
so that, overall, the net result is a contribution of order $1$.  If the Sakharov 
phases are improperly treated we might even come to the conclusion that {\em any} post-inflationary evolution (including radiation) may reduce the suppression 
of the magnetic power spectra. We instead demonstrated that the magnetogenesis requirements are only relaxed if the post-inflationary evolution expands at a rate that is slower than radiation but the opposite is true when the post-inflationary expansion rate is faster than radiation. In this second instance the power spectra are even more suppressed than during a radiation stage. After combining the present findings with the evolution of the gauge coupling it turns out that these results are consistent with a magnetogenesis scenario where the gauge coupling is always perturbative throughout all the stages of its evolution while, in the dual case, the same requirements cannot be satisfied. In this respect the present results clarified that 
the power spectra during inflation do not coincide with the one evaluated 
after inflation and, in particular, at the crossing time $\tau_{k}$. The electromagnetic 
power spectra must therefore be computed from the correct 
post-inflationary mode functions that are related to their inflationary 
counterpart via an appropriate mixing matrix whose 
properties are essential for the correctness of the late-time results.

\section*{Acknowledgements}
It is a pleasure to thank T. Basaglia, A. Gentil-Beccot, 
S. Rohr and J. Vigen of the CERN Scientific Information Service for their kind assistance. 

\newpage

\begin{appendix}

\renewcommand{\theequation}{A.\arabic{equation}}
\setcounter{equation}{0}
\section{The limits the power spectra}
\label{APPA}
In this paper we assumed that the gauge coupling  evolves during inflation 
and then freezes later on. Since the continuity of the description is essential 
to infer the correct limit of the Sakharov phases, after inflation 
the evolution of $g$ is pametrized by $\delta$ (in the case 
of increasing gauge coupling) or by $\overline{\delta}$ in the case 
of decreasing gauge coupling (see, for instance, Eqs. (\ref{POST1}) and (\ref{POST9})). In the bulk of the paper we determined the simplified 
form of the Sakharov phases by first taking the limit $x_{1} \ll 1$ and by 
then requiring $\delta \ll 1$ (see, in this respect, Eqs. (\ref{POST7})--(\ref{POST8a0})). Inverting the order of the limits the power spectra should not change
and this is the aspect specifically analyzed here. For fixed $x_{1}$ we can then take the limit  $\delta \to 0$ and the result for the elements of the transition matrix of Eqs. (\ref{DIAG1})--(\ref{DIAG2}) is:
\begin{eqnarray}
&&\lim_{\delta \to 0} {\mathcal T}_{f\, f}(k, \tau_{1}, \tau) = \cos{(x+ x_{1})}, \qquad\qquad
 \lim_{\delta \to 0} {\mathcal T}_{f\, g}(k, \tau_{1}, \tau) = \sin{(x+ x_{1})},
\nonumber\\
&&\lim_{\delta \to 0} {\mathcal T}_{g\, f}(k, \tau_{1}, \tau) = -\sin{(x+ x_{1})}, \qquad\qquad
 \lim_{\delta \to 0} {\mathcal T}_{g\, g}(k, \tau_{1}, \tau) = \cos{(x+ x_{1})},
\label{POST8a}
\end{eqnarray}
where, by definition, $x_{1} = k \tau_{1}$ and $x = k \tau$. Equation (\ref{POST8a})
is consistent with the general form of Eqs. (\ref{DIAG1})--(\ref{DIAG2}) for $\tau = - \tau_{1}$;
as it can be explicitly seen ${\mathcal T}_{f\, f}(k, \tau_{1}, -\tau_{1}) =  {\mathcal T}_{g\, g}(k, \tau_{1}, -\tau_{1}) =1$
and ${\mathcal T}_{f\, g}(k, \tau_{1}, - \tau_{1}) = {\mathcal T}_{g\, f}(k, \tau_{1}, -\tau_{1}) =0$.
Thanks to Eq. (\ref{POST8a}) the comoving power spectra of Eqs. (\ref{TWENTY1})--(\ref{TWENTY2}) and (\ref{TWENTY3}) can then be written as: 
\begin{eqnarray}
P_{E}(k,\tau) &=& \frac{k^3}{2\pi^2} \biggl| - k\, \sin{(x+ x_{1})} \, f_{k}(- \tau_{1}) + \cos{(x+ x_{1})}\,g_{k}(-\tau_{1})\biggr|^2,
\label{POSTa1}\\
P_{B}(k,\tau) &=& \frac{k^5}{2\pi^2} \biggl| \cos{(x+ x_{1})} f_{k}(- \tau_{1}) + \sin{(x+ x_{1})} \frac{g_{k}(-\tau_{1})}{k}\biggr|^2,
\label{POSTa2}
\end{eqnarray}
where $f_{k}(- \tau_{1})$ and $g_{k}(-\tau_{1})$ are the inflationary mode functions 
evaluated for $\tau \to - \tau_{1}$ and they depend on the values of $\gamma$:
\begin{eqnarray}
f_{k}(-\tau_{1}) &=& \frac{N_{\mu}}{\sqrt{2 k}} \sqrt{x_{1}} \, H_{\mu}^{(1)}(x_{1}), \qquad \qquad \mu = | \gamma -1/2|, 
\qquad \qquad \bigl| N_{\mu} \bigr| = \sqrt{\frac{\pi}{2}},
\label{POSTa3}\\
g_{k}(-\tau_{1}) &=& N_{\mu} \,\sqrt{\frac{k}{2}}\,\sqrt{x_{1}} \, H_{\mu+1}^{(1)}(x_{1}),\qquad\qquad \gamma > \frac{1}{2},
\label{POSTa4}\\
g_{k}(-\tau_{1})  &=& - N_{\mu} \,\sqrt{\frac{k}{2}} \,  \sqrt{x_{1}} \, H_{\mu-1}^{(1)}(x_{1}),\qquad\qquad 0< \gamma < \frac{1}{2}.
\label{POSTa5}
\end{eqnarray}
Note that $H_{\alpha}^{(1)}(z) = J_{\alpha}(z) + i \, Y_{\alpha}(z)$ are the Hankel functions of the first kind 
with argument $z$ and index $\alpha$ \cite{abr1,abr2}. For $\gamma >1/2$ Eqs. (\ref{POSTa1})--(\ref{POSTa2}) imply 
the following two explicit expressions of the comoving power spectra that we may regard directly 
as a function of $x$ and $x_{1}$:
\begin{eqnarray}
P_{E}(x, x_{1}) &=& \frac{a_{1}^4\, H_{1}^4}{8 \pi} \, x_{1}^{5} \,\bigl| H_{\mu+1}^{(1)}(x_{1})\bigr|^2 \,\,
\biggl| \cos{(x+ x_{1})} - \sin{(x + x_{1})} \frac{H_{\mu}^{(1)}(x_{1})}{H_{\mu+1}^{(1)}(x_{1})}\biggr|^2,
\label{POSTa6}\\
P_{B}(x, x_{1}) &=& \frac{a_{1}^4\, H_{1}^4}{8 \pi} \, x_{1}^{5} \,\bigl| H_{\mu}^{(1)}(x_{1})\bigr|^2 \,\,
\biggl| \cos{(x+ x_{1})} - \sin{(x + x_{1})} \frac{H_{\mu+1}^{(1)}(x_{1})}{H_{\mu}^{(1)}(x_{1})}\biggr|^2.
\label{POSTa7}
\end{eqnarray}
As expected, in spite of the value of $x$, the term containing the sine dominates in $P_{B}(x, x_{1})$
while the term containing the cosine dominates in $P_{E}(x, x_{1})$; the reason is that the limit 
$x_{1}$ is always verified for all the modes of the spectrum and, in this limit,
\begin{equation}
\frac{H_{\mu+1}^{(1)}(x_{1})}{H_{\mu}^{(1)}(x_{1})} = \frac{2 \, \Gamma(\gamma + 1/2)}{\Gamma(\gamma -1/2) \, x_{1}}\,\gg\, 1.
\label{POSTa8}
\end{equation}
The ratio of Eq. (\ref{POSTa8}) dominates inside the square modulus of Eq. (\ref{POSTa7}) while 
its inverse is subleading inside the square modulus of Eq. (\ref{POSTa6}). It therefore follows
that:
\begin{eqnarray}
P_{E}(k, \tau) &=&  \frac{a_{1}^4\, H_{1}^4}{8 \pi} \, x_{1}^{5} \, H_{\mu+1}^{(1)}(x_{1}) \cos^2{(x+ x_{1})} = 
a_{1}^4 \, H_{1}^4 {\mathcal C}(\gamma+1/2) \biggl(\frac{k}{a_{1} \, H_{1}}\biggr)^{4 - 2 \gamma} \cos^2{k\tau},
\label{POSTa9}\\
P_{B}(k, \tau) &=&  \frac{a_{1}^4\, H_{1}^4}{8 \pi} \, x_{1}^{5} \, H_{\mu+1}^{(1)}(x_{1}) \sin^2{(x+ x_{1})} = 
a_{1}^4 \, H_{1}^4 {\mathcal C}(\gamma+1/2) \biggl(\frac{k}{a_{1} \, H_{1}}\biggr)^{4 - 2 \gamma} \sin^2{k\tau},
\label{POSTa10}
\end{eqnarray}
where we simply took into account that, in spite of the value of $\tau$, $ \tau > |\tau_{1}|$.
The results are valid, strictly speaking, for $\gamma > 1/2$ however exactly 
the same conclusion follows when $0 < \gamma < 1/2$ since, in this case, we should just 
replace $H_{\mu +1}^{(1)}(x_{1})$ by $H_{\mu - 1}^{(1)}(x_{1})$ and, as a result,
the relevant ratio corresponding to Eq. (\ref{POSTa8}) is now given by:
\begin{equation}
\frac{H_{\mu-1}^{(1)}(x_{1})}{H_{\mu}^{(1)}(x_{1})} = \frac{\Gamma(1/2 + \gamma)}{\Gamma(1/2-\gamma) }\, \biggl(\frac{2}{x_{1}}\biggr)^{2 \gamma}\,\gg\, 1.
\label{POSTa11}
\end{equation}
Equations (\ref{POSTa9})--(\ref{POSTa10}) do not assume a 
specific range of $x = k\tau$ except for the requirement $ \tau > |\tau_{1}|$ and
if we consider the expression of Eq. (\ref{POSTa7}) in the two concurrent limits:
\begin{equation}
x_{1} \ll 1, \qquad \qquad x \ll 1, \qquad \qquad \frac{x}{x_{1}} = \frac{\tau}{\tau_{1}} \gg 1,
\label{POSTa12}
\end{equation}
the final outcome is simply given by:
\begin{equation}
P_{B}(x, x_{1}) = \frac{a_{1}^4 H_{1}^4}{8 \pi} 2^{2 \mu} \Gamma^2(\mu) x_{1}^{5 - 2 \mu} \biggl[ 1 + 2 \mu \biggl(\frac{x}{x_{1}}\biggr)\biggr] = a_{1}^4 H_{1}^4 {\mathcal C}(\gamma+1/2) \biggl(\frac{k}{a_{1} H_{1}}\biggr)^{ 4 - 2 \gamma}
\,\, | k \tau|^2,
\label{POSTa13}
\end{equation} 
where ${\mathcal C}(x) = 2^{2 x -3} \, \Gamma^2(x)/\pi^3$  has been 
already introduced in the bulk of the paper (see Eq. (\ref{PB1}) and discussion 
thereafter). The last expression of Eq. (\ref{POSTa13}) coincides with Eq. (\ref{POSTa10}) evaluated for $ k \tau <1$ where $\sin{k \tau} = k\tau + {\mathcal O}(k^2 \tau^2)$. There are different ways of writing Eq. (\ref{POSTa13}) like, for instance,
\begin{equation}
P_{B}(k,\tau) = a_{1}^4 H_{1}^4 {\mathcal C}(\gamma+1/2) \biggl(\frac{k}{a_{1} H_{1}}\biggr)^{ 6 - 2 \gamma} \biggl(\frac{\tau}{\tau_{1}}\biggr)^2.
\label{POSTa14}
\end{equation}
The ratio $\tau/\tau_{1}$ can be expressed in various different ways; for instance we could write that
\begin{equation}
\frac{\tau}{\tau_{1}} =  \sigma \biggl(\frac{{\mathcal H}_{1}}{{\mathcal H} }\biggr)= \sigma \biggl(\frac{a_{1} \, H_{1}}{a\, H}\biggr),
\label{POSTa15}
\end{equation}
where $\sigma$ is a numerical factor that depends on the post-inflationary evolution and the 
same parametrization has been used in section \ref{sec4} (see, in particular, Figs. \ref{FIGU2} and \ref{FIGU3}). If the post-inflationary phase is dominated by radiation (as in Fig. \ref{FIGU1}), $\sigma\to 1$.
For a generic barotropic fluid we have instead that $\sigma \to 2/(3 w+1)$, and more details 
on the logic behind the post-inflationary evolution can be found in section \ref{sec5} (see also, along a more technical persepctive, in the discussion of appendix \ref{APPB}).
Again Eq. (\ref{POSTa13}) is only valid in the limit $ k \tau \ll 1$ and it is comparatively 
less general than the ones of Eqs. (\ref{POSTa9})--(\ref{POSTa10}). As a result Eq. (\ref{POSTa13}) 
could be also expressed as
\begin{equation}
P_{B}(k, \tau) = a_{1}^4 \, H_{1}^4\, {\mathcal C}(\gamma -1/2)  x_{1}^{6 - 2 \gamma} \biggl|1 + \sigma (2 \gamma -1)
\biggl(\frac{a_{1} \, H_{1}}{a\, H} \biggr) \biggl|^2.
\label{POSTa16}
\end{equation}
Equation (\ref{POSTa16}) is actually misleading since the $1$ inside the square modulus is 
always irrelevant in comparison with the second term in the range of applicability 
of the expression (i.e. for wavelengths larger than the Hubble radius). All in all the results 
of Eqs. (\ref{POSTa9})--(\ref{POSTa10}) are more general and also more transparent from the viewpoint of the duality symmetry that must be enforced at every step of the calculation. We note, incidentally, that Eq. (\ref{POSTa15}) is not completely accurate but it is well justified for $\tau > -\tau_{1}$. In general we actually have that if ${\mathcal H}_{1}$ is evaluated at the end of inflation and ${\mathcal H}$ is computed {\em after} the end of inflation the ratio ${\mathcal H}_{1}/{\mathcal H}$
\begin{equation}
\frac{{\mathcal H}_{1}}{{\mathcal H}} = \frac{1}{\sigma (1 -\epsilon)} \biggl(\frac{\tau}{\tau_{1}} +1\biggr) + 1,
\label{POSTa17}
\end{equation}
where, as usual, $\epsilon$ denotes the slow-roll parameter.
For $\tau = - \tau_{1}$ we have that ${\mathcal H} \to {\mathcal H}_{1}$ and the right-hand side of Eq. (\ref{POSTa17}) correctly gives $1$. For $\tau \gg - \tau_{1}$ the result of Eq. (\ref{POSTa15}) is recovered. This factor is immaterial and has been omitted from the explicit equations is included in the numerical estimates.
\newpage
\renewcommand{\theequation}{B.\arabic{equation}}
\setcounter{equation}{0}
\section{Multiple phases}
\label{APPB}
If the post-inflationary expansion history consists of $N$ successive stages expanding at different rates the final value of the physical power spectra are not necessarily enhanced, as claimed by some. In this appendix we list some of the relevant equations where $a_{1}$ denotes the scale factor at the end of inflation and $a_{N} = a_{r}$ corresponds instead to the onset of the radiation-dominated stage of expansion. In the conformal time coordinate the expansion rate is characterized by $a_{i}(\tau) \sim \tau^{\sigma_{i}}$, it is practical to measure the duration of each phase from the ratios of the curvature scales during two successive 
stages of expansion denoted here by $\xi_{i}$:
\begin{equation}
\xi_{i} = \frac{H_{i+1}}{H_{i}} < 1, \qquad \xi_{r} = \prod_{i}^{N-1} \xi_{i} =  \frac{H_{r}}{H_{1}} <1.
\label{APPB1}
\end{equation}
While in Eq. (\ref{APPB1}) $\xi_{i}$ gives instead the ratio of the expansion rates between two successive stage, $\xi_{r}$ ultimately denotes the ratio between the Hubble rates at the onset of the radiation-dominated stage (i.e. $H_{r}$) and at the end of the inflationary phase (i.e. $H_{1}$). By definition, both $\xi_{i}$ and $\xi$ are smaller than $1$ since 
$H_{i+1} < \, H_{i}$ and $H_{r} < H_{1}$. Thanks to Eq. (\ref{APPB1}) 
we can show that Eq. $k/(a_{1} \, H_{1})$ depends upon $\xi_{i}$ and $\sigma_{i}$
in the following peculiar manner:
\begin{equation}
\frac{k}{a_{1}\, H_{1}} = 10^{-23.24} \prod_{i =1}^{N-1} \, \, \xi_{i}^{\frac{\sigma_{i} -1}{2(\sigma_{i} +1)}}\,  \biggl(\frac{k}{\mathrm{Mpc}^{-1}}\biggr)\,\, \biggl(\frac{r_{T}}{0.06}\biggr)^{-1/4} \,  
\biggl(\frac{{\mathcal A}_{{\mathcal R}}}{2.41\times 10^{-9}}\biggr)^{-1/4} \, \biggl(\frac{h_{0}^2 \, \Omega_{R\,0}}{4.15 \times 10^{-5}}\biggr)^{-1/4}.
\label{APPB2}
\end{equation}
Unlike the result of Eq. (\ref{APPB2})  $k/(a_{r} \,H_{r})$ depends on the various $\xi_{i}$ 
but not on $\sigma_{i}$:
 \begin{equation}
 \frac{k}{a_{r}\, H_{r}} = \frac{k}{a_{N}\, H_{N}} = \prod_{j = 1 }^{N-1} \, \sqrt{\xi_{j}} \,\,\, 
 \biggl(\frac{k}{\mathrm{Mpc}^{-1}}\biggr)\,\, \biggl(\frac{r_{T}}{0.06}\biggr)^{-1/4} \,  
\biggl(\frac{{\mathcal A}_{{\mathcal R}}}{2.41\times 10^{-9}}\biggr)^{-1/4} \, \biggl(\frac{h_{0}^2 \, \Omega_{R\,0}}{4.15 \times 10^{-5}}\biggr)^{-1/4}.
  \label{APPB3}
 \end{equation}
We stress that the second equality appearing in Eq. (\ref{APPB3}) follows since, by definition, the product of 
all the different $\xi_{i}$ coincides in fact with $\xi_{r}$:
\begin{equation}
\prod_{j = 1 }^{N-1} \, \xi_{j}= \xi_{1}\,\xi_{2} \,.\,.\,. \,\xi_{N-2}\, \xi_{N-1} = \xi_{r}.
\label{APPB4}
\end{equation}
Equations (\ref{APPB3})--(\ref{APPB4}) demonstrate, 
as anticipated above, that while $k/(a_{1}\, H_{1})$ have  is sensitive to the 
whole expansion history,  $k/(a_{r} \, H_{r})$ only depends upon $\sqrt{\xi_{r}}$ (where  $\xi_{r}= H_{r}/H_{1}$). For all the other intermediate wavenumbers between 
$k/(a_{1}\, H_{1})$ and $k/(a_{r} \, H_{r})$ the dependence upon $\xi_{i}$ 
and $\sigma_{i}$ interpolates between the ones of Eqs. (\ref{APPB2}) and (\ref{APPB3}):
\begin{eqnarray}
\frac{k}{a_{m} H_{m}} &=& \prod_{j=1}^{m-1} \sqrt{\xi_{j}}\,\, \prod_{i =m}^{N- 1} \, \, \xi_{i}^{\frac{\delta_{i} -1}{2(\delta_{i} +1)}}\, \biggl(\frac{k}{\mathrm{Mpc}^{-1}}\biggr)\,\, \biggl(\frac{r_{T}}{0.06}\biggr)^{-1/4} \,  
\biggl(\frac{{\mathcal A}_{{\mathcal R}}}{2.41\times 10^{-9}}\biggr)^{-1/4} 
\nonumber\\
&\times& \, \biggl(\frac{h_{0}^2 \, \Omega_{R\,0}}{4.15 \times 10^{-5}}\biggr)^{-1/4}, \qquad\qquad m = 2,\, 3,\, .\,.\,.\, N-2,\, N-1.
\label{APPB5}
\end{eqnarray}
When all the different stages of expansion collapse to a single phase expanding exactly like radiation we have  $\sigma_{i} \to 1$ for all the $i= 1, \, .\, ., \,. \, N$ and the standard results 
illustrated in Fig. \ref{FIGU1} are recovered. When $N=2$ in Eqs. (\ref{APPB2})--(\ref{APPB3}) 
we have a single post-inflationary stage characterized by the two parameters, i.e. $\xi_{r}$ 
and $\sigma$. The next level of complication (illustrated for instance in Fig. \ref{FIGU6} ) stipulates that $N=3$ and since $a_{3} \to a_{r}$,  Eq. (\ref{APPB1}) also implies $\xi_{r} = \xi_{1} \, \xi_{2} = H_{r}/H_{1}$ while the relevant expansion rates in each stage are given by $\sigma_{1}$ and $\sigma_{2}$.

\end{appendix}


\newpage

\begin{thebibliography}{99}
\itemsep -2pt

\bibitem{one} F. Hoyle, Proc. of Solvay Conference {\it ``La structure et l'evolution de l'Univers''}, (ed. by R. Stoop, Brussels) p. 59 (1958). 

\bibitem{two} Ya. B. Zeldovich,  Sov. Phys. JETP {\bf 21}, 656 (1965) [Zh. Eksp. Teor. Fiz., {\bf 48} 986 (1965)].

\bibitem{three} E. R. Harrison, Phys. Rev. Lett. {\bf 18}, 1011 (1967).

\bibitem{four} K. S. Thorne, Astrophys. J. {\bf 148}, 51 (1967).

\bibitem{five} A. Lichnerowicz, {\it Relativistic hydrodynamics and magnetohydrodynamics},  (W. A. Benjamin Inc, New York, 1967).

\bibitem{six} M. P. Ryan and L. C. Shepley {\it Homogeneous Relativistic Cosmologies} (Princeton, NJ, PrincetonUniversity Press, 1975).

\bibitem{seven} M. Giovannini, Class. Quantum Grav. {\bf 35} 084003 (2018).

\bibitem{eight}  D.~N.~Spergel {\it et al.} [WMAP Collaboration],  Astrophys.\ J.\ Suppl. {\bf 148}, 175 (2003).

\bibitem{nine}  D.~N.~Spergel {\it et al.} [WMAP Collaboration],  Astrophys.\ J.\ Suppl.  {\bf 170}, 377 (2007).

\bibitem{ten} Y.~Akrami {\it et al.} [Planck Collaboration], Astron. Astrophys. {\bf 641}, A10 (2020).

\bibitem{eleven} J. D. Barrow, Phs. Rev. D {\bf 55}, 7451 (1997).

\bibitem{twelve} M. Giovannini, Phys. Rev. D {\bf 62}, 067301 (2000).

\bibitem{thirteen} K. Yamamoto,  Phys. Rev. D {\bf 85}, 043510 (2012).

\bibitem{elevena}  A.~A.~Starobinsky,  JETP Lett.\  {\bf 37}, 66 (1983). 

\bibitem{elevenb}  R.~M.~Wald,  Phys.\ Rev.\ D {\bf 28}, 2118 (1983).

\bibitem{fivea} S. Deser and C. Teitelboim, Phys. Rev. D {\bf 13}, 1592 (1976).

\bibitem{fiveb}  S. Deser, J. Phys. A  {\bf 15}, 1053 (1982).

\bibitem{elevend} J.~D.~Barrow, Phys. Rev. D \textbf{35}, 1805 (1987).

\bibitem{elevene}  J.~K.~Webb, V.~V.~Flambaum, C.~W.~Churchill, M.~J.~Drinkwater and
 J.~D.~Barrow, Phys. Rev. Lett. \textbf{82}, 884 (1999).

\bibitem{elevenf} J. D. Barrow, Phys. Rev. D \textbf{71}, 083520 (2005).

\bibitem{eleveng} C.~M.~Will, Living Rev. Rel. \textbf{17}, 4 (2014).

\bibitem{fourteen} R. D. Peccei and H. R. Quinn, Phys. Rev. Lett. {\bf 38}, 1440  (1977).

\bibitem{fourteena}   R. D. Peccei and H. R. Quinn, Phys. Rev. D {\bf 16}, 1791 (1977).

\bibitem{fifteena} J. Kim, Phys. Rep. {\bf 150}, 1 (1987); H.-Y. Cheng, {\em ibid}., {\bf 158}, 1 (1988).

\bibitem{fifteenb} G. G. Raffelt, Phys. Rep. {\bf 198}, 1 (1990);  Lect.\ Notes Phys.\  {\bf 741}, 51 (2008).

\bibitem{sixteen}  S. Carroll, G. Field and R. Jackiw, Phys. Rev. D {\bf 41},  1231 (1990).

\bibitem{seventeen}  W. D. Garretson, G. Field and S. Carroll, Phys. Rev. D {\bf 46}, 5346 (1992).

\bibitem{eighteen} G. Field and S. Carroll Phys.Rev.D {\bf 62}, 103008 (2000).

\bibitem{nineteen} M.~Giovannini, Phys.\ Rev.\ D {\bf 61}, 063502 (2000); Phys.\ Rev.\ D {\bf 61}, 063004 (2000).

\bibitem{twentytwo} M.~Giovannini and M.~E.~Shaposhnikov,  Phys.\ Rev.\ D {\bf 57}, 2186 (1998); 
M.~Giovannini,  Phys.\ Rev.\ D {\bf 88}, 083533 (2013);  Phys.\ Rev.\ D {\bf 89}, 063512 (2014).

\bibitem{twentythree} D. Kharzeev, L. McLerran and H. Warringa, Nucl. Phys. A {\bf 803}, 227 (2008); 
K.~Fukushima, D.~Kharzeev and H.~Warringa, Phys.\ Rev.\ D {\bf 78}, 074033 (2008).

\bibitem{twentythreea} D.~Kharzeev,  Annals Phys.\  {\bf 325}, 205 (2010).

\bibitem{twentyfour} B. Ratra,  Astrophys.\, J.\, Lett.  {\bf 391}, L1 (1992).

\bibitem{twentyfive} M.~Gasperini, M.~Giovannini, and G.~Veneziano, Phys. Rev. Lett. {\bf 75}, 3796 (1995); 
M.~Giovannini,  Phys.\ Rev.\ D {\bf 56}, 3198 (1997).

\bibitem{CC0} M. Giovannini, Phys.\ Rev.\  D {\bf 64}, 061301 (2001).

\bibitem{CC1}  A.~Akhtari-Zavareh, A.~Hojjati and B.~Mirza, Prog. Theor. Phys. {\bf 117}, 803 (2007).
 
\bibitem{CC2} D.~Seery, JCAP {\bf 08}, 018 (2009).

\bibitem{CC3} V.~Demozzi, V.~Mukhanov and H.~Rubinstein, JCAP {\bf 08}, 025 (2009).

\bibitem{CC4} M.~Karciauskas and D.~H.~Lyth, JCAP {\bf 11}, 023 (2010). 

\bibitem{CC5} I. Brown, Astrophys. J. {\bf 733}, 83 (2011).

\bibitem{CC6} T.~Fujita and S.~Mukohyama,  JCAP {\bf 1210}, 034 (2012).

\bibitem{CC7} F. Membiela, Nucl. Phys. B {\bf 885}, 196  (2014).

\bibitem{CC8} L.~Campanelli, Phys. Rev. D {\bf 93}, 063501 (2016).

\bibitem{CC9} P.~Qian, Y.~F.~Cai, D.~A.~Easson and Z.~K.~Guo, Phys. Rev. D {\bf 94} , 083524 (2016).

\bibitem{CC10}  R.~Koley and S.~Samtani, JCAP {\bf 04}, 030 (2017).

\bibitem{CC11} A.~Talebian, A.~Nassiri-Rad and H.~Firouzjahi, Phys. Rev. D {\bf 102}, 103508 (2020).

\bibitem{CC12} O. Sobol, A. Lysenko, E. Gorbar and S. Vilchinskii, Phys. Rev. D {\bf 102}, 123512 (2020).

\bibitem{CC13} M.~Giovannini, Phys. Rev. D \textbf{104}, 123509 (2021).

\bibitem{CC14} E.~V.~Gorbar, K.~Schmitz, O.~O.~Sobol and S.~I.~Vilchinskii, Phys. Rev. D \textbf{105},  043530 (2022).

\bibitem{fortytwo}  A. D. Sakharov,  Sov. Phys. JETP {\bf 22}, 241 (1966) [Zh. Eksp. Teor. Fiz. {\bf 49}, 345 (1965)].

\bibitem{fortythree} H. Jorgensen, E. Kotok, P. Naselsky, and I Novikov, Astron. Astrophys. {\bf 294}, 639 (1995).

\bibitem{SAK2} P.~J.~E.~Peebles and J.~T.~Yu,  Astrophys.\ J.\  {\bf 162} 815 (1970).
 
\bibitem{fortyseven} M.~Giovannini, JCAP {\bf 04}, 003 (2010).
 
\bibitem{fortyeight} S. Weinberg, Phys. Rev. D {\bf 77}, 123541 (2008).

\bibitem{fortynine}  M.~Giovannini, Phys. Lett. B {\bf 819}, 136444 (2021).

\bibitem{fifty} I. T. Drummond and S. J. Hathrell, Phys. Rev. D {\bf 22}, 343 (1980).

\bibitem{fiftyone} T. J. Hollowood and G. M. Shore, Nucl. Phys. B {\bf 795}, 138 (2008).

\bibitem{fiftytwo} M. Turner and L. Widrow, Phys. Rev. D {\bf 37}, 2743 (1988).

\bibitem{fiftythree} G. Feinberg and J. Sucher, Phys. Rev. A {\bf 2}, 2395 (1970).

\bibitem{fiftyfour} G. Feinberg and J. Sucher,  Phys. Rev. D {\bf 20}, 1717 (1979).

\bibitem{fiftyfive} M. Giovannini,   Phys. Rev. D {\bf 92}, 043521 (2015).

\bibitem{RT0} P.~Ade {\it et al.} [BICEP2 and Keck Array Collaborations],  Phys.\ Rev.\ Lett.\  {\bf 116}, 031302 (2016).

\bibitem{RT2}  N.~Aghanim \textit{et al.} [Planck Collaboration], Astron. Astrophys. {\bf 641}, A6 (2020).

\bibitem{abr1}  A. Erdelyi, W. Magnus, F. Oberhettinger, and F. R. Tricomi {\em Higher Trascendental Functions} (Mc Graw-Hill, New York, 1953).

\bibitem{abr2} M. Abramowitz and I. A. Stegun, {\em Handbook of Mathematical Functions} (Dover, New York, 1972).

\bibitem{mg04} M.~Giovannini, Int. J. Mod. Phys. D \textbf{13}, 391-502 (2004).

\bibitem{MAC} M.~Giovannini and N.~Q.~Lan, Phys. Rev. D \textbf{80}, 027302 (2009).

\bibitem{MAC2} H. K. Moffat, {\it Magnetic Field Generation in Electrically Conducting Fluids} (Cambridge University Press, Cambridge 1978).

\bibitem{MAC3}  E. Parker, {\it Cosmical Magnetic Fields} (Oxford University Press, Oxford, 1979).

\bibitem{FFF} T. Kobayashi and M. Sloth, Phys. Rev. D {\bf 100}, 023524 (2019).

\bibitem{SCH1a}  C. Stahl, E. Strobel, and S. S. Xue, Phys. Rev. D {\bf 93}, 025004 (2016).

\bibitem{SCH1b} R. Sharma and S. Singh, Phys. Rev. D {\bf 96}, 025012 (2017).

\bibitem{SCH1c} E. Bavarsad, S. P. Kim, C. Stahl, and S. S. Xue, Phys. Rev. D {\bf 97}, 025017 (2018).

\bibitem{SCH2} M. Giovannini, Phys. Rev. D {\bf 97}, 061301(R) (2018).

\bibitem{SCH3} M.~Giovannini, Class. Quant. Grav. \textbf{38}, 135018 (2021).

\bibitem{SCH4} M.~Giovannini, Eur. Phys. J. C \textbf{81},  503 (2021).


 \end{thebibliography}
\end{document}